\documentclass{cernyrep} 
\usepackage{texnames}
\usepackage[T1]{fontenc}
\usepackage{graphicx}
\usepackage{epstopdf}

\def\be{\begin{equation}}
\def\ee{\end{equation}}
\def\ba{\begin{eqnarray}}
\def\ea{\end{eqnarray}}

\def\lsim{\raise0.3ex\hbox{$\;<$\kern-0.75em\raise-1.1ex\hbox{$\sim\;$}}}
\def\gsim{\raise0.3ex\hbox{$\;>$\kern-0.75em\raise-1.1ex\hbox{$\sim\;$}}}

\begin{document}

\title{High-energy astroparticle physics}
 
\author{D. Semikoz}

\institute{APC, Paris, France}

\maketitle 

\begin{abstract}
In these three lectures I discuss the present status of high-energy astroparticle physics 
including Ultra-High-Energy Cosmic Rays (UHECR), high-energy gamma rays, 
and neutrinos. The first lecture is devoted to ultra-high-energy cosmic rays.  After a brief introduction to UHECR I discuss the acceleration of charged particles to highest energies in the astrophysical objects, their propagation in the intergalactic space, recent observational results by the Auger and HiRes experiments, anisotropies of UHECR arrival directions, and secondary gamma rays produced by UHECR. In the second lecture I review recent results on TeV gamma rays. After a short introduction to detection techniques, I discuss recent exciting results of the H.E.S.S., MAGIC, and Milagro experiments on the point-like and diffuse sources of TeV gamma rays. A special section is devoted to the detection of extragalactic magnetic fields with TeV gamma-ray measurements. Finally, in the third lecture I discuss Ultra-High-Energy (UHE) neutrinos. I review three different UHE neutrino detection techniques and show the present status of searches for diffuse neutrino flux and point sources of neutrinos.

\end{abstract}

\section{Ultra-high-energy cosmic rays}
\label{chapter:UHECR}
 
\subsection{Introduction}

\begin{figure}[htb]
\includegraphics[width=0.5\textwidth]{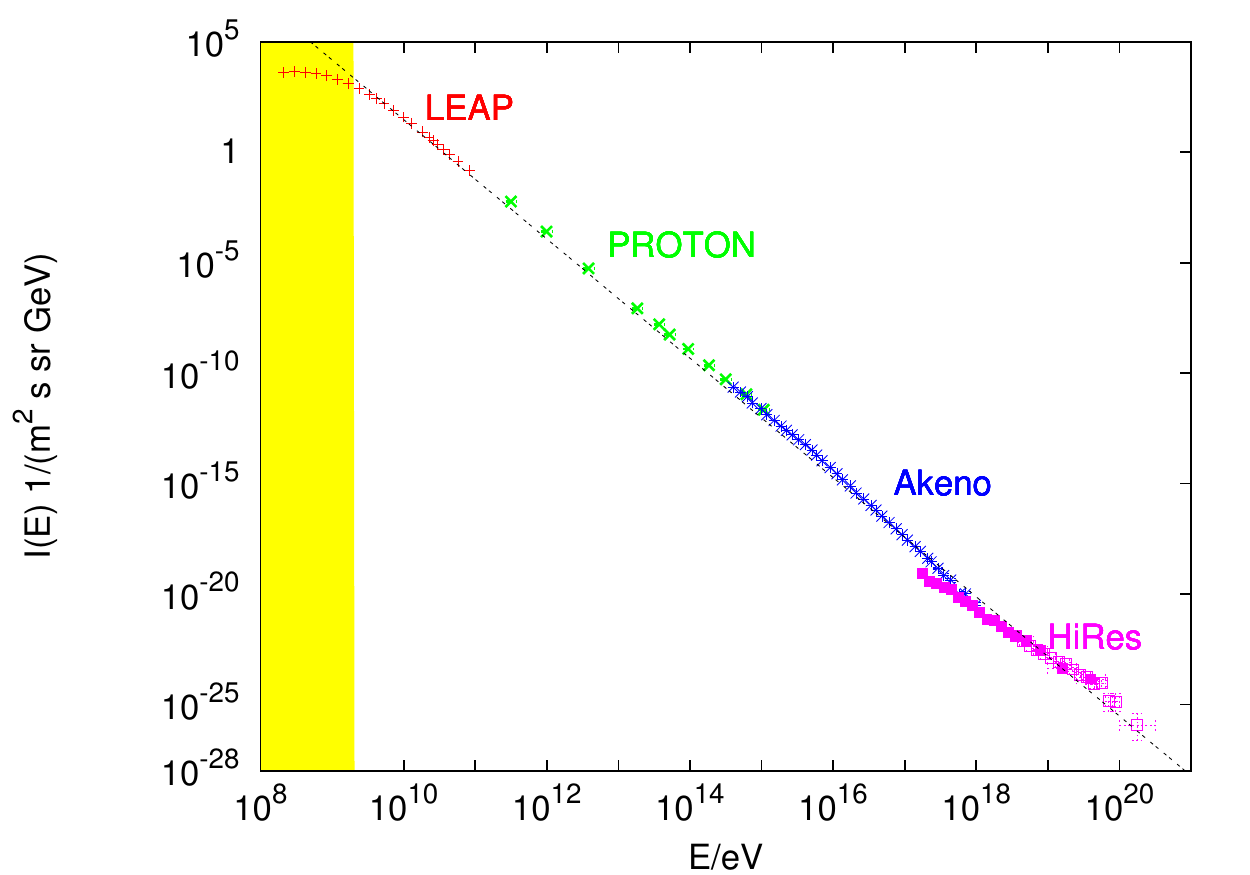}
\includegraphics[width=0.46\textwidth]{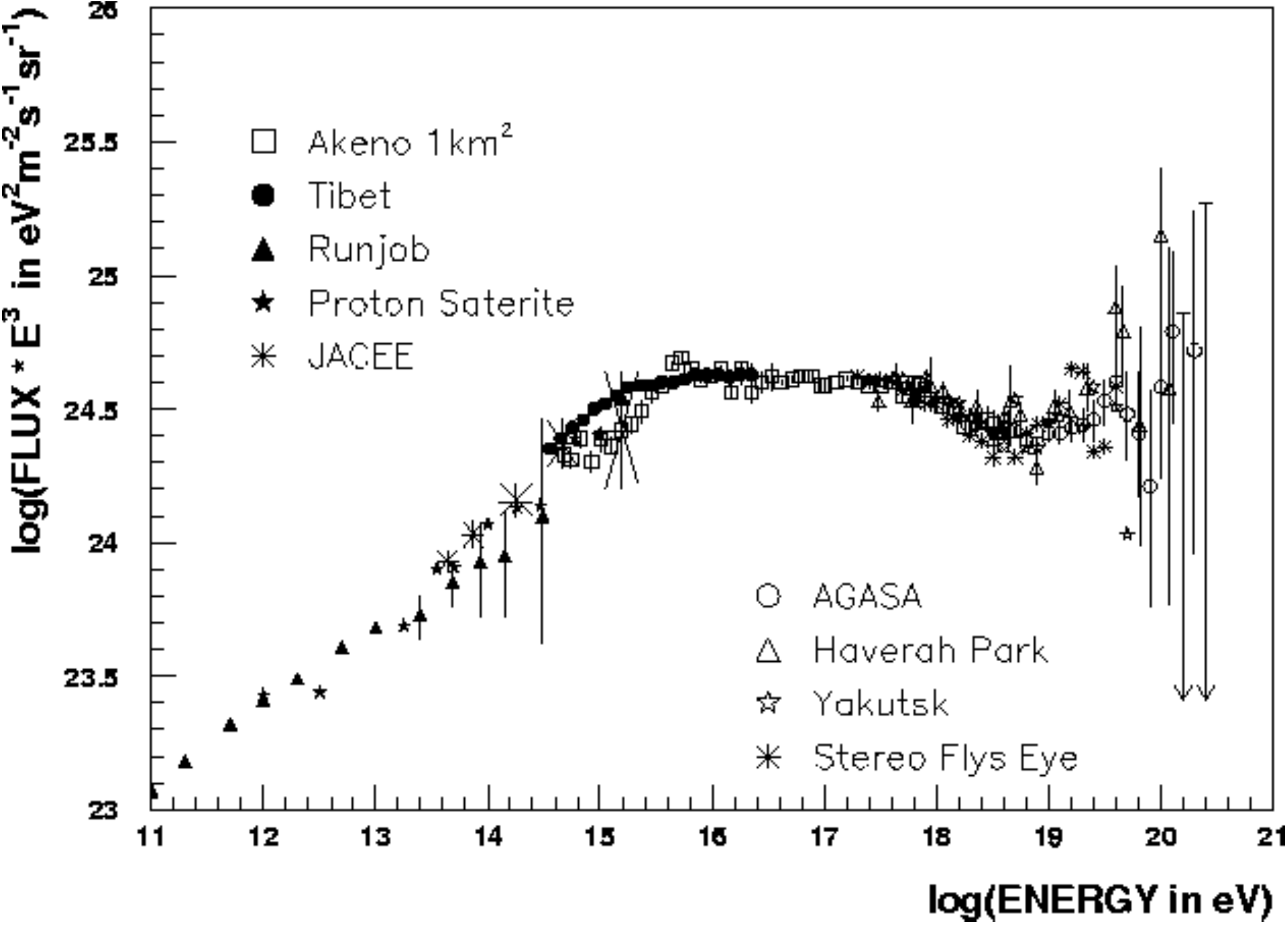}
\vspace{-0.3cm}
\caption{\label{cosmic ray}
{\bf Left:} The cosmic ray spectrum $I(E)$ as function of kinetic energy $E$, 
compiled using results 
from the LEAP, proton, Akeno, and HiRes experiments~\cite{exp,spectrum_HiRes}. 
The energy region 
influenced by the Sun is marked in yellow and an $1/E^{2.7}$ power-law is
also shown. {\bf Right:} The same spectrum at high energies $E>10^{11}$ eV multiplied by $E^3$~\cite{NaganoWatson}.
Spectrum changes are called the `knee' at $10^{15}$ eV and the `ankle' at $10^{19}$ eV. 
}
\end{figure}

 Particles coming from space to the atmosphere of the Earth historically were called 
cosmic rays. 
Most cosmic rays,  however, are not `rays' or photons, but charged particles, protons and nuclei.
 Real high-energy gamma rays coming from space to the Earth
  are only a small  fraction of total flux, and they will be discussed in Section~\ref{chapter:gamma}. 
The measured spectrum of cosmic rays from 100 GeV  to highest energies $E>10^{20}$ eV 
is presented in Fig.~\ref{cosmic ray} (left).  The yellow strip at low energies presents the contribution of the Sun.
The remaining spectrum can be fitted with a single power law $1/E^{2.7}$ up to highest energies. The main contribution to
it above 100 GeV gives galactic sources.
After multiplication of the spectrum on the energy cube, one can see changes of power law in 
 Fig.~\ref{cosmic ray} (right).  At $E>10^{15}$ eV the spectrum  becomes steeper. This change in the spectrum called the `knee'
 and associated energy $E=10^{15}$ eV is the maximum energy up to which  galactic sources  accelerate cosmic rays.
 The next  change of the spectrum is located at $E=3 \cdot 10^{18}$~eV and has two possible interpretations. 
 Either this is the place where extragalactic sources start to dominate or it is the result of pair- production energy loss by extragalactic protons
 (see Section \ref{sec:propag}). At the end of the spectrum there is a cutoff, which was not seen in the old experiments
 presented in Fig.~\ref{cosmic ray} (right) due to small statistics, 
 but it was observed recently by the HiRes ~\cite{spectrum_HiRes} and Auger ~\cite{auger_spectrum2009} experiments.

In this lecture I briefly discuss the theory and observations of 
Ultra-High Energy Cosmic Rays (UHECR), the highest-energy particles 
 measured on Earth with energy $E>10^{18}$ eV.
Such particles, protons and nuclei,  can be accelerated in astrophysical objects,
propagate through intergalactic space, losing energy in the interactions with 
Cosmic Microwave Background (CMB).  UHECR are charged particles. Therefore they are also  deflected in the Galactic and intergalactic magnetic fields on the way from the source to the Earth. 
For a more detailed introduction to UHECR I recommend recent lectures by M.~Kachelriess~\cite{Kachelriess:lectures}.

There are several important scales commonly used in astroparticle physics. Distance is usually measured in parsecs,
$\rm{1~pc} = 3 \cdot 10^{18}$~cm. Corresponding larger units are kiloparsec $\rm{1~kpc} = 10^3 \rm{~pc}$ and megaparsec $\rm{1~Mpc} = 10^6 \rm{~pc}$. Energy at highest energies is usually expressed in units of  $\rm{EeV} = 10^{18}$~eV.

The plan of this lecture is as follows. In Section \ref{sec:accel} I shall discuss possible acceleration mechanisms of cosmic rays and astrophysical objects which potentially can be their sources. In Section \ref{sec:propag} I present the 
main energy loss processes for UHECR particles and briefly discuss their deflection in the magnetic fields.
In Section \ref{sec:observ} I sum up recent observational results from the Pierre Auger Observatory and other experiments.
In Section \ref{sec:anisot} results on anisotropy at highest energy are discussed. In Section \ref{sec:second} I review expectations on  secondary photons and neutrinos from UHECR protons.  Results are summed up in Section \ref{sec:UHECRsum}.

\subsection{Acceleration}
\label{sec:accel}

There are several possible acceleration mechanisms 
that can work in astrophysical objects. These include first-order 
Fermi acceleration on the shocks in plasma or acceleration in
the potential difference, which we call one-shot acceleration below.
However, in any case, 
the Larmor radius of a particle does not exceed the accelerator size,
otherwise the particle escapes from the accelerator and cannot gain energy
further. This  criterion is called the Hillas condition~\cite{Hillas:1985is} and sets the limit
\begin{equation}
{\cal E} \le {\cal E}_{\rm H}=qBR
\label{eq:Hillas}
\end{equation}
for the energy ${\cal E}$ gained by a particle with charge $q$ in the
region of size $R$ with the magnetic field $B$.

\begin{figure}[htb]
\includegraphics[width=0.45\textwidth]{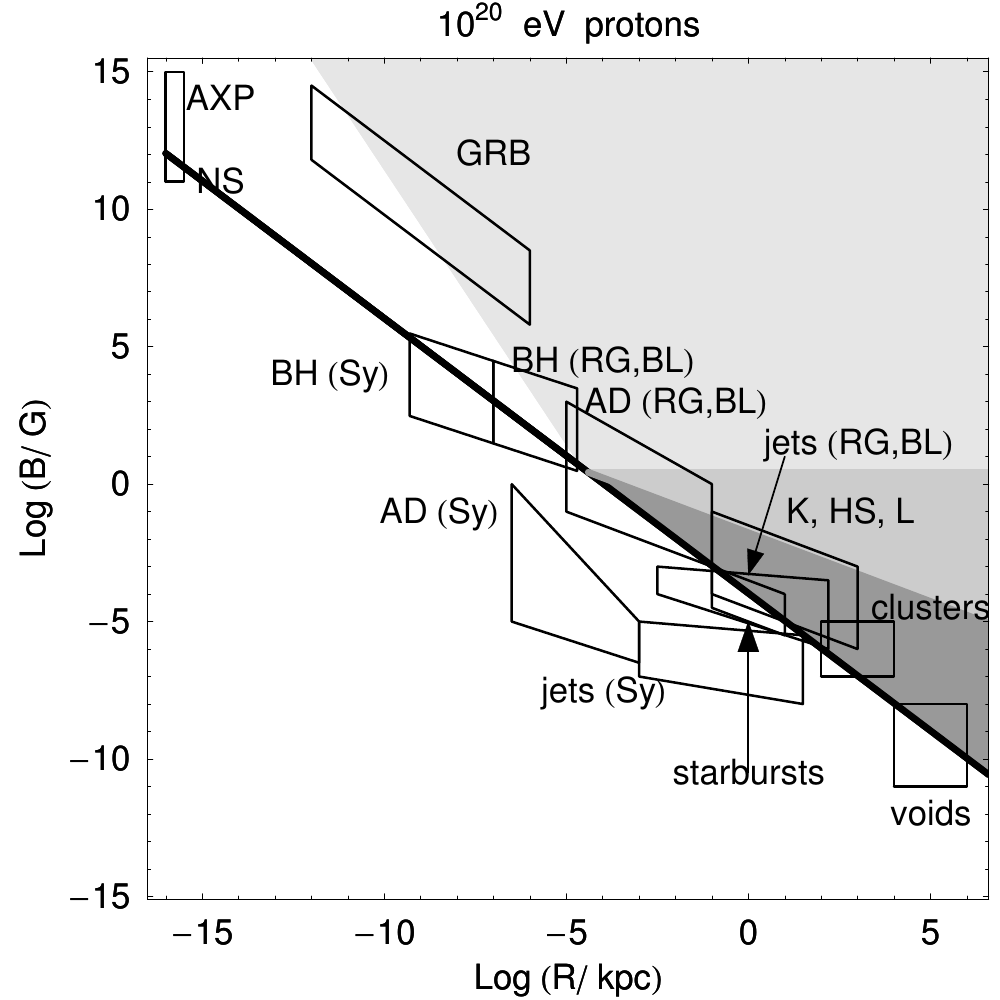}
\includegraphics[width=0.45\textwidth]{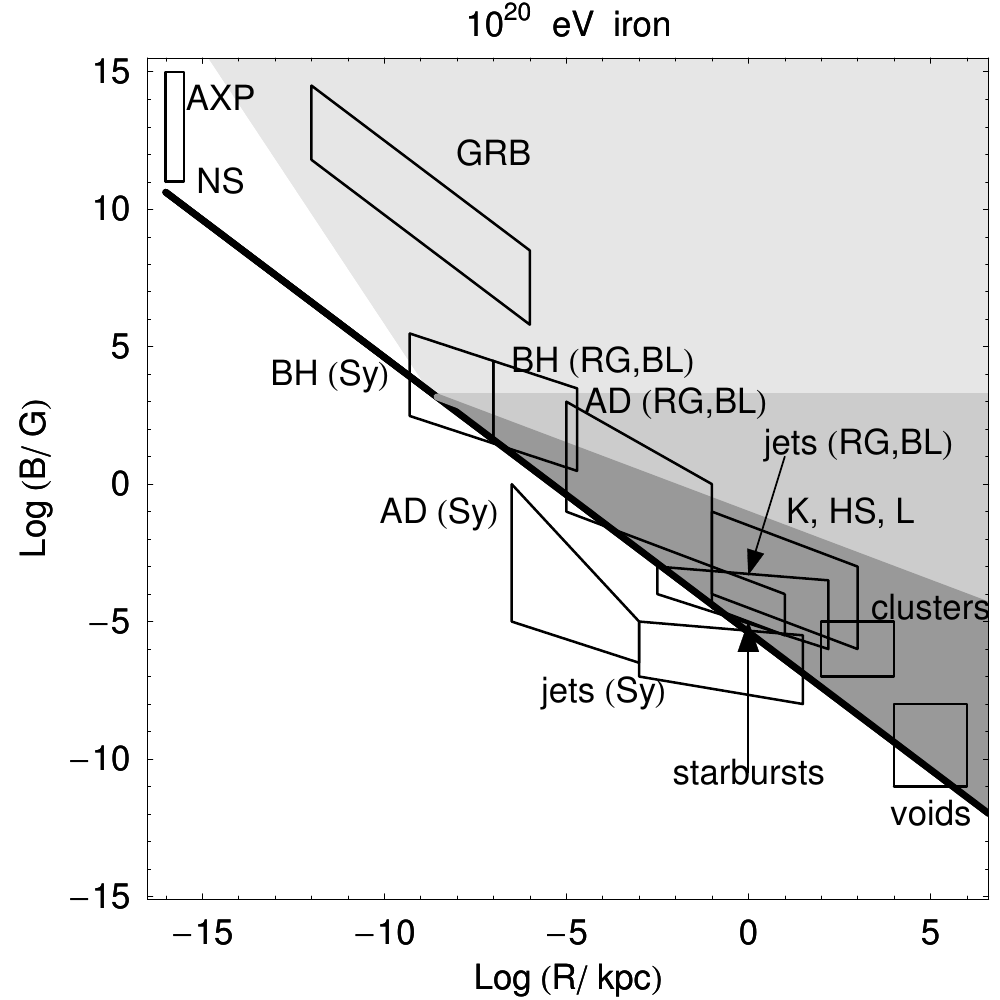}
\vspace{-0.3cm}
\caption{\label{accel}
The Hillas plot with constraints from geometry and radiation losses for
$10^{20}$~eV protons (left) and iron (right). The thick line represents the lower boundary of the
area allowed by the Hillas criterion, Eq.~(\ref{eq:Hillas}). Shaded areas are allowed by the
radiation-loss constraints as well: light grey corresponds to one-shot
acceleration in the curvature-dominated regime only; grey allows also for
one-shot acceleration in the synchrotron-dominated regime; dark grey allows
for both one-shot and diffusive (e.g.,\ shock) acceleration.  
}
\end{figure}

The maximum energy of the accelerated particle  can be restricted even more  than required by 
Eq.~(\ref{eq:Hillas}) if one takes into account energy losses during acceleration. 
Unavoidable losses come from particle emission in the external magnetic field, 
which can be either synchrotron-dominated if the velocity of the particle is not parallel to the magnetic field,
or curvature-dominated in the opposite case. 

In Fig.~\ref{accel}  in
 the plane magnetic field versus acceleration region size, 
   the Hillas condition ~Eq.~(\ref{eq:Hillas}) is shown by a thick black line. The left figure is for protons and the right one for 
   iron nuclei.  
  Possible acceleration in different  astrophysical objects is shown with thin solid figures.
  Notations are the following: NS are neutron stars, GRB are gamma-ray bursts, BH are black holes, AD are accretion disks, 
  jets are jets in active galaxies, K and HS are knots and hot spots in the jets, L are lobes of radio galaxies,
  clusters  are clusters of galaxies, starbursts are starburst galaxies,  voids are
  voids in large-scale structure. Additional notations in brackets are subtypes of active galaxies: Sy for  Seyfert  galaxies,
  BL for BL Lac galaxies and RG for radio galaxies. 
Only objects above the Hillas line have the potential possibility to accelerate particles to $10^{20}$ eV.  This is a necessary
condition, but not enough for a specific acceleration mechanism. As seen from Fig.~\ref{accel}, for example, neutron stars
cannot accelerate particles to highest energies under any condition, while shock acceleration 
would work only for objects presented in the dark grey corner of this plot.

\subsection{Propagation}
\label{sec:propag}

Owing to expansion of the Universe, particles which come from sources at redshift $z$
lose their energy as
\begin{equation}
E_P \rightarrow    E'_P = E_P/(1+z)~.
\label{redshift}
\end{equation}
A typical energy loss distance, i.e., distance at which particles lose a significant part of their energy 
 for this process, is of the order of $z \sim 1$ (50\% of energy), 
i.e., $R \sim  3~\mbox{Gpc} = 10^{28}$ cm.

As well as during propagation in the intergalactic space, protons 
lose energy due to two other main processes of interactions with Cosmic Microwave Background (CMB) photons.
Those are electron--positron pair production and pion production. In both processes massive particles have to be produced and they have threshold energy. Since the typical energy of CMB photons is very small, 
$\epsilon_{CMB}=6 \times 10^{-4}$ eV, the threshold for those processes is very high.
Only at energies above $E_{th} =  m_e^2/\epsilon_{CMB} \sim 10^{15}$~eV does the electron--positron pair-production process
become important:

\begin{equation}
P + \gamma_{CMB} \rightarrow    P + e^+ + e^- .
\label{pair_production}
\end{equation}

The typical energy loss distance for this process is 

\begin{equation}
R= \frac{M_P}{2m_e} \frac{1}{ \sigma_{P{e^+e^-}}n_{CMB}}  = 600 ~\mbox{Mpc} = 2 \times 10^{27}~ \mbox{cm}~,
\end{equation}
where $n_{CMB}=400/\mbox{cm}^3$ is the density of CMB photons, $\sigma_{P{e^+e^-}} \approx 10^{-27}/\mbox{cm}^2$
is the proton-pair production cross-section. The factor $M_P/2m_e$ comes here from the fact that in every interaction a
proton loses only a tiny fraction of its energy proportional to the proton/electron mass ratio.

\begin{figure}[ht]
\begin{center}
\includegraphics[width=0.45\textwidth,angle=0]{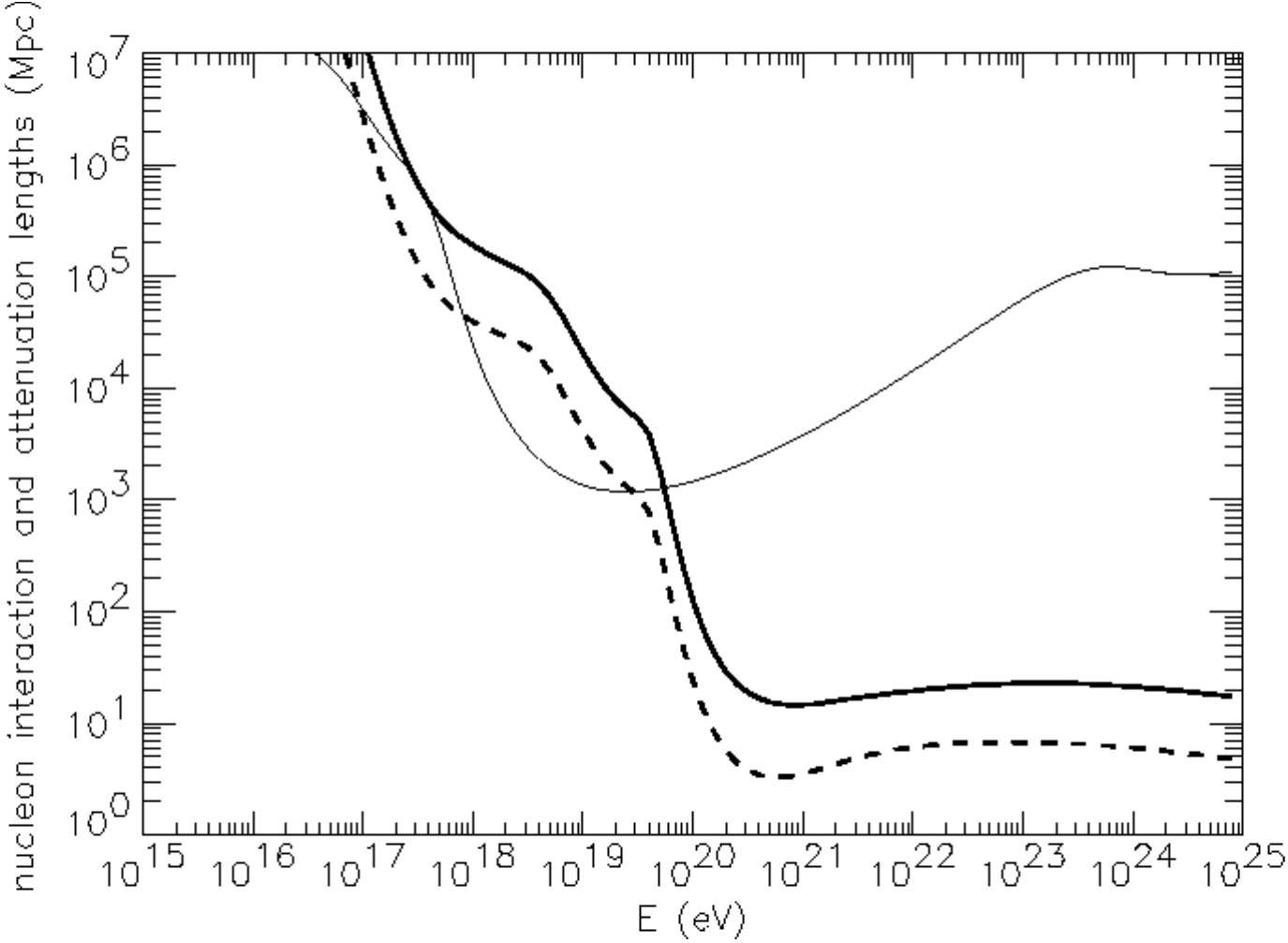}
\includegraphics[width=0.47\textwidth,angle=0]{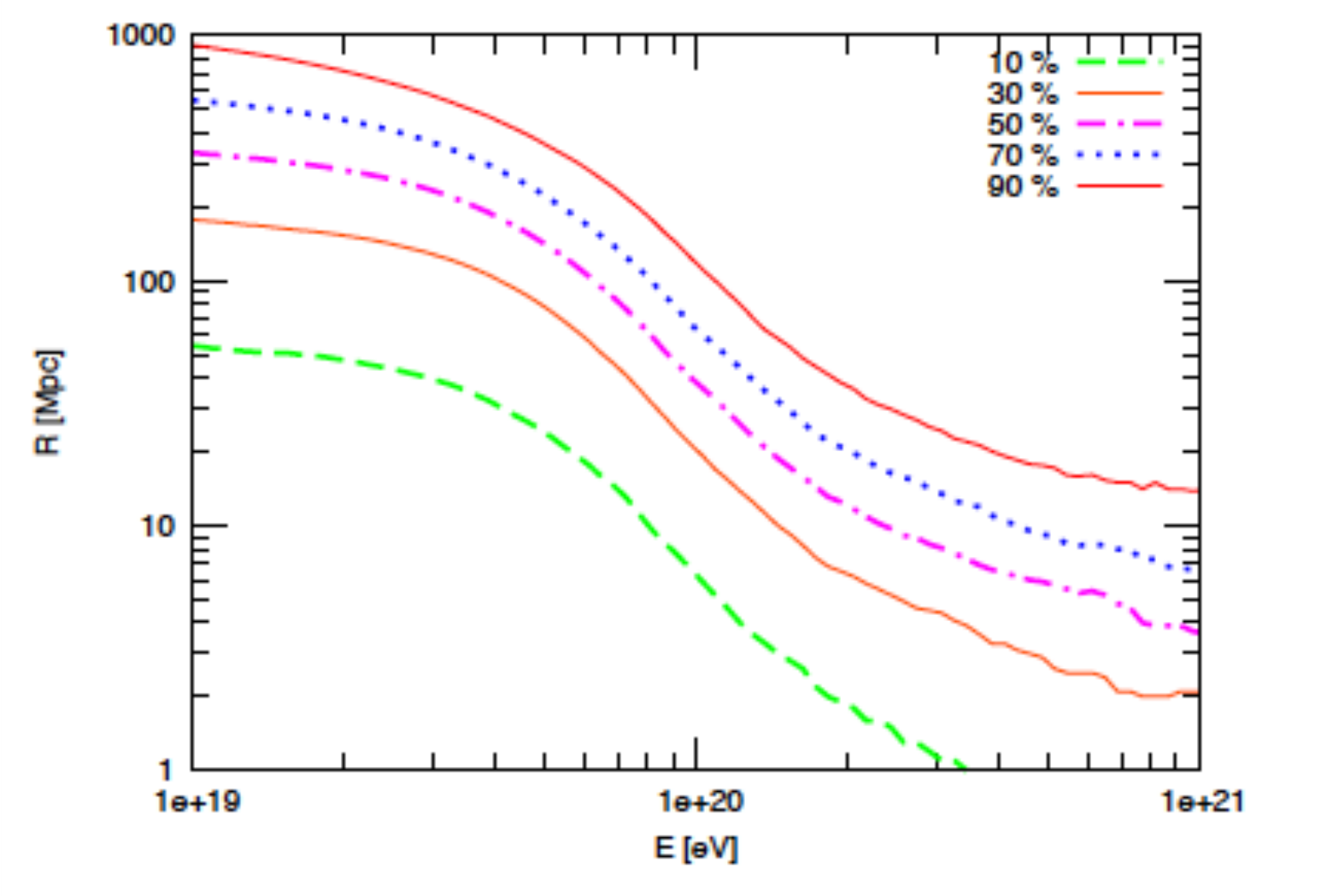}
\end{center}
\vspace{-0.3cm}
\caption{\label{horizon}
{\bf Left:} Nucleon interaction length as function of energy from Ref.~\cite{sigl_review}.  Attenuation due to pion production Eq.~(\ref{pion_production}) presented by the thick solid line, the same for pair production Eq.~(\ref{pair_production}) presented by the 
thin solid line.
{\bf Right:} Horizon (maximal distance) from which  protons with given or higher  energy can arrive.
Lines for 10\%, 30\%, 50\%, 70\% and 90\% of events are shown~\cite{Kachelriess:2007bp}. 
}
\end{figure}

At energies above threshold $E_{th} \approx m_\pi M_P/\epsilon_{CMB} = 10^{20}$ eV, the pion production
process dominates energy losses. This process for cosmic rays was first considered by Greizen, Zatsepin, and Kuzmin in 1966~\cite{GZK}  and is now named the  GZK process.
\begin{equation}
P + \gamma_{CMB} \rightarrow  \left\{ \begin{array}{c} P + \pi^0  + \sum_i \pi_i\\ N+ \pi^+ + \sum_i \pi_i
\end{array} \right.
\label{pion_production}
\end{equation}

The typical energy loss distance for this process is 

\begin{equation}
R= \frac{M_P}{m_\pi} \frac{1}{ \sigma_{P\gamma}n_{CMB}}  = 100 ~\mbox{Mpc} = 3 \times 10^{26}~ \mbox{cm}~,
\label{p_atten}
\end{equation}
where  $\sigma_{P\gamma} \approx 6 \times 10^{-28}/\mbox{cm}^2$
is the proton pion production cross-section. The factor $M_P/m_\pi$ comes here from the fact that in every interaction the 
proton loses only 15--20\% of its energy proportional to the proton/pion mass ratio.
Note that at higher energies the dominating process in Eq.~(\ref{pion_production}) is multi-pion production,
in which the proton loses 50\% of its energy in every interaction, however, the  cross-section for this process 
$\sigma_{\sum \pi} = 10^{-28}/\mbox{cm}^2$ is a factor 6 lower than single pion production.

None of the above processes allows a proton with high energy to come from a very large distance. 
The distance from which protons can come as a function of energy is presented in Fig.~\ref{horizon} (left)~\cite{sigl_review}.
The interaction length for pion production [Eq.~(\ref{pion_production})] is shown by the dashed line.
 Attenuation due to pion production [Eq.~(\ref{p_atten})] is presented by the thick solid line, the same for pair production [Eq.~(\ref{pair_production})] is presented by the thin solid line. Figure~\ref{horizon} (left) shows the  average distance travelled 
 by a single particle. However, for searches of UHECR sources the important question is the maximum distance or horizon 
from which UHECR can come to the detector. In Fig.~\ref{horizon} (right) we present the horizon as a function
of minimal proton energy. The lines 10\%, 30\%, 50\%, 70\% and 90\% show the fraction of events
which come from a given distance. For example, 90\% of events with $E>10^{20}$ eV should come  from distances
$R<100$ Mpc. This distance is sometimes called the GZK distance, because energy losses in this case are dominated 
 by the GZK process of Eq.~(\ref{pion_production}).

The dominant loss process for nuclei of energy $E\gsim 10^{19}\,$eV
is photodisintegration $A+\gamma\to (A-1)+N$ in the CMB and the infrared
background due to the giant dipole resonance~\cite{stecker69}. The threshold for this
reaction follows from the binding energy per nucleon, $\sim 10\:$MeV.
Photo-disintegration leads to a suppression of the flux of nuclei
above an energy that varies between $3\times 10^{19}\:$eV for He and
$8\times 10^{19}\:$eV for Fe.

\begin{figure}[htb]
\begin{center}
\includegraphics[width=0.38\textwidth,angle=0]{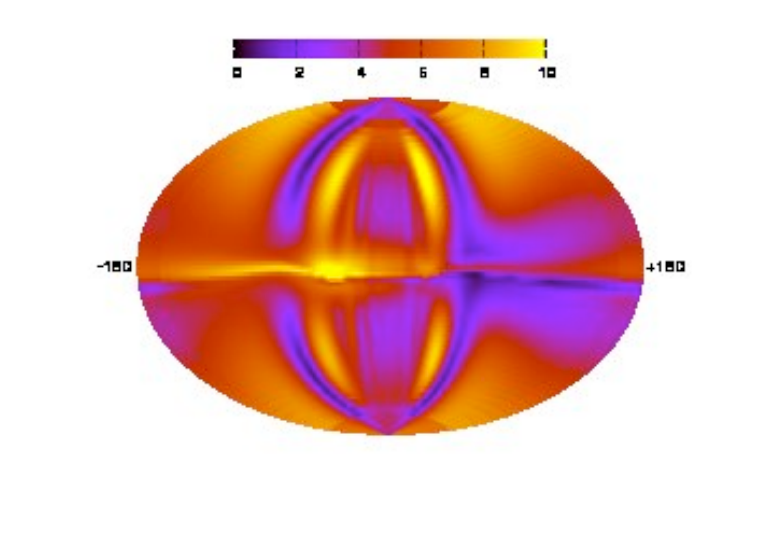}
\includegraphics[width=0.58\textwidth,angle=0]{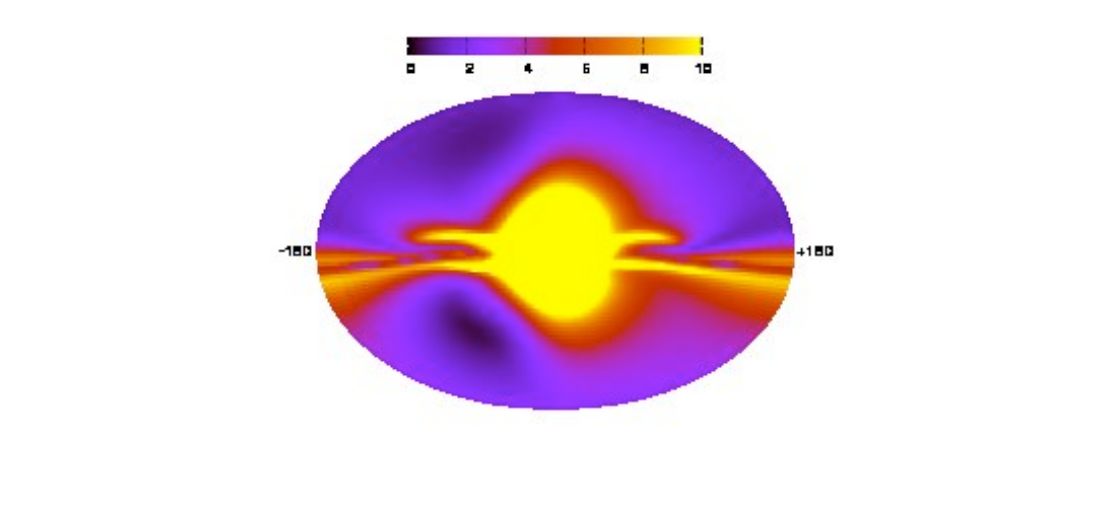}
\end{center}
\vspace{-0.3cm}
\caption{\label{gal_mf}
Sky map of the UHECR proton deflections for energy $E=40$ EeV in two different models of Galactic magnetic field from Ref.~\cite{{galactic}}.   The colours show deflections from 0 to 10 degrees.
}
\end{figure}

\begin{figure}[htb]
\begin{center}
\includegraphics[width=0.46\textwidth,angle=0]{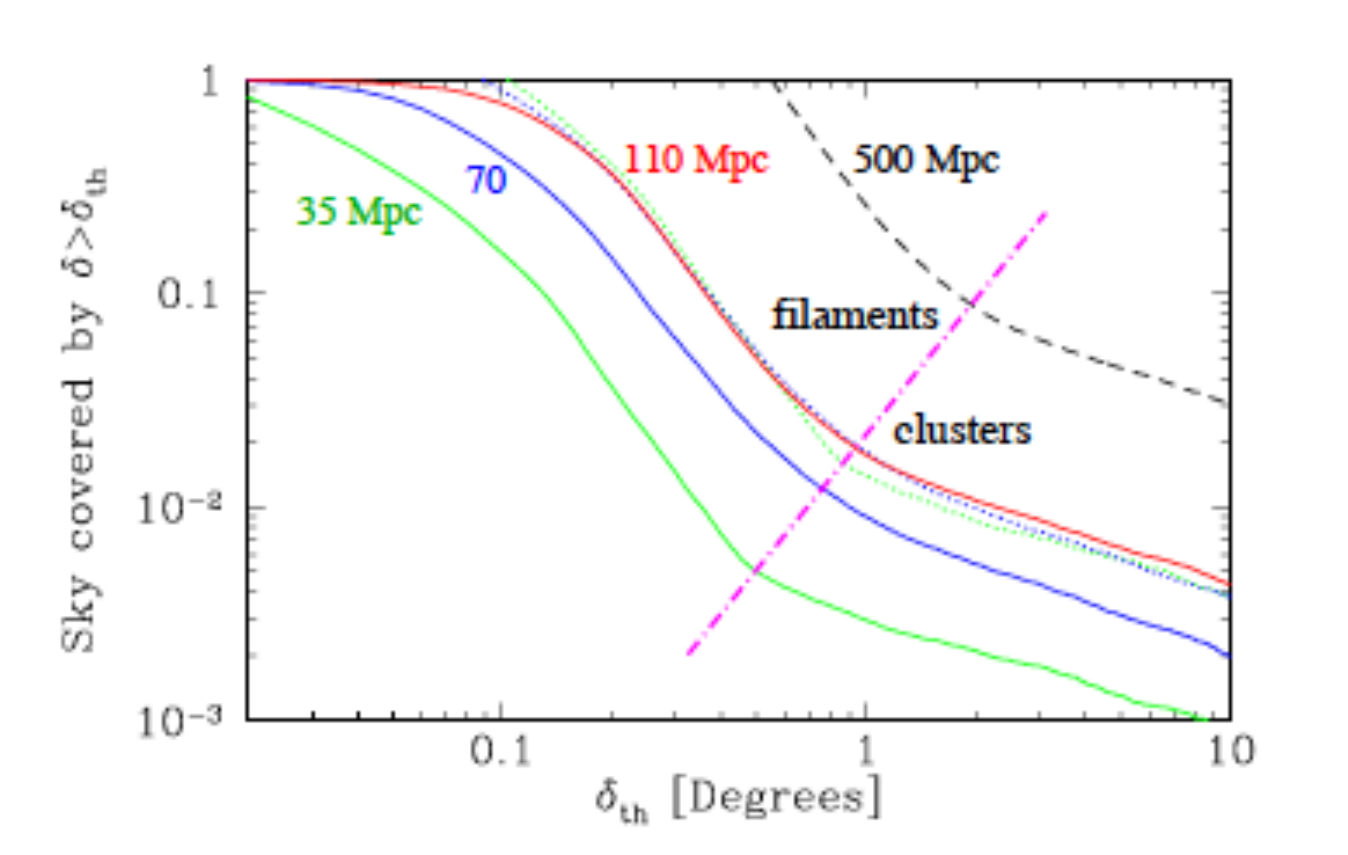}
\includegraphics[width=0.4\textwidth,angle=0]{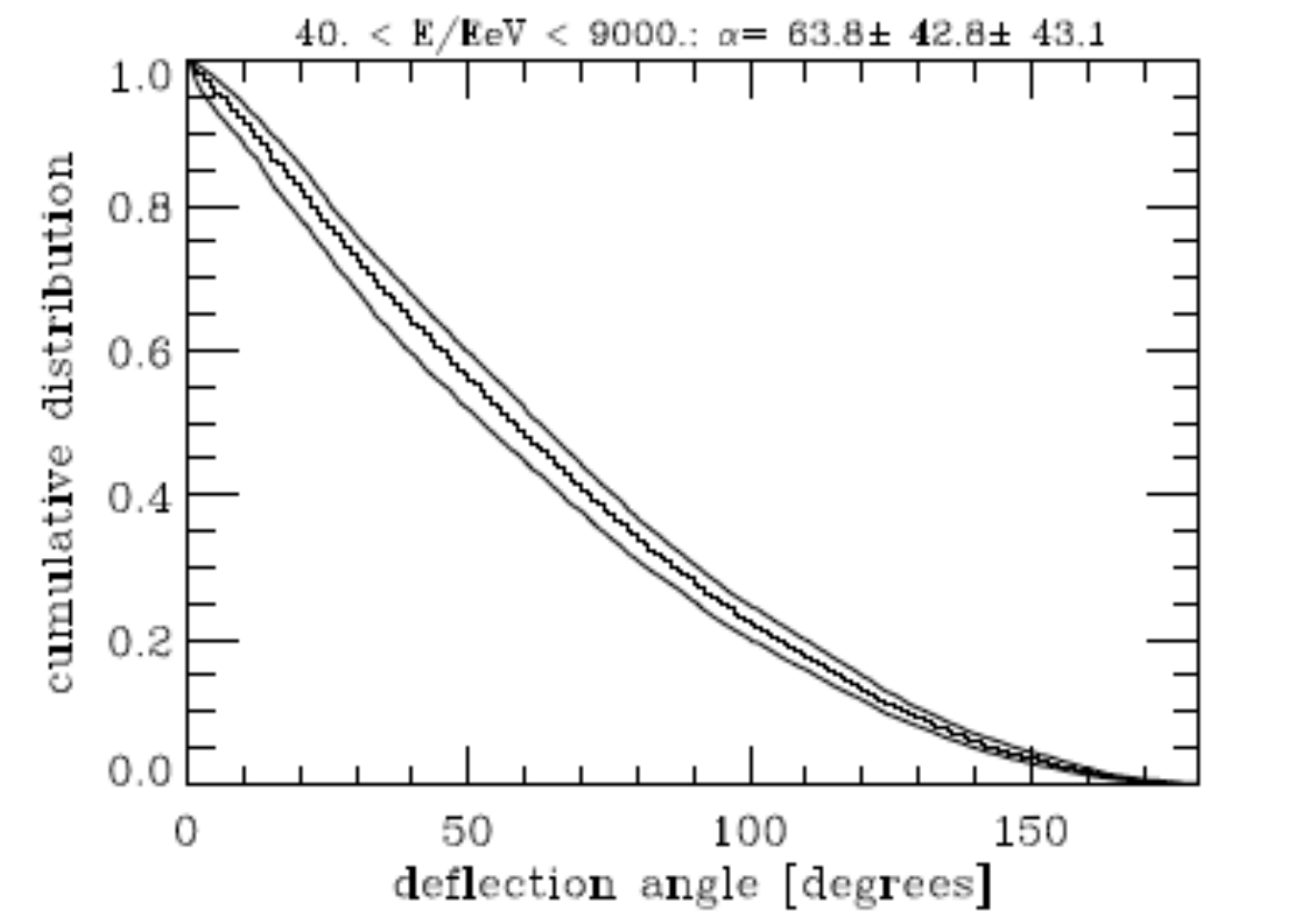}
\end{center}
\vspace{-0.3cm}
\caption{\label{extragal_mf}
Fraction of the sky in which the deflection in the extra-Galactic magnetic field is bigger than the given value. {\bf Left:} Constraint simulation of K.~Dolag et al. ~\cite{Dolag:2004kp}.  {\bf Right:} Simulation of Sigl et al.~\cite{Sigl:2004yk}.
}
\end{figure}

Since UHECR are charged particles, they not only lose energy in the interactions with background photons, but also when deflected by Galactic and intergalactic magnetic fields. 

The magnetic field of the Milky Way galaxy is conventionally modelled as a sum of the regular and turbulent components of the field in the disk and halo of the Galaxy. This means that the deflection in the Galactic field $\theta_{\rm Gal}$ is a superposition of at least four terms:
\begin{equation}
\label{UHECR_gal}
\theta_{\rm Gal} =  \theta_{\rm Disk}^{\rm regular} + \theta_{\rm Disk}^{\rm turbulent} +\theta_{\rm Halo}^{\rm regular}+\theta_{\rm Halo}^{\rm turbulent}~.
\end{equation}

The deflection angle of UHECR in a regular magnetic field after propagation of distance $D$ is given by:
\begin{equation}
\label{UHECR_reg}
\theta^{\rm regular} \simeq \frac{ZeB_{\bot} D}{E_{\rm UHECR}}  \simeq 5^\circ Z\left[\frac{E_{\rm UHECR}}{4 \cdot 10^{19}\mbox{ eV}}\right]^{-1}\left[\frac{B_\bot}{2 \cdot 10^{-6}\mbox{ G}}\right]\left[\frac{D}{2\mbox{ kpc}}\right]~,
\end{equation}
where where $B_\bot$ is the magnetic field component orthogonal to the line of sight, $E_{\rm UHECR}$ is the particle energy, and $Z$ is the atomic charge. 
In the case of deflection by the turbulent field on the distance $D$ much larger than the correlation length of the field $\lambda_B$
and where the deflection angle is small, the deflection is given by
\begin{eqnarray}
\label{UHECR_turb}
\theta^{\rm turb}&\simeq&\frac{1}{\sqrt{2}} \frac{ZeB_{\bot} \sqrt{D \lambda_B}}{E_{\rm UHECR}}\simeq 1.2^\circ Z\left[\frac{E_{\rm UHECR}}{4 \cdot 10^{19}\mbox{ eV}}\right]^{-1} \left[\frac{B_\bot}{4 \cdot 10^{-6}\mbox{ G}}\right]  \left[\frac{D}{2\mbox{ kpc}}   \right]^{1/2}\left[\frac{\lambda_B}{50\mbox{ pc}}\right]^{1/2}
\end{eqnarray}

The deflection  angles by regular and turbulent components of Galactic Disk and Halo, $\theta^{\rm Disk}_{\rm regular}$ and $\theta^{\rm Disk}_{\rm turbulent}$ are  given by  Eqs. (\ref{UHECR_reg}), (\ref{UHECR_turb}). Contributions of the Halo fields   $\theta^{\rm Halo}_{\rm regular}$ and $ \theta^{\rm Halo}_{\rm turbulent}$ are less certain, but the result in deflections is usually assumed to be less than the one for disk fields.

 Deflections of UHECR by the regular field in the disk $\theta_{\rm Disk}^{\rm regular}$ have been studied in many theoretical models. 
 A sky map of the deflections of  UHECR with  $E=40$ EeV in two different models  is presented in Fig.~\ref{gal_mf}. 
Despite both models being consistent on average with the expectation of Eq.~(\ref{UHECR_reg}), predictions in any given direction are 
strongly model-dependent.

Extragalactic magnetic fields are unknown except in the centres of galaxy clusters. Therefore one has to use theoretical models for the 
evolution of the magnetic fields. In such models magnetic fields follow the formation of large-scale structures.
Because the structure of extragalactic magnetic fields is very non-trivial with very large fields near large-scale structures and tiny fields in the voids,
one cannot use Eq.~(\ref{UHECR_turb}) everywhere. Instead one can introduce a fraction of the sky with deflections lower than a given value.  Unfortunately, modern models give a very broad range of predictions. In Fig.~\ref{extragal_mf} we show calculations by two groups
that show very different results.   The group of K.~Dolag et al. made constraint simulations of local large-scale structures within 100 Mpc around the Earth~\cite{Dolag:2004kp}.
This means that all big structures such as clusters of galaxies are located in exactly the same places as in the real sky. 
Also the density of points in this simulation is adaptive with more points at clusters and fewer on filaments. 
 The results of this simulation are shown in Fig.~\ref{extragal_mf} (left). According to this simulation, at 100 Mpc from the Earth 
 only in 2\% of the sky are deflections bigger than $\theta^{EGMF} = 1^\circ$. Those places are centres of  galaxy clusters.
In contrast the simulation of G.~Sigl et al.~\cite{Sigl:2004yk} uses a uniform grid with more points on filaments and fewer on clusters.
Unfortunately this simulation is not constrained and thus cannot be directly compared to local large-scale structures. 
In this simulation $\theta^{EGMF} > 50^\circ$ in  60\% of the sky. 

Let us note here that propagation energy losses are extremely important for the understanding of experimental 
results on spectrum and composition discussed in the next section. Deflections in turn are a key issue 
in the anisotropy studies in Section \ref{sec:anisot}.

\subsection{Observations}
\label{sec:observ}

 In this section we shall discuss the present status of UHECR observations.
\begin{figure}[ht]
\begin{center}
\includegraphics[width=0.5\textwidth,angle=0]{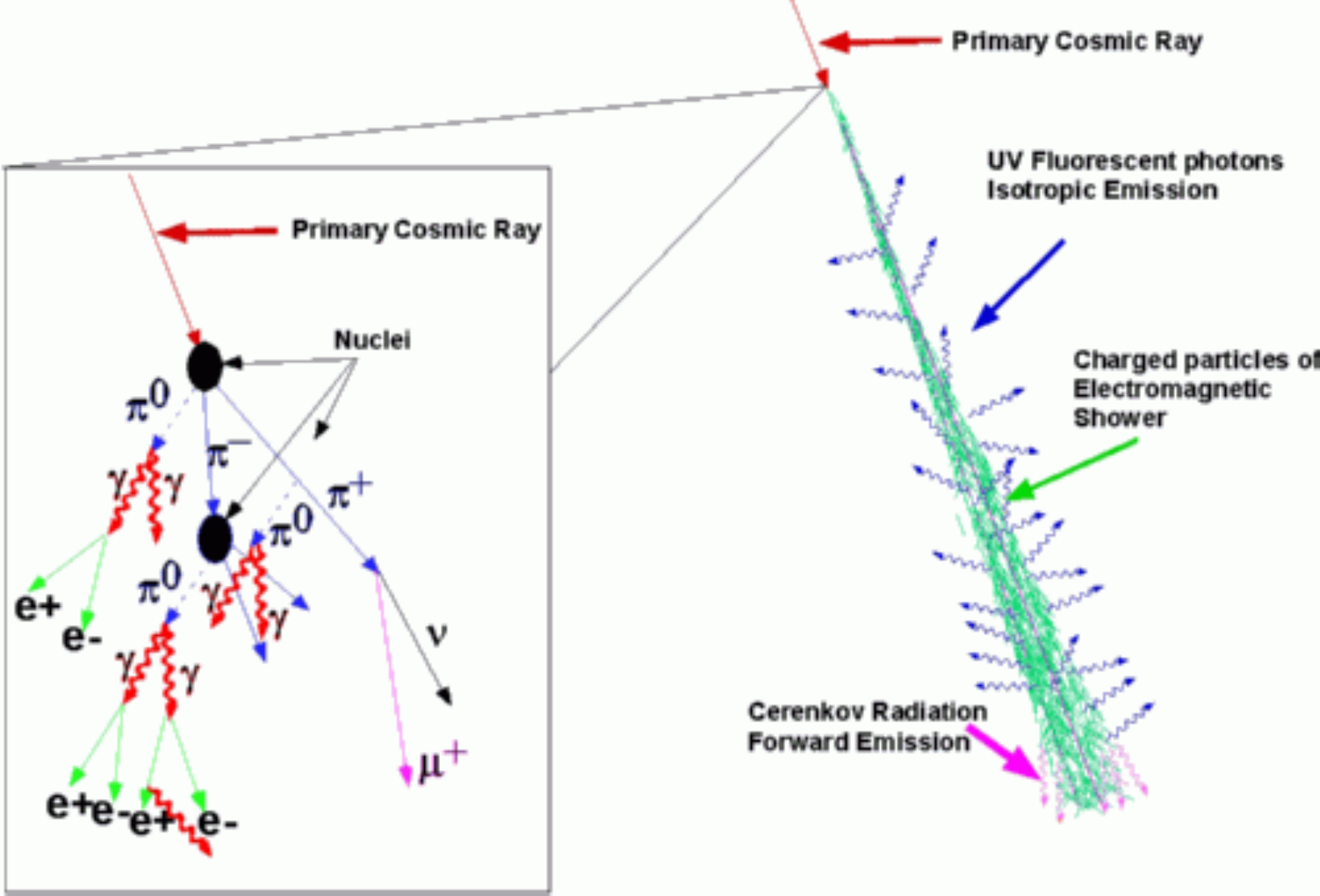}
\end{center}
\vspace{-0.3cm}
\caption{\label{shower}
Extensive air shower produced by a UHECR particle in the atmosphere}
\end{figure}

We start with the detection of UHECR in the atmosphere. 
The typical column density of the atmosphere is 1000~g/cm$^2$ and in 1g  there are $N=10^{24}$ protons,
while a strong cross-section is $\sigma_{PP} \sim 10^{-25} ~\mbox{cm}^2$.  Thus UHECR protons or nuclei
should interact within the atmosphere many times before they reach the Earth's surface.
 In these interactions it would produce extensive air showers.
An example of such a shower is illustrated in Fig.~\ref{shower}.
 After first interaction the
primary proton or nuclei  would produce a large number of pions. Neutral pions would start an electromagnetic cascade,
while charge pions would produce muons. At the maximum of shower development one expects $N=10^{9-10}$ particles
distributed in an area with a radius of a few kilometres. At this point the shower mostly consists of 10~MeV electrons and photons
and only 5--10\% of its energy is in muons. If the shower is not vertical, the column density increases as $1/\cos(\theta_{zenith})$
and reaches 2000 g/cm$^2$ for $\theta_{zenith}=60^\circ$. At such a depth all electromagnetic components of the shower
disappear and the shower would consist of muons only.  Another way to detect the shower is to look for fluorescent UV light of nitrogen atoms in the atmosphere. This method is called `calorimetric' and gives a 3-dimensional image of the shower.
The main problem with this method is that detection is possible only on a moonless night, making the duty cycle of such 
detectors possible only 10--15\% of the time. Finally one can detect direct Cherenkov light of the charged particles,
but since this light is concentrated only within the central kilometre of the shower, one cannot use this
technique at highest energies.

\begin{figure}[ht]
\begin{center}
\includegraphics[width=0.70\textwidth,angle=0]{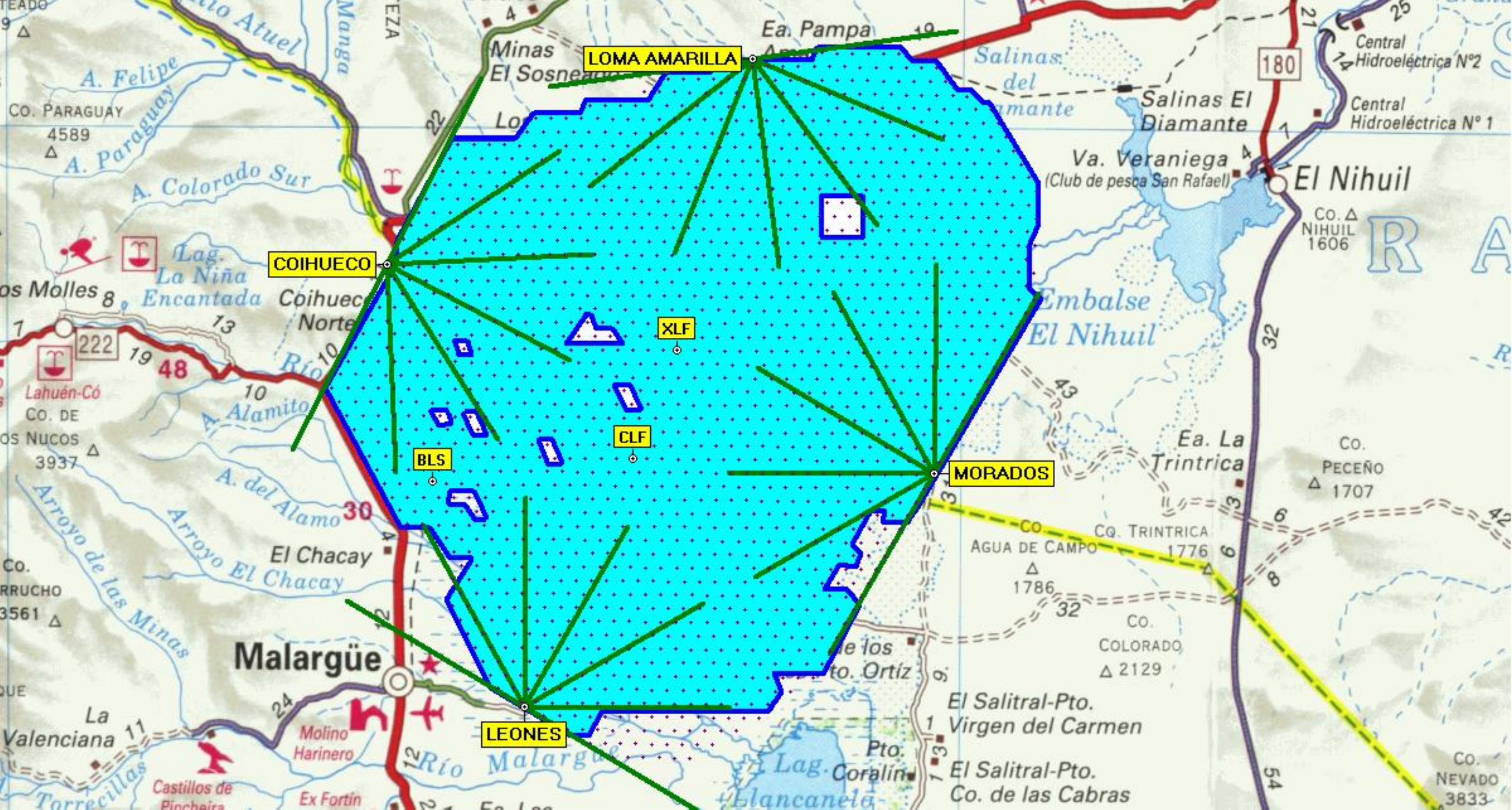}
\end{center}
\vspace{-0.3cm}
\caption{\label{PAO}
Pierre Auger Observatory detector with more than 1600 water tanks and 4 fluorescence telescopes
(see Ref.~\cite{auger_detector} for details) }
\end{figure}

\begin{figure}[ht]
\begin{center}
\includegraphics[width=\textwidth,angle=0]{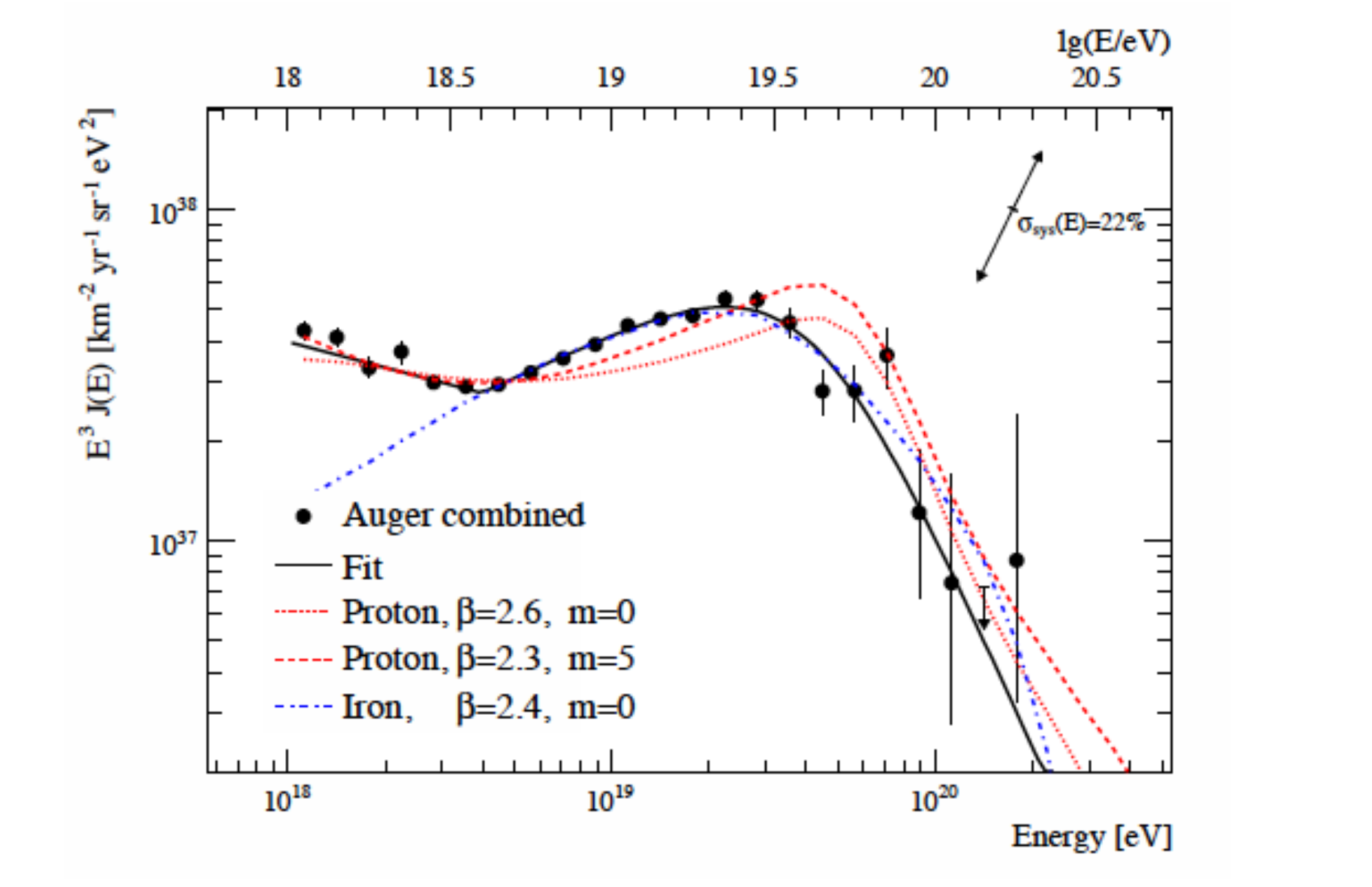}
\end{center}
\vspace{-0.3cm}
\caption{\label{spectrum2009}
Energy spectrum of UHECR as a function of energy measured by the
Pierre Auger Observatory and model predictions for iron nuclei (blue) and protons (red)~\cite{auger_spectrum2009}}
\end{figure}

The Pierre Auger Observatory (Auger) is the largest UHECR detector in the world at the moment with an area of 3000 km$^2$.
Such a big area is required to collect enough statistics with  UHECR at highest energies $E>60$ EeV, because the flux 
of such UHECR is tiny, one  particle per 100 km$^2$ per year.
Auger is  located on the plateau at an altitude of 1000 metres in the Mendoza province of Argentina.
The ground detector consists of 1600 water tanks distributed 1.5 km from each other 
as presented in Fig.~\ref{PAO}.
Also there are four fluorescence telescopes pointed at the atmosphere above the ground detectors as shown in Fig.~\ref{PAO}. Detection
of 10\% of showers both by  fluorescence detectors (FD) and by ground detectors guarantees a good quality of events and 
at the same time allows one to calibrate the ground detectors by FD.

In Fig.~\ref{spectrum2009} we show a recent energy spectrum which was measured by Auger before 31 March 2009~\cite{auger_spectrum2009}. 
The steeply falling flux of UHECR is multiplied by $E^3$ in order to show details of the spectrum.    The total systematic 
energy error is 22\% and is shown in the top right corner of the figure.  For energy bins with $E<3 \cdot 10^{19}$ eV
statistical errors are not important, while at highest energies $E>60$~EeV the shape of the spectrum is still uncertain
and more statistics are needed. On the other hand, the suppression of the spectrum is statistically significant and is
clearly seen in Fig.~\ref{spectrum2009}. 

This is an important experimental result, since it is independent confirmation of 
similar observations made by the HiRes experiment~\cite{spectrum_HiRes}.  Thus cutoff in the energy spectrum exists.
However, there are several questions to be answered before one can tell that this really is a GZK cutoff.
First, is this cutoff due to the maximum energy of sources, or to energy losses? In Section \ref{sec:accel}
we have seen that indeed the maximum energy for many types of sources is close to $10^{20}$ eV. The 
ultimate answer to this question would be the detection of several sources at different distances with cutoff following
expectations of energy losses.  

\begin{figure}[ht]
\begin{center}
\includegraphics[width=0.46\textwidth,angle=0]{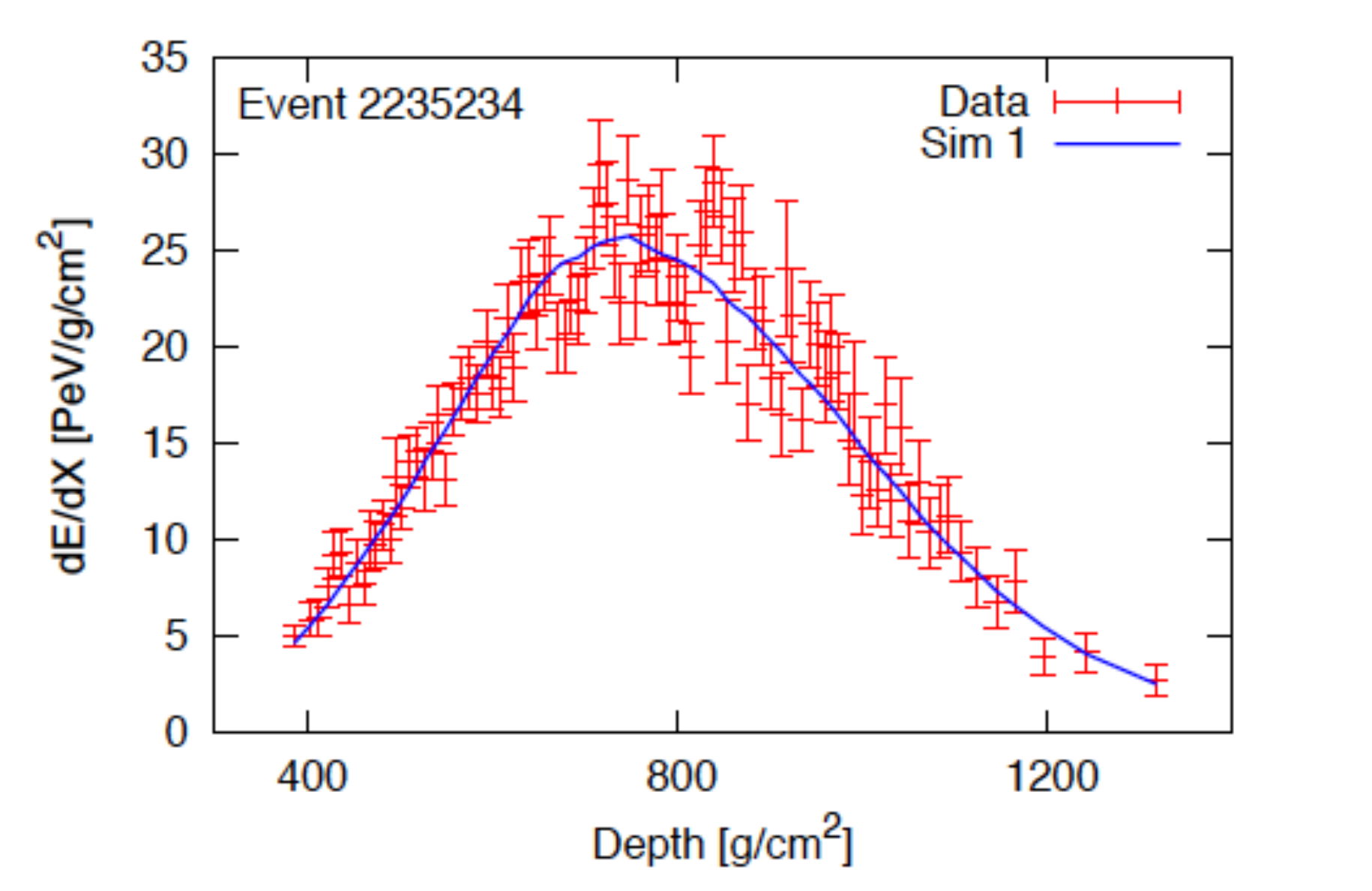}
\includegraphics[width=0.46\textwidth,angle=0]{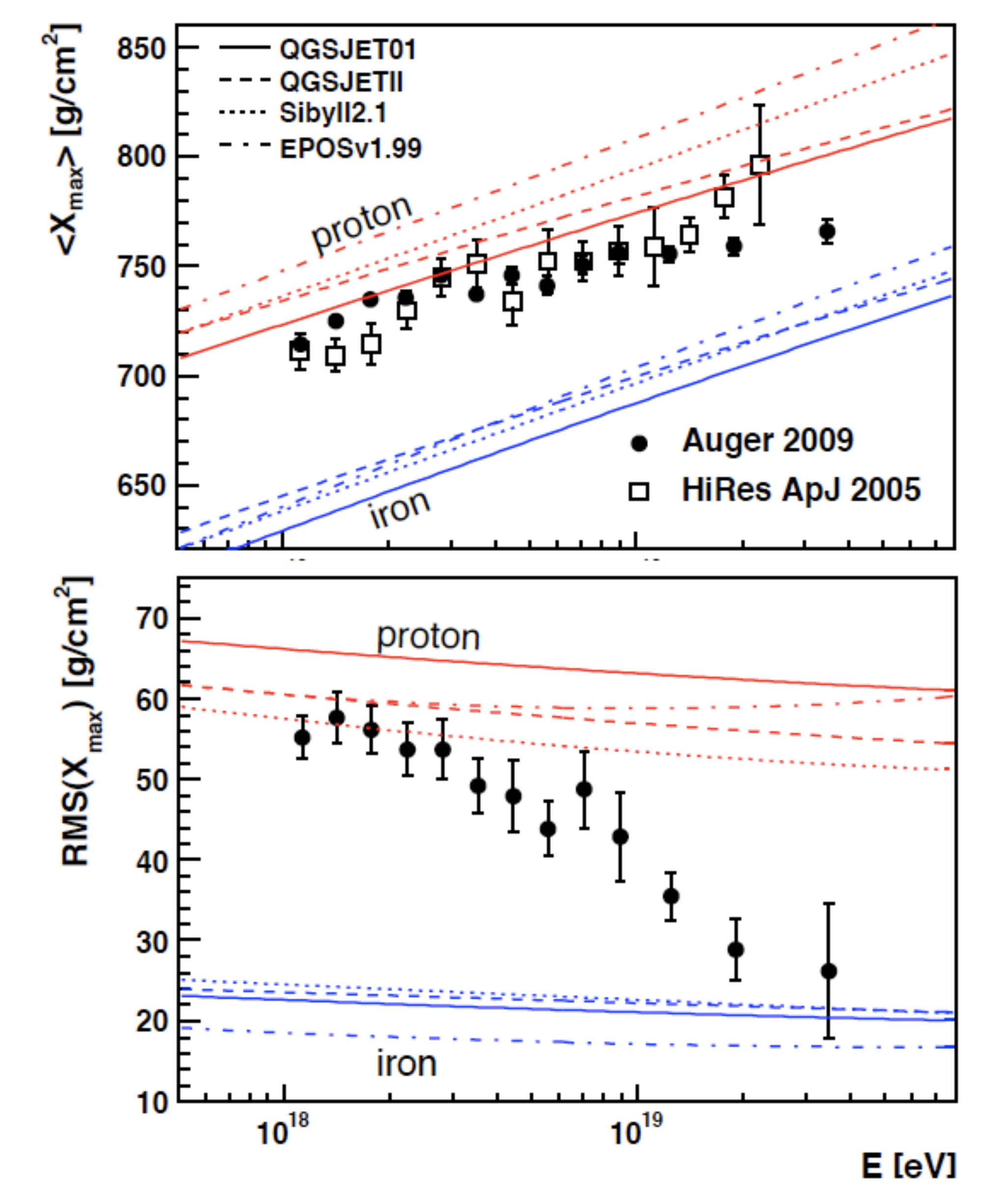}
\end{center}
\vspace{-0.3cm}
\caption{\label{composition2009} 
{\bf Left:} Measurement of shower development by signals in the fluorescence detectors as a function of depth in the atmosphere.
The maximum of shower development in this example is $X_{max}=753 ~g/\mbox{cm}^2$~\cite{auger_composition0}.
{\bf Right:} Average $X_{max}$ of showers measured by HiRes and Auger and $RMS$ of  $X_{max}$ 
measured by Auger in 2009~\cite{auger_composition}.
}
\end{figure}

Second, is the chemical composition of UHECR proton-dominated at those energies? Since in our Galaxy all elements
up to iron are accelerated to energies around the knee $E=10^{15}$ eV, the same situation can exist in  astrophysical objects which 
accelerate to highest energies.
Experimental answers to this question can be found
in the future even by Auger, but already current data show that the composition becomes heavy at high energies.
Indeed, recent Auger results from Refs.~\cite{auger_composition0,auger_composition} are shown in Fig.~\ref{composition2009}.
In Fig.~\ref{composition2009} (left) we present the shape of the shower development in the atmosphere as seen by a fluorescence telescope.  
The signal is proportional to the number of electrons and positrons in the shower. Signals grow due to the development of electromagnetic cascades. The 
maximum of the signal corresponds to the maximum development of the cascade in the atmosphere. After that the shower loses its energy due to dissipation effects.  The depth of the atmosphere corresponding to the maximum of the shower development is called $X_{max}$. 
For the example presented, this maximum is at $X_{max}$~=~753~g/cm$^2$ and the energy of the event is $E=1.6 \cdot 10^{19}$ eV~\cite{auger_composition0}.  At the same energy, protons on average interact much deeper in the atmosphere than heavy nuclei. 

\begin{figure}[ht]
\begin{center}
\includegraphics[width=0.46\textwidth,angle=0]{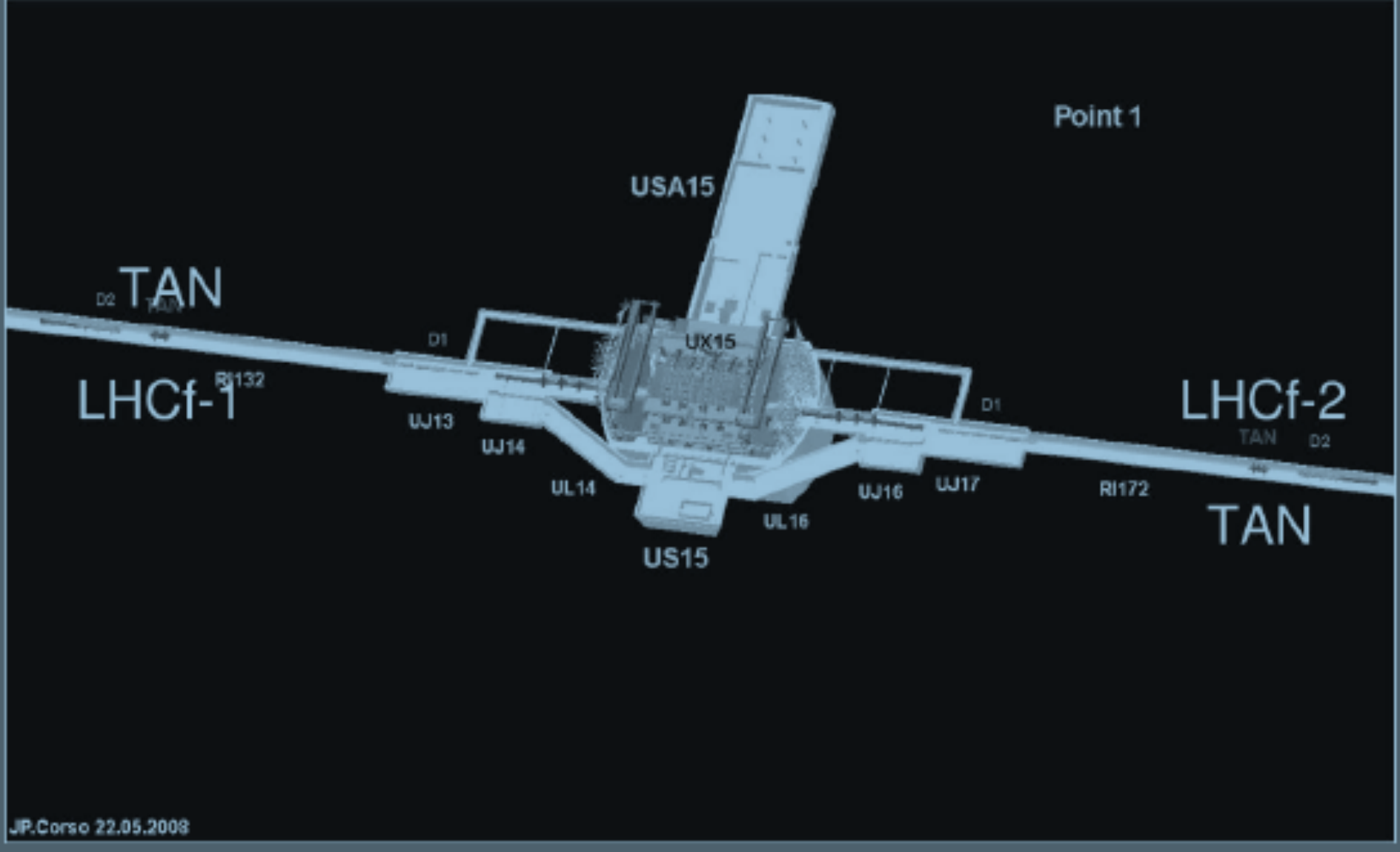}
\end{center}
\vspace{-0.3cm}
\caption{\label{lhc_exp} 
The layout around the Interaction Point 1 (IP1) of the LHC. The structure at the centre indicates the ATLAS detector surrounding the collision point. The
LHCf detectors are installed in the instrumentation slot of the TANs located $\pm 140$~m from IP1. Two independent detectors, LHCf Arm1 and
LHCf Arm2 are installed at either side of IP1\cite{LHCf}.
}
\end{figure}

On the top panel of Fig.~\ref{composition2009} (right) one can see the results of the most common hadronic models presented with red lines for protons
and with blue lines for iron. The example event in Fig.~\ref{composition2009} (left) is definitely proton-like. The averaged $X_{max}$ values in each bin are presented in the same figure for both the Auger and HiRes experiments. Both results are consistent with each other showing a relatively light
composition from $10^{18}$ eV to $10^{19}$ eV. However, Auger shows heavier  composition at highest energies. 

The main problem when measuring the composition with $X_{max}$ is its strong model dependence, as seen on the top panel of Fig.~\ref{composition2009} (right). There are two complementary ways out. One is to use a composition-sensitive parameter that weakly depends on the model choice. Such a parameter is $RMS(X_{max}) = \sqrt {\langle X_{max}^2\rangle - \langle X_{max}\rangle^2}$, presented in the lower panel of Fig.~\ref{composition2009} (right).
One can see that according to this measurement the composition becomes heavier at high energy. Another important way is to test models
and find the best one. For this purpose a dedicated experiment LHC forward (LHCf) was constructed at CERN.  The idea of this experiment is to measure the neutral particles emitted in the very forward region of LHC collisions at low luminosity. The configuration of this experiment is presented in Fig.~\ref{lhc_exp}. Data required for testing the hadronic models will be collected in the first scientific runs of the LHC~\cite{LHCf}.
Thus in the near future we shall have better knowledge of hadronic models and more understanding of the composition of 
UHECR at highest energy. At present the Auger results indicate heavy composition; this was not confirmed by independent 
measurements and the fraction of light nuclei in the data remains uncertain. 

This is a very important question for searches of UHECR sources, which we shall discuss in the next section.

\subsection{Anisotropy}
\label{sec:anisot}

\begin{figure}[htb]
\includegraphics[width=0.35\textwidth,angle=270]{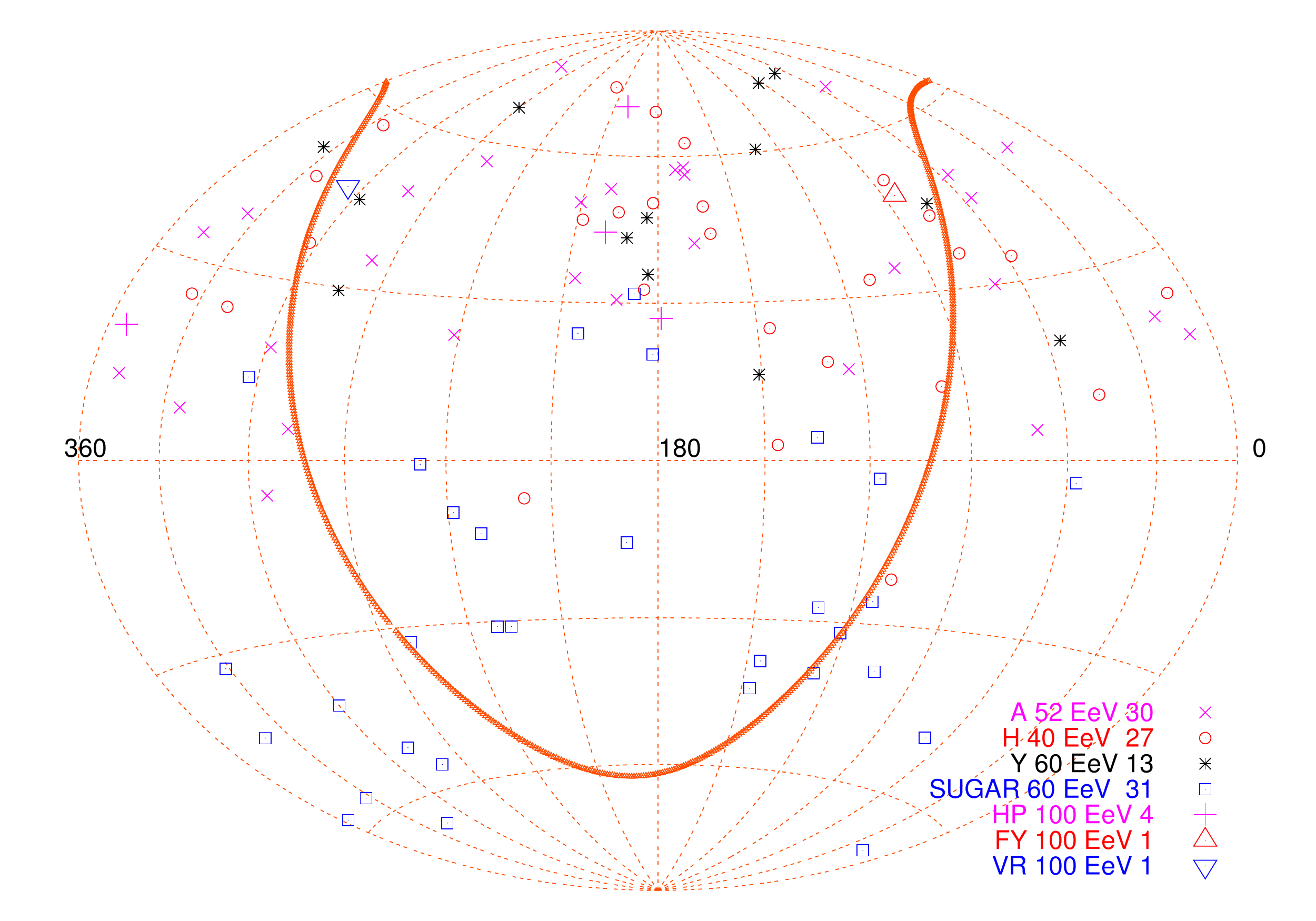}
\includegraphics[width=0.35\textwidth,angle=270]{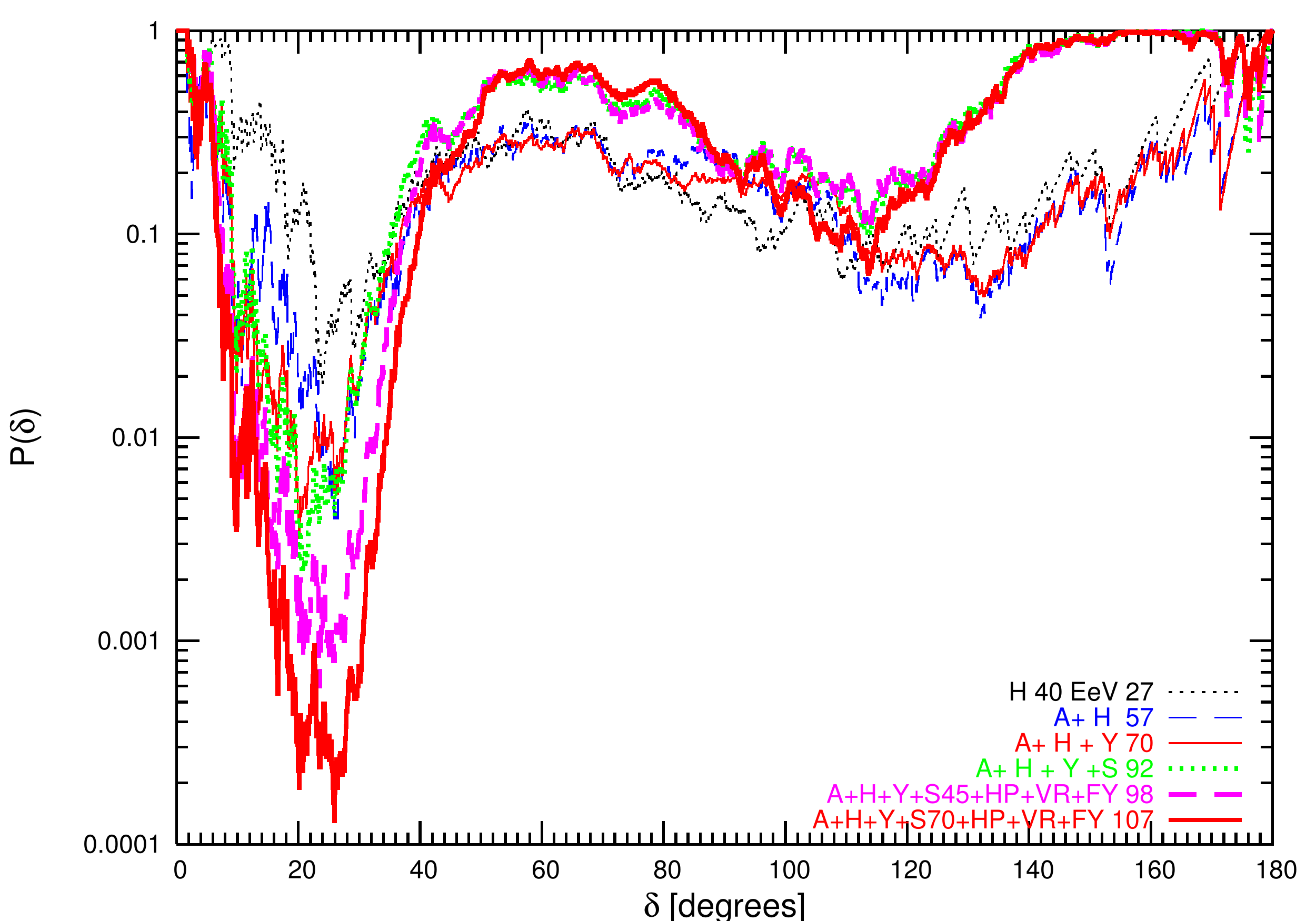}
\vspace{-0.3cm}
\caption{\label{LSS} 
{\bf Left:} Sky map of arrival directions of UHECR with $E>40$~EeV in old experiments.
{\bf Right:} Probability that this anisotropy is a function of angular distance between events~\cite{Kachelriess:2005uf}.
}
\end{figure}

Since for every UHECR event the arrival direction is detected, for tens of years 
many attempts were made to find sources of UHECR in the experimental data. Unfortunately none 
of them has  been confirmed so far.  There are two ways to look for the sources. One is to look for the data itself
and try to find anisotropy in autocorrelation factions. The second is to pick up a catalogue of possible 
sources and look for the  cross-correlations with this catalogue. This second way always requires confirmation 
by an independent data set, since completeness of the catalogue is a very complicated issue and it is difficult
to estimate the probability due to the parameter choice  a posteriori. 

Here we start  with autocorrelations. In the left panel of Fig.~\ref{LSS}  one  can see the sky map with the arrival direction of events with $E>40$ EeV
in several old experiments, including SUGAR, AGASA, HiRes, Yakutsk, Havera Park, Volcano Ranch, and Fly's Eye. On the right panel of the same figure one can see the probability  that autocorrelations between selected events are by chance within a given angle. One can see that
the probability is minimal at angles 20--25 degrees. After penalization on the choice of angle, the probability that this happened by chance is $P=0.3$\%~\cite{Kachelriess:2005uf}. This clustering of events on rather moderate scales can be due to the location of the sources in the Large Scale Structure.

\begin{figure}[htb]
\begin{center}
\includegraphics[width=0.6\textwidth,angle=0]{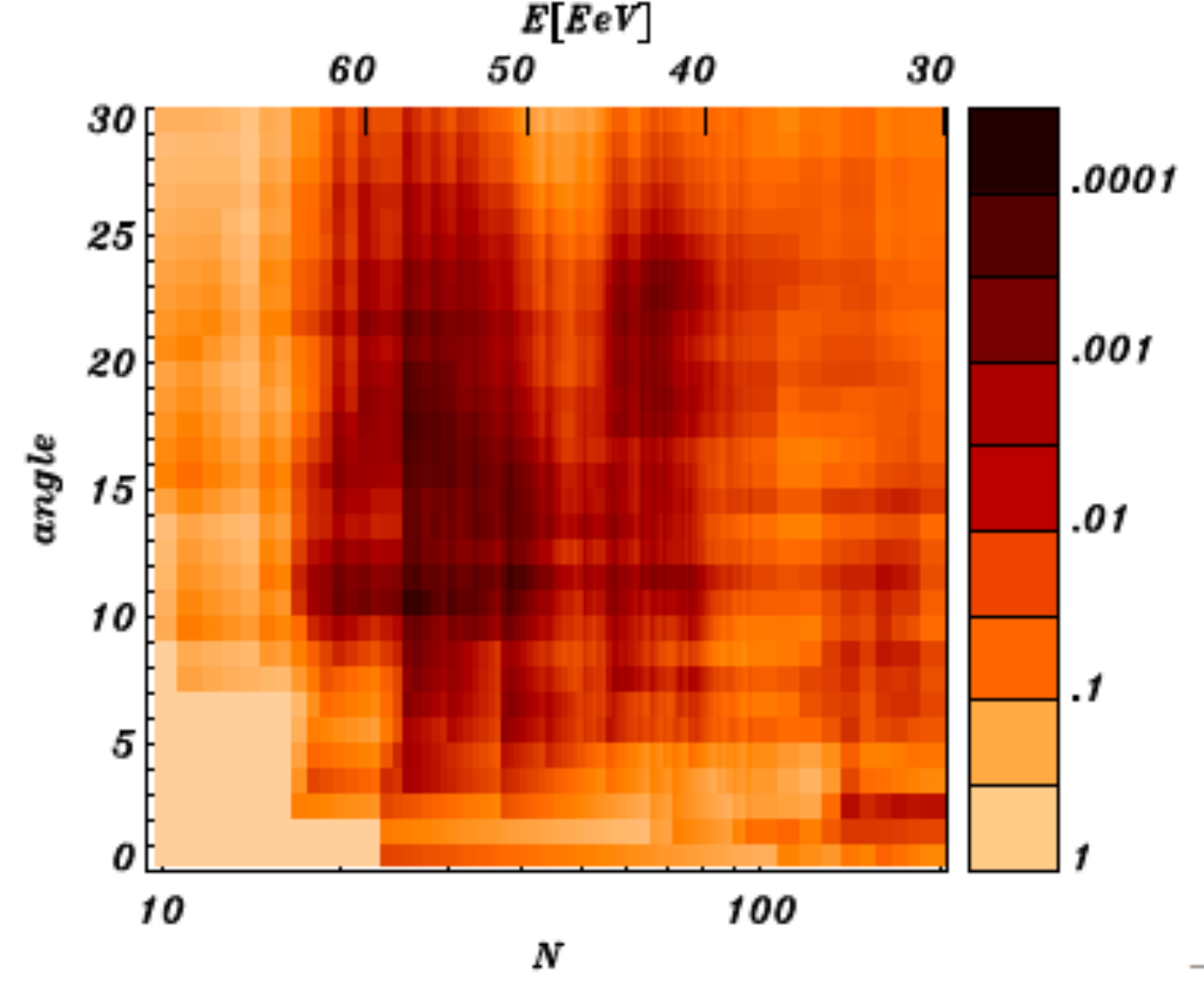}
\end{center}
\vspace{-0.3cm}
\caption{\label{autocorr}
Probability of autocorrelations as a function of energy 
and angular distance between events, see Ref.~\cite{auger_auto}
}
\end{figure}

The same probability in the first Auger data is presented in Fig.~\ref{autocorr} as a function of both energy and angle. One can see that
for exactly the same energy $E=40$ EeV and angle $\theta=20^\circ$--$25^\circ$ the probability is $P \sim 10^{-2}$.  However, recent 
results with larger statistics did not show more significant anisotropy at such energies~\cite{auger_anisotropy}. This makes the situation with 
anisotropy in the data less clear.

\begin{figure}[htb]
\begin{center}
\includegraphics[width=0.5\textwidth,angle=0]{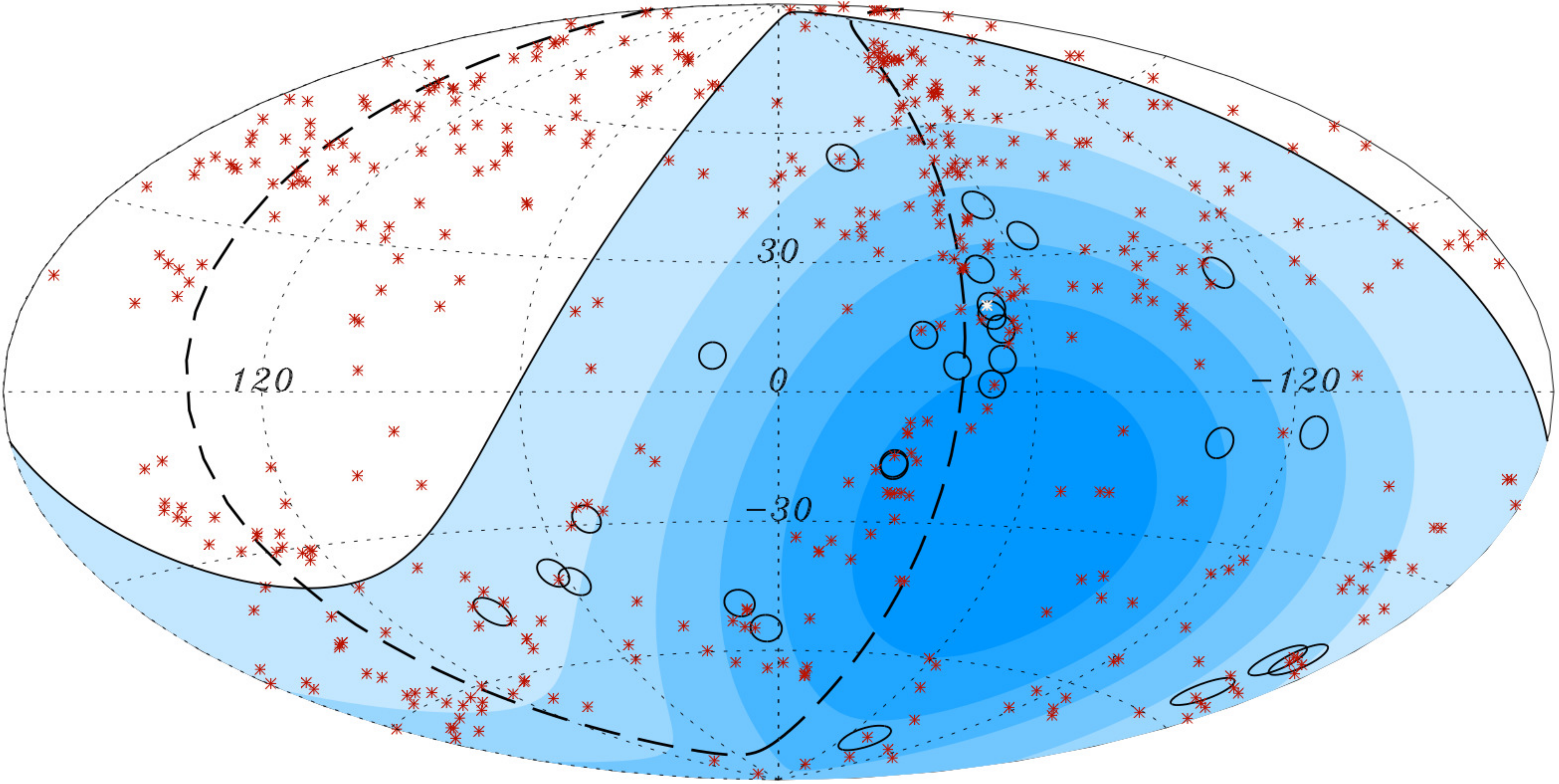}
\includegraphics[width=0.4\textwidth,angle=0]{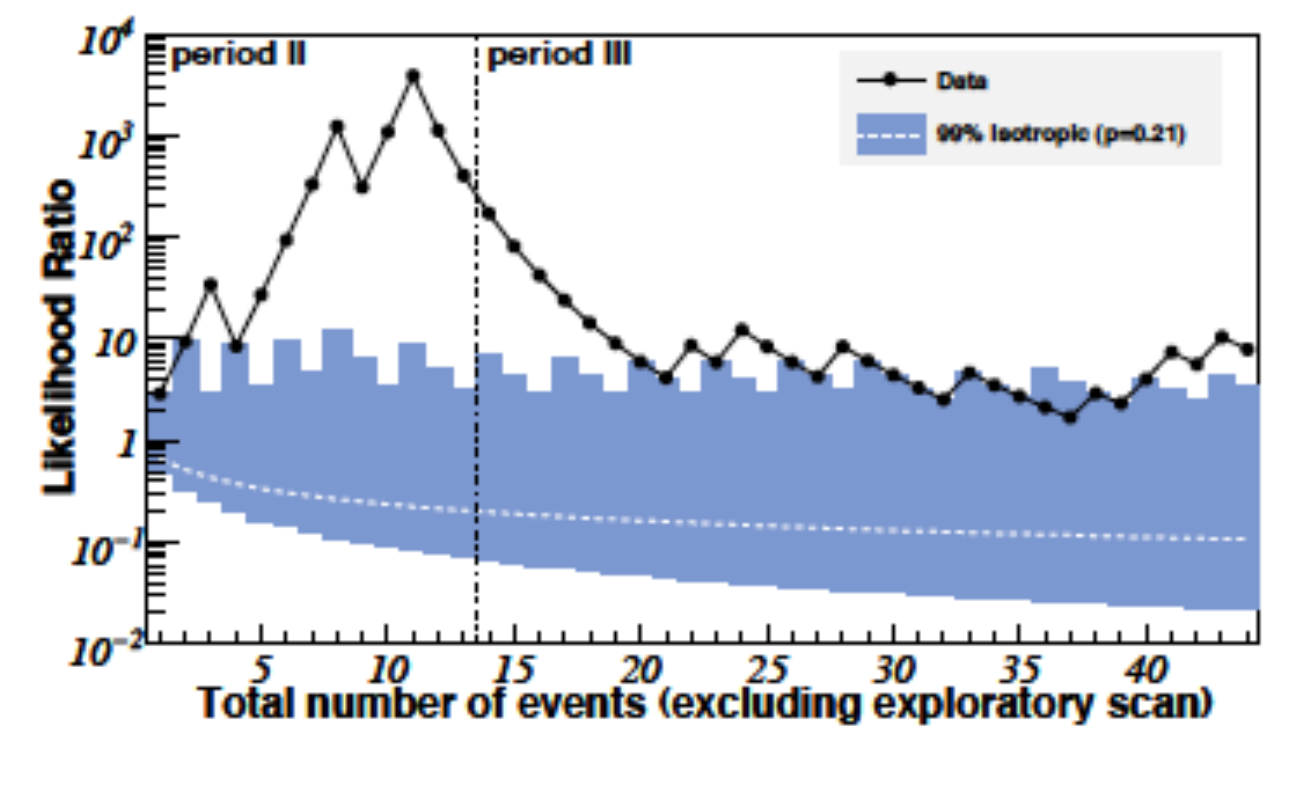}
\end{center}
\vspace{-0.3cm}
\caption{\label{sky2007}
{\bf Left:} Sky map of arrival directions of 27 UHECR with $E>57$ EeV measured by the Pierre Auger Observatory before August 2007
  in galactic coordinates  (circles) and 472 nearby AGNs (red stars)~\cite{auger_science}. Blue contours show the Auger exposure.
{\bf Right:} Likelihood ratio  for events after formulation of the prescription. Period II is for data on the left panel. Period III is for new data up to March 2009~\cite{auger_anisotropy}.  
}
\end{figure}

Now let us discuss correlations with astrophysical objects. First Auger data have shown strong correlations with nearby active galaxies  
called Active Galactic Nuclei (AGN).  Namely, 12 out of 14 events with E > 57~ EeV were correlated within $\theta<3.1^\circ$ from
472 AGNs from the Veron catalogue with distances $R<75$ Mpc.  This correlation was considered by the Auger Collaboration as a
formal way to study the deviation of cosmic rays from isotropic distribution. Data from Period I was tested with the prescription during Period II,
where 13 new events were detected, out of which 9 obeyed prescription parameters. The prescription was fulfilled, i.e., the observed 
sky was considered anisotropic at the 99\% confidence level~ \cite{auger_science}. Data used in this publication and shown in Fig.~\ref{sky2007}
 (left) correspond to Periods I (not shown) and  II shown in Fig.~\ref{sky2007} (right) before the vertical line. Unfortunately this 
 correlation was not confirmed in the later data [Period III in Fig.~\ref{sky2007} (right)]. 
 
 \begin{figure}[htb]
\begin{center}
\includegraphics[width=0.5\textwidth,angle=0]{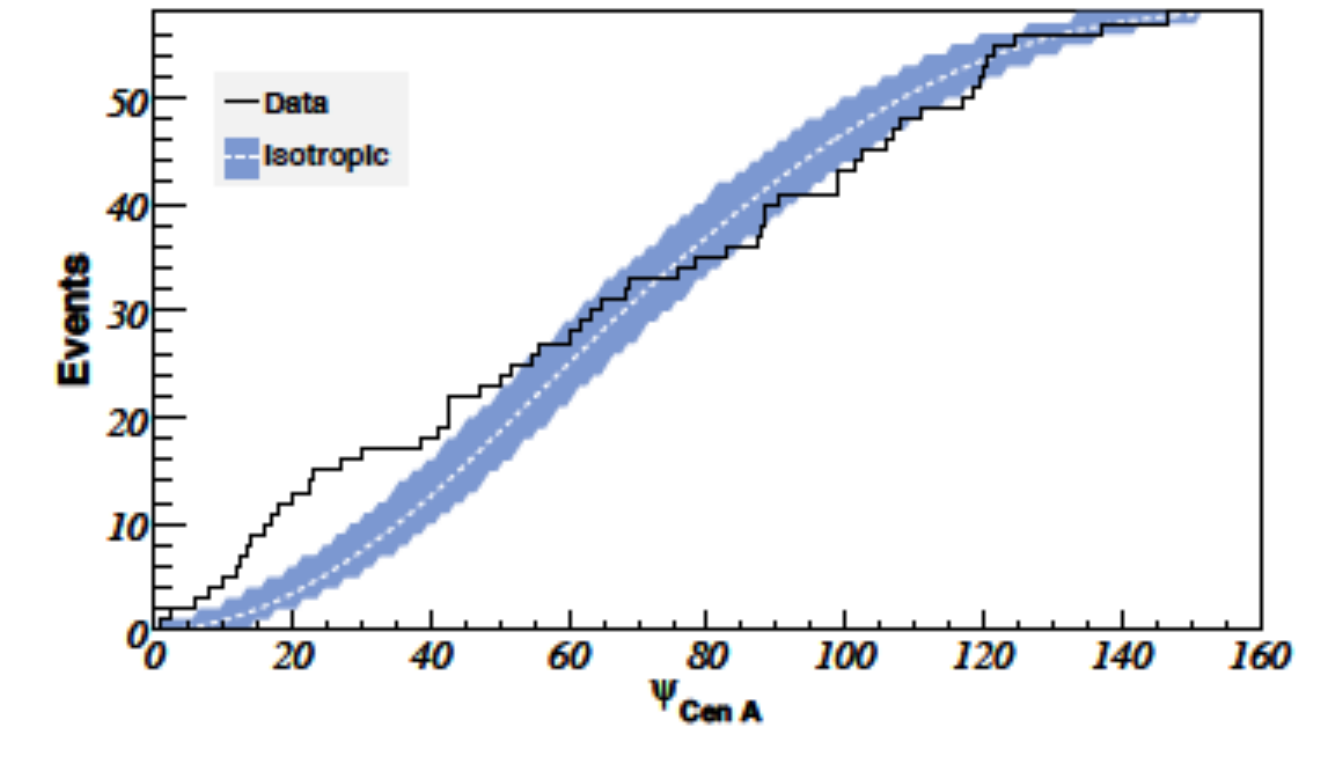}
 \end{center}
\vspace{-0.3cm}
\caption{\label{CenA}
Angular distribution of events around the Cen A galaxy in Auger data compared to isotropic ones 
}
\end{figure}
It does not mean that all anisotropy signals in Auger have completely disappeared. 
There is still a remaining excess of events around the Cen A galaxy on scales of 20 degrees,
see Fig.~\ref{CenA}. This anisotropy has to be tested by future data.

\subsection{Secondary photons and neutrinos from UHECR}
\label{sec:second}

 \begin{figure}[htb]
\begin{center}
\includegraphics[width=0.4\textwidth,angle=0]{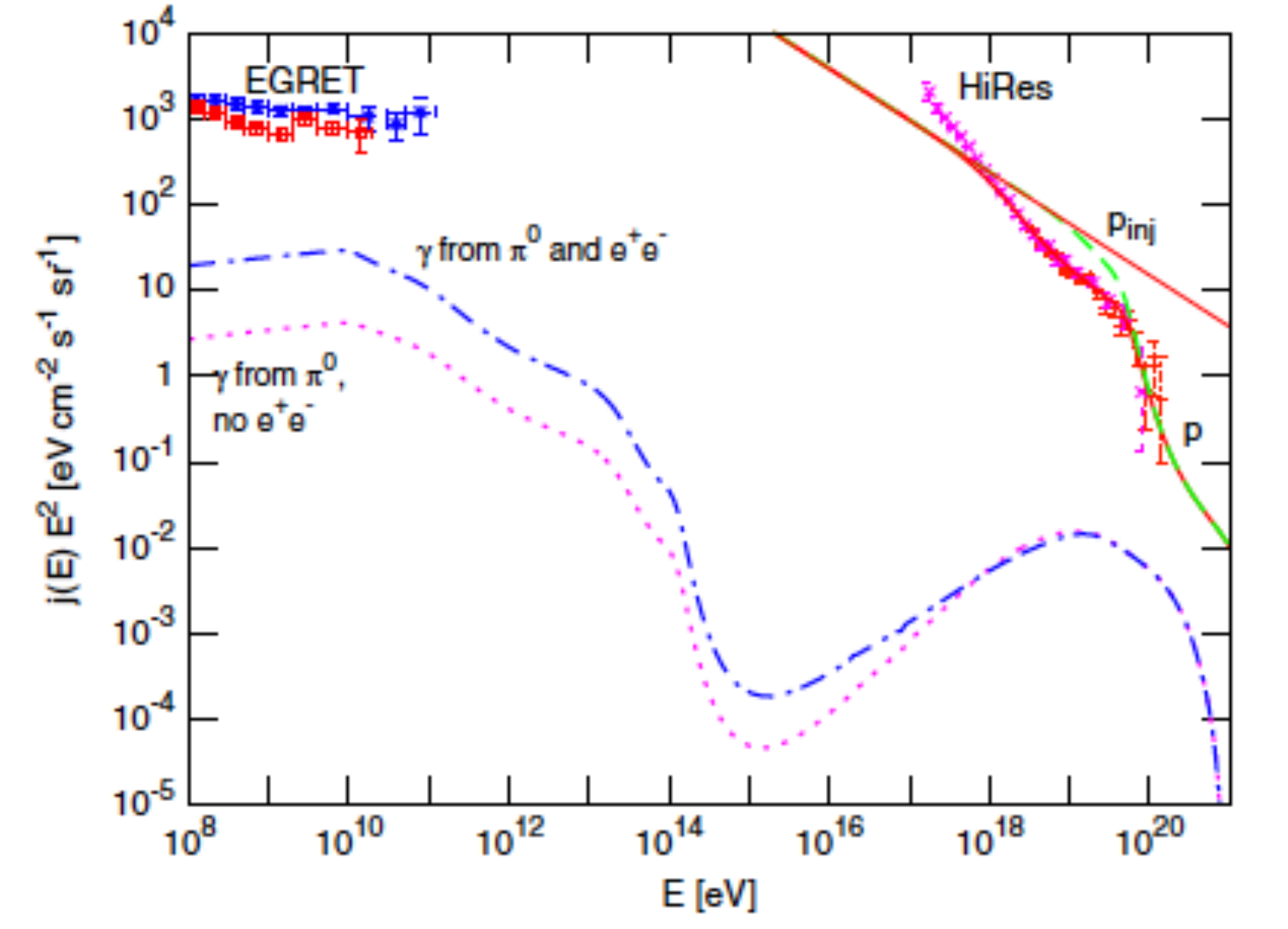}
\includegraphics[width=0.4\textwidth,angle=0]{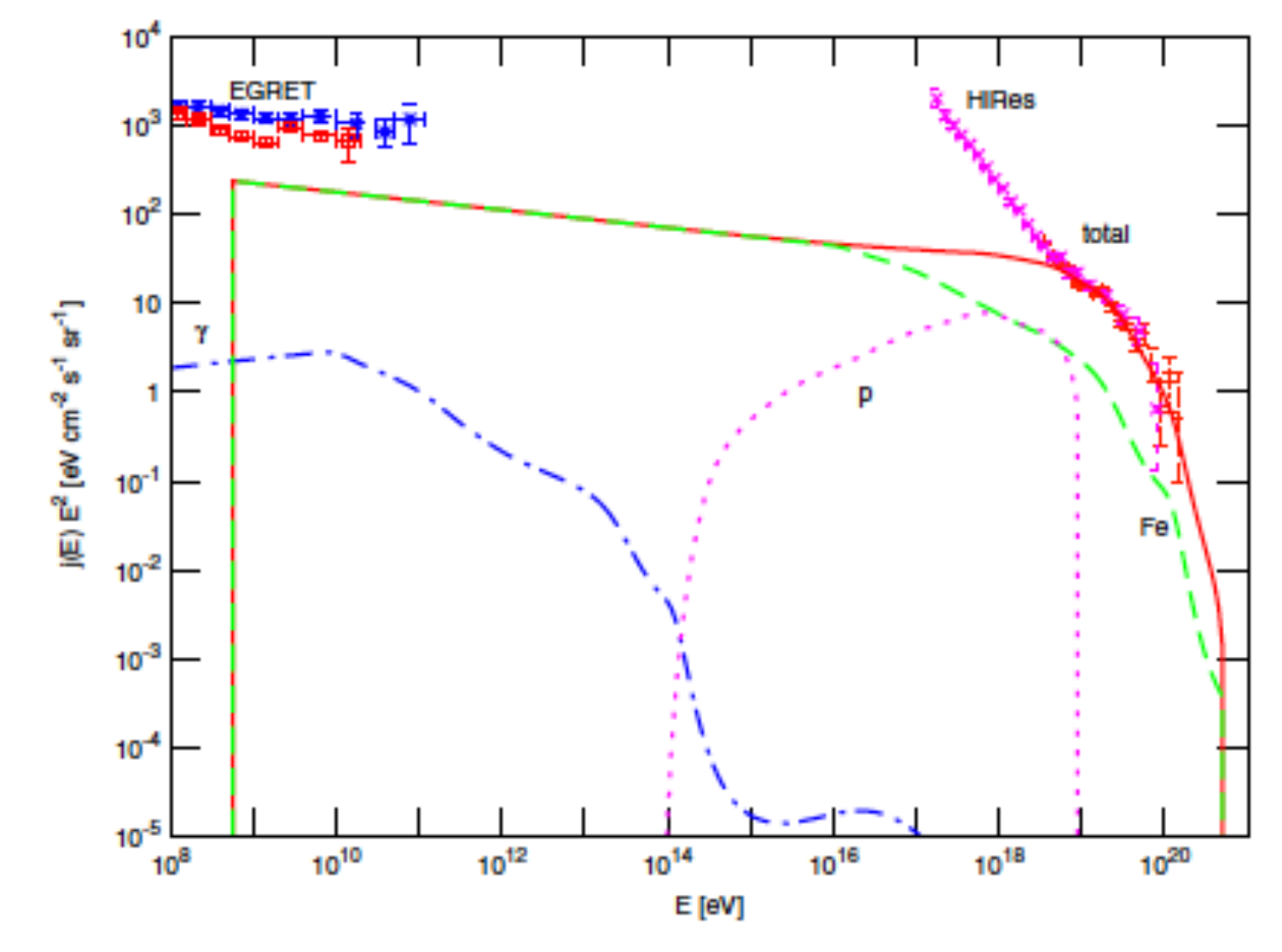}
 \end{center}
\vspace{-0.3cm}
\caption{\label{photon_fluxes}
{\bf  Left: }
Fluxes of protons and secondary photons as a function of energy.
Primary protons with spectrum $1/E^{2.6}$  and maximum energy $E_{max}=10^{21}$ eV 
are shown by the thin red line. Secondary protons fit the UHECR spectrum 
from $E>10^{18}$ eV (thick red line).
Secondary photons  from all reactions are shown by the blue dashed line and from pion production only, Eq.~(\ref{pion_production}) by the magenta line.
{\bf Right:}
Fluxes of UHECR and secondary photons in the case of iron nuclei primaries with spectrum $1/E^{2.1}$ and maximum energy $E_{max}=10^{21}$ eV.
The remaining iron nuclei are shown by the green line. Secondary protons by the magenta line. Secondary photons by the blue line.
}
\end{figure}

 \begin{figure}[htb]
\begin{center}

\includegraphics[width=0.6\textwidth,angle=0]{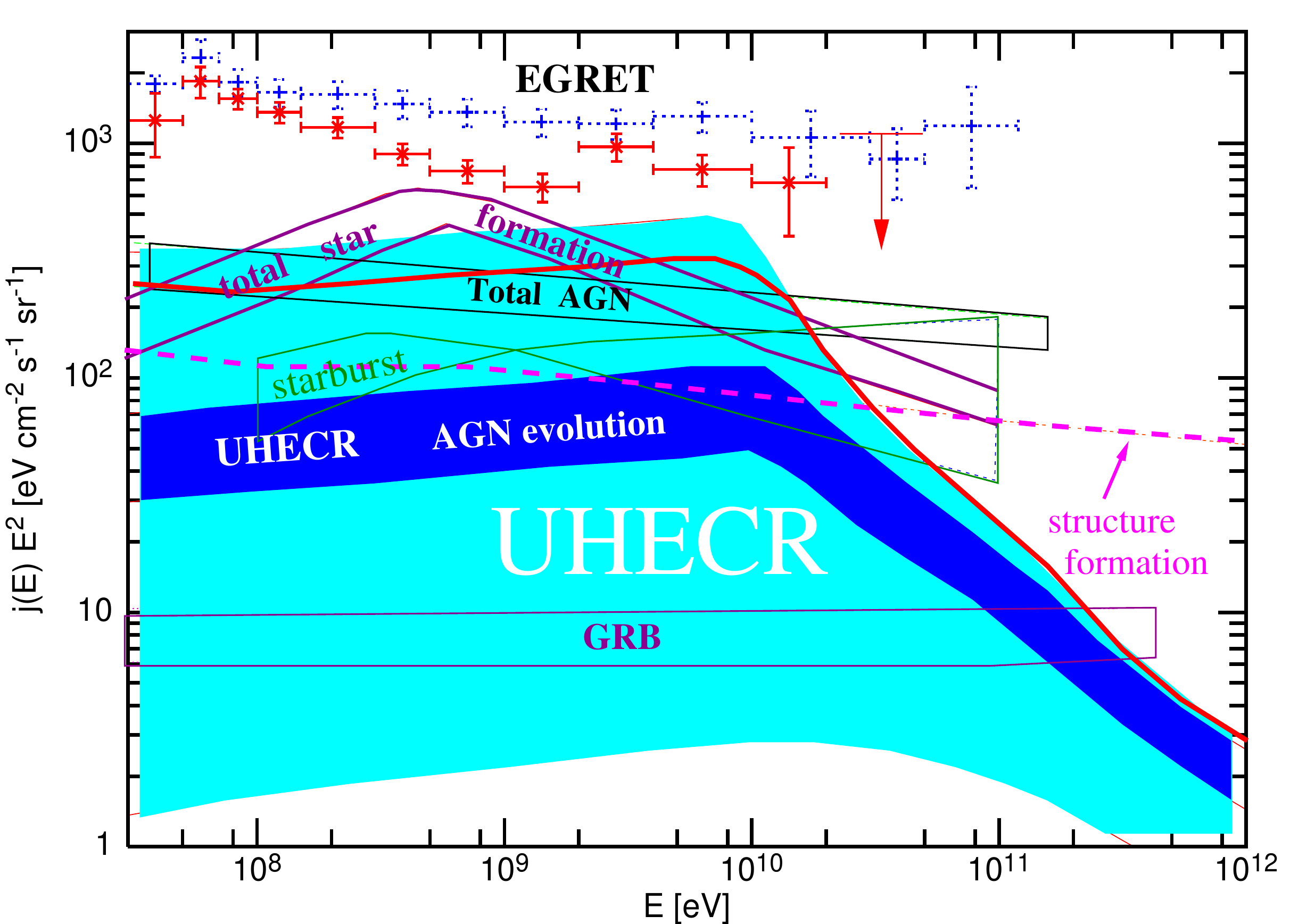}
 \end{center}
\vspace{-0.3cm}
\caption{\label{photons_EGRET}
Contribution of secondary photons from UHECR to the extragalactic gamma-ray background as a function of energy
and other possible sources which contribute to the same background~\cite{photons_cascade}
}
\end{figure}

As was discussed in Section \ref{sec:propag}, protons lose their energy in pair production and pair production reactions.
Since secondary pions quickly decay, secondary photons and neutrinos are produced. Neutrinos propagate to the Earth without 
interactions on the way, but photons cannot. They start to interact with background photons and produce pairs.
Electrons and positrons in turn up-scatter CMB photons or produce synchrotron radiation:

\begin{eqnarray}
\label{cascade}
\gamma + \gamma_{background}  &\rightarrow& e^+ + e^-  \nonumber \\
e^{\pm} + \gamma_{background}  &\rightarrow& e^\pm + \gamma \\
e^{\pm} + B  &\rightarrow& e^\pm + \gamma_{synch} \nonumber
\end{eqnarray}

The sequence  of processes in Eq.~(\ref{cascade}) is called an electromagnetic cascade. At energies above $10^{15}$~eV the cascade proceeds 
on the CMB background (400/cm$^3$), but at lower energies pair production on CMB is impossible. At such energies the cascade continues on a much less abundant   infrared background (1/cm$^3$) and at lower energies  on optical background (0.01/cm$^3$). Then it stops at the multi-GeV
energies of gamma rays.

In Fig.~\ref{photon_fluxes} we plot primary cosmic-ray and secondary photon fluxes from primary protons (left) and iron (right) from Ref.~\cite{photons_cascade}.
Secondary protons after interaction fit the UHECR spectrum from $E>10^{18}$~eV in Fig.~\ref{photon_fluxes} (left). Secondary photons
cascade down to the GeV region.  Only a small fraction of photons come from the pion production reaction (magenta dotted line).
Most of the photons generated are from the $e^+e^-$ production reaction with total flux shown by the dash-dotted blue line. 
The number of secondary protons is much lower in the case of iron primaries, as shown by the dotted magenta line in Fig.~\ref{photon_fluxes} (right).
As a result, the secondary photon flux in the GeV region is much smaller in this case, on the level of 0.2\% of the EGRET measurement. Also 
very high energy photons are absent in this case due to low maximum proton energy.

 \begin{figure}[htb]
\begin{center}
\includegraphics[width=0.44\textwidth,angle=0]{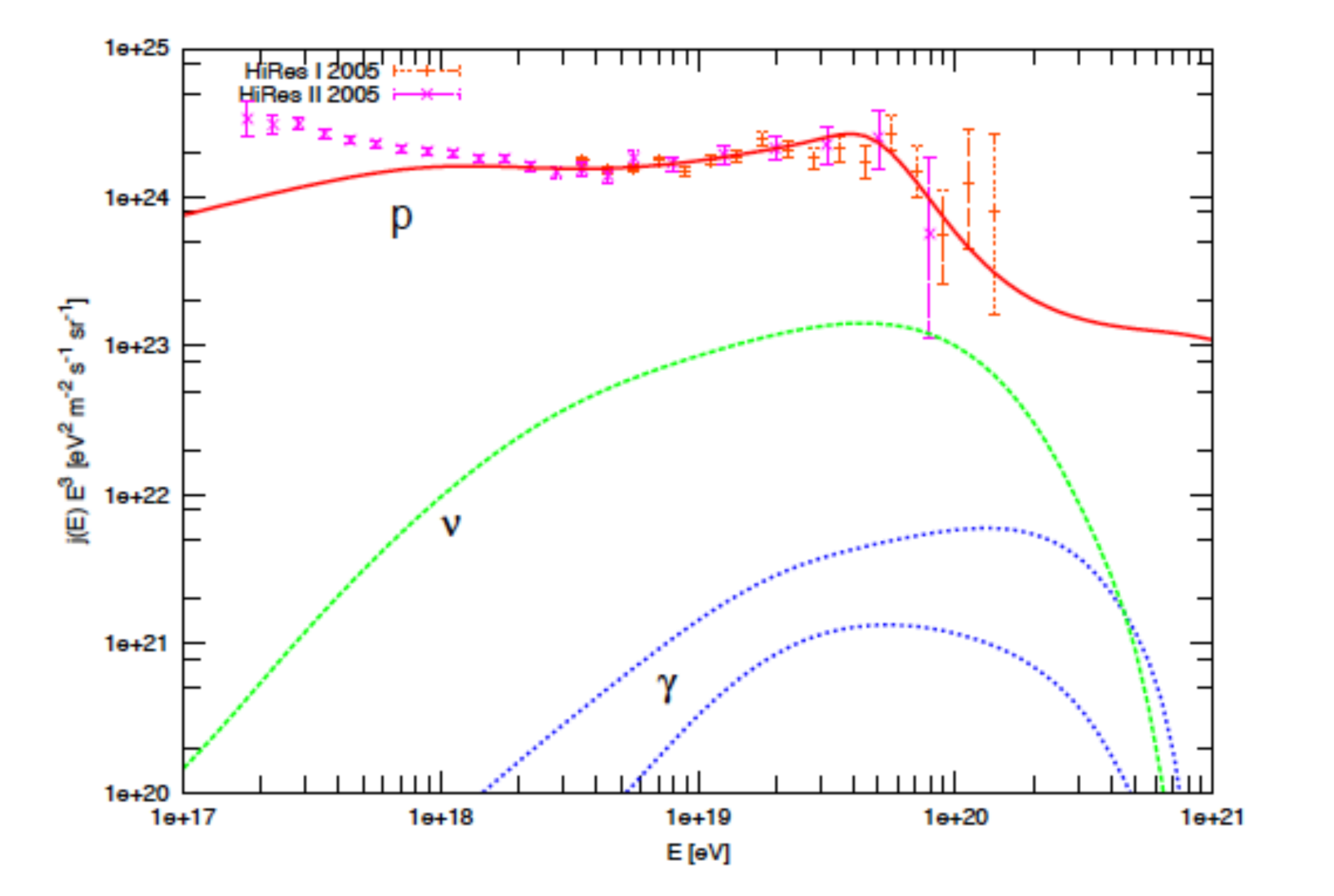}
\includegraphics[width=0.4\textwidth,angle=0]{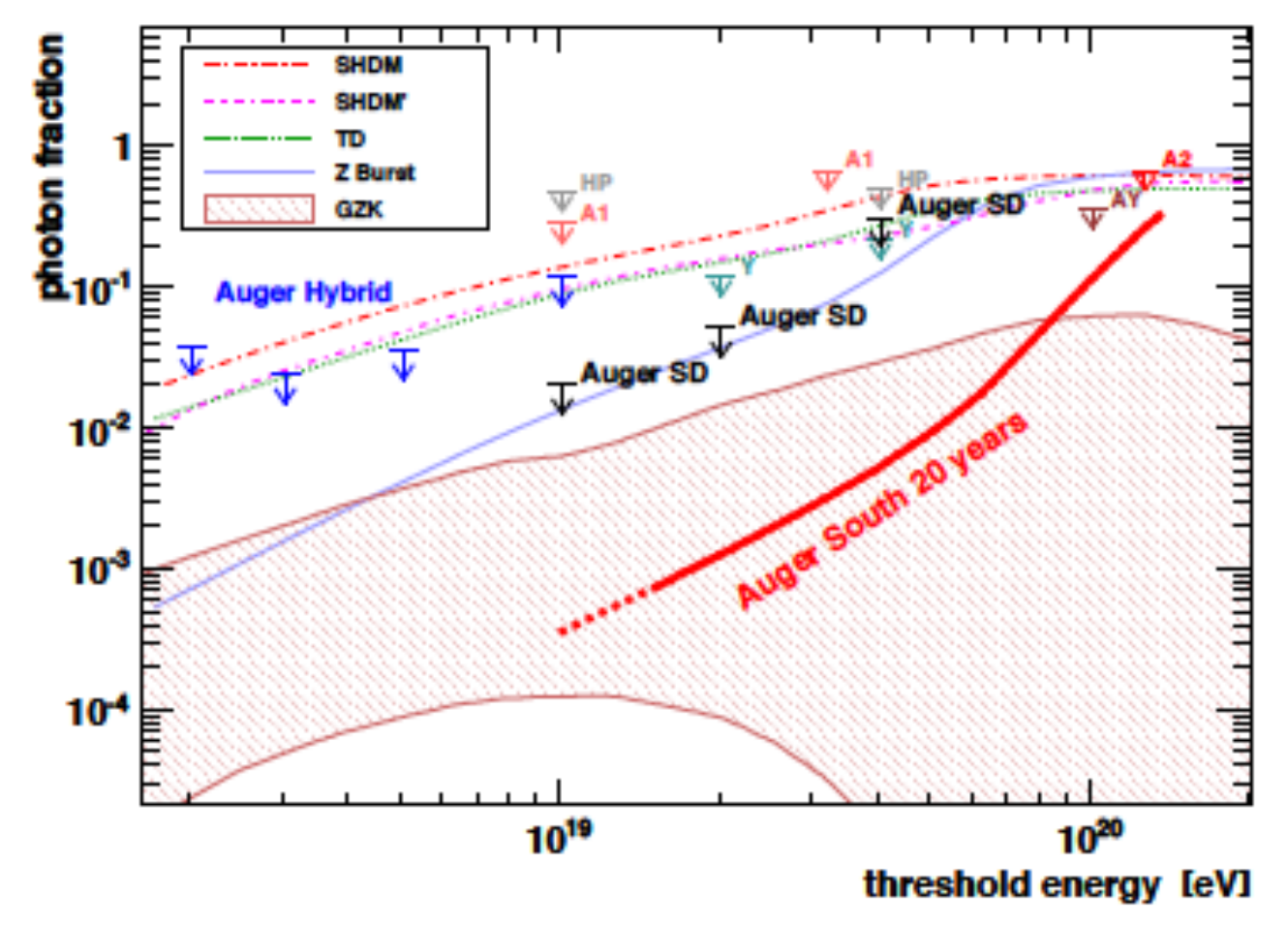}
 \end{center}
\vspace{-0.3cm}
\caption{\label{photon_limits}
{\bf Left:} Example of GZK photon flux from Ref.~\cite{gzk_photons}. UHECR protons fit the HiRes spectrum. Secondary neutrinos are shown by a green line. The remaining secondary 
photons are in the range between the blue lines.
{\bf Right:} Experimental upper limits on the photon fraction in the UHECR spectrum from Ref.~\cite{auger_anisotropy}.
}
\end{figure}

In Fig.~\ref{photons_EGRET} 
we compare the range of the electromagnetic cascade fluxes
from UHECR with other possible astrophysical contributions
in the EGRET band. Note that most of the
uncertainty of the UHECR cascade flux comes from an
unknown source evolution. The scatter for a given class of sources
is thus much smaller, as seen from 
Fig. ~\ref{photons_EGRET} for the case of AGNs.

In Fig.~\ref{photon_limits} (left) we plot the possible range of GZK gamma-ray fluxes
for a given proton flux which fit the UHECR spectrum. The range of fluxes comes from the variation
of possible values of the extragalactic magnetic field and the range of the models for extragalactic 
radio background. Also on the same figure the corresponding  neutrino flux is shown by a green line.
In Fig.~\ref{photon_limits} (right) we show the experimental upper limits on the fraction of photons
in the UHECR flux.  The range of possible GZK photon fluxes corresponds to protons with a
range of power law injection spectra and source evolution fitting the UHECR spectrum.
One can note that the current best upper limits of Auger are still above the range of
expected theoretical values. On the  other hand, existing limits already exclude some exotic models.

\subsection{Summary}
\label{sec:UHECRsum}

In the first lecture we briefly discussed many aspects of UHECR physics. 

Observed cosmic rays have energies up to $10^{20}$ eV.  
Acceleration in astrophysical objects to such energies is a very non-trivial task and
there are no objects in our Galaxy which can do this job. There are very few 
classes of exceptionally powerful objects in the Universe, some of which can be 
real sources of UHECR.
Accelerated particles lose their energy in interactions with the CMB background and are 
also deflected by electromagnetic fields during their propagation from sources 
to the Earth. 

There are three important experimental challenges in UHECR physics: the 
spectrum of cosmic rays, the chemical composition of cosmic rays, and the search for 
anisotropies in the sky with the ultimate goal of finding UHECR sources.

The cutoff in the energy spectrum at highest energies $E>6 \cdot 10^{19}$ eV 
has now been established by two independent experiments,  HiRes and Auger.  
 
The most striking result of 2009 was evidence of heavy composition, shown 
by the Auger experiment at highest energies, Fig.~\ref{composition2009}.
This result still needs independent confirmation. 
Also the interpretation of composition measurements is affected by uncertainty in the hadronic models. This question 
can be clarified in the near future by the LHCf experiment.

Finally, most challenging is the search for UHECR sources. The last result in this direction 
was made by Auger in 2007. They found that the sky is anisotropic at the highest energies, at least at the  99\% C.L.,
by looking at the correlations with nearby AGNs. Unfortunately those correlations were not confirmed in the new data,
and the only anisotropy excess remaining in the Auger data at the highest energies is an excess around the Cen A galaxy, see Fig.~\ref{CenA}.

During energy losses the UHECR protons produce secondary photons and neutrinos.
Most of the secondary photons cascade down to the GeV energies, where this contributes to the 
diffuse extragalactic background. An experimental search for the remaining gamma rays at highest energies $E>10^{18}$ eV
is challenging and existing upper limits are just above theoretical predictions, see Fig.~\ref{photon_limits}.

Thus there are many unsolved problems in UHECR physics.  They require both theoretical and experimental 
efforts in the near and more distant future.

%
%
%
%
%

\section{High-energy gamma rays}

\label{chapter:gamma}

\subsection{Introduction}

In this lecture I shall discuss the theory of TeV gamma rays and recent observations made in this field. 
I shall give a brief introduction to the experimental detection techniques
and present some selected results on the subject. For more detailed study I would  like to recommend  the recent review by  F.~Aharonian, J.~Buckley, T.~Kifune, and G.~Sinnis ~\cite{aharonian_review}.

Relativistic particles can travel with a speed larger than the speed of light in the medium $V>V_M=c/n$. Here $n>1$ is the 
refractive index of the medium.  This index in the air is $n_{\mbox{a}}=1.008$ and in water $n_{\mbox{w}}=1.33$.  

The charged particles polarize the molecules of the medium, 
which then return rapidly to their ground state,
 emitting prompt radiation called  Cherenkov radiation. This radiation
 is emitted under a constant Cherenkov angle  with the particle trajectory, given by
\begin{equation}
\cos \delta= \frac{V_M}{V} = \frac{c}{nV} = \frac{1}{\beta n} ~.
\label{Cherenkov_angle}
\end{equation}

\begin{figure}[ht]
\begin{center}
\includegraphics[width=0.5\textwidth,angle=0]{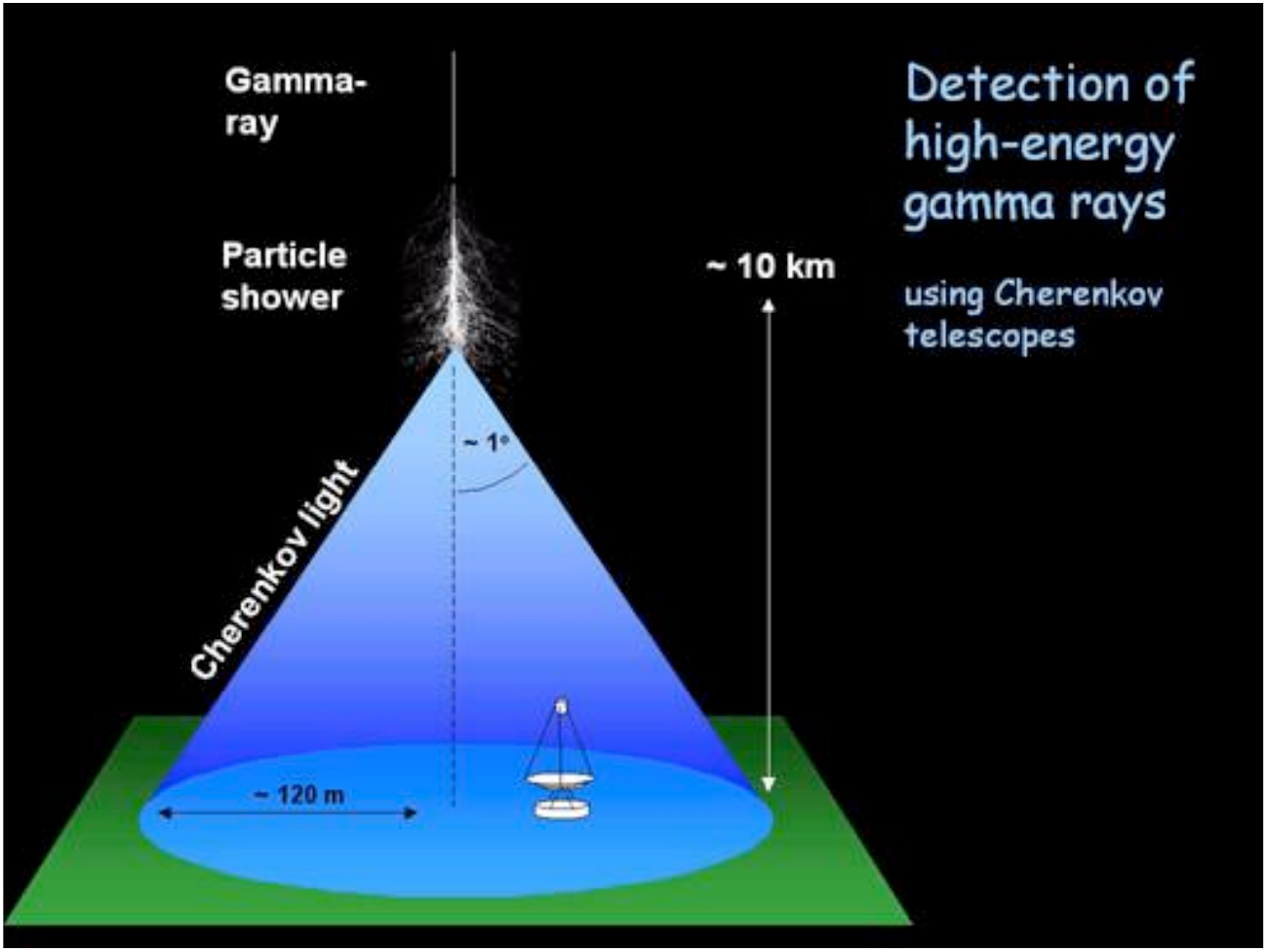}
\includegraphics[width=0.3\textwidth,angle=0]{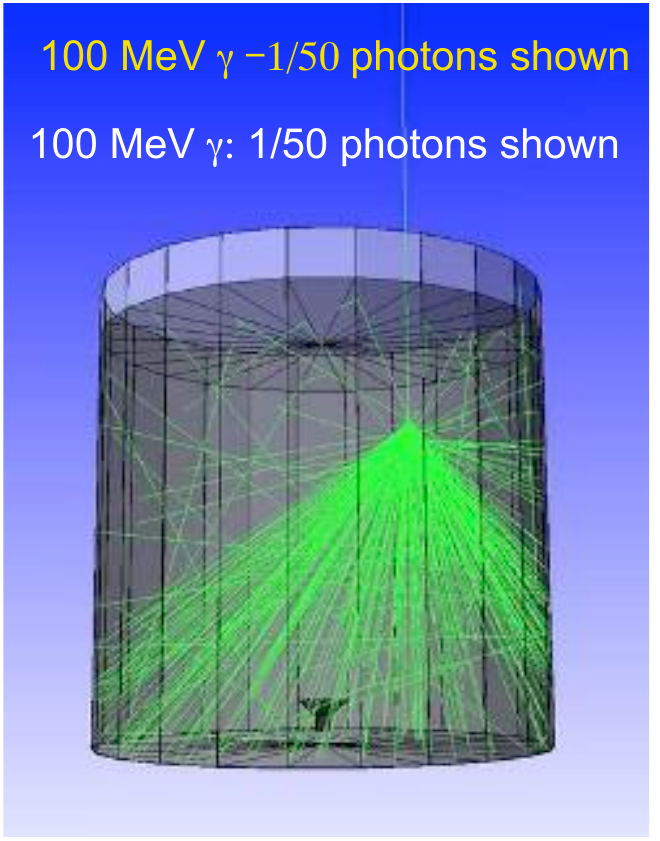}
\end{center}
\vspace{-0.3cm}
\caption{\label{cherenkov} 
Detection of high-energy gamma rays by Cherenkov telescopes  in air (left) and in water  (right)}
\end{figure}
\begin{figure}[ht]
\begin{center}
\includegraphics[width=0.4\textwidth,angle=0]{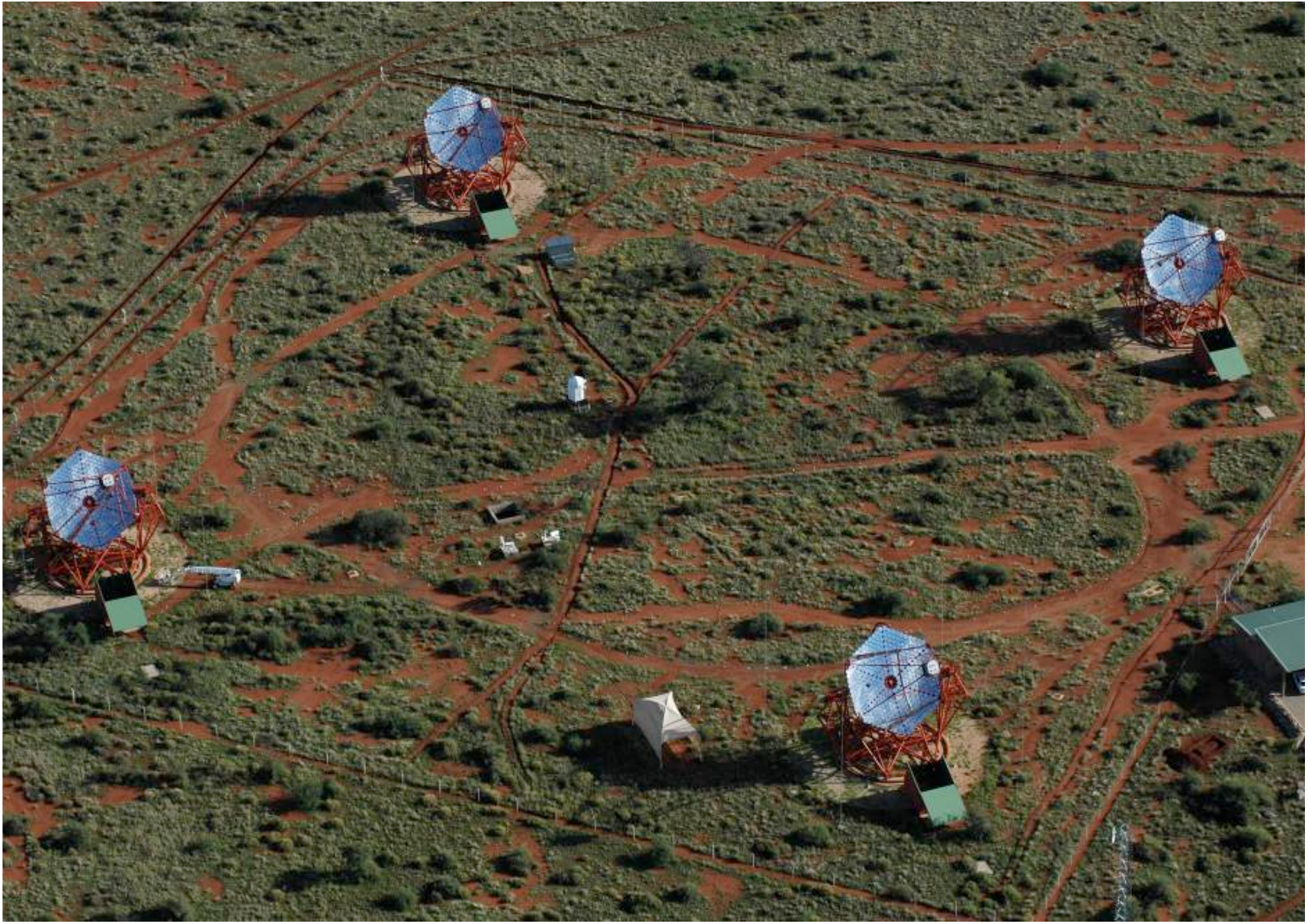}
\includegraphics[width=0.38 \textwidth,angle=0]{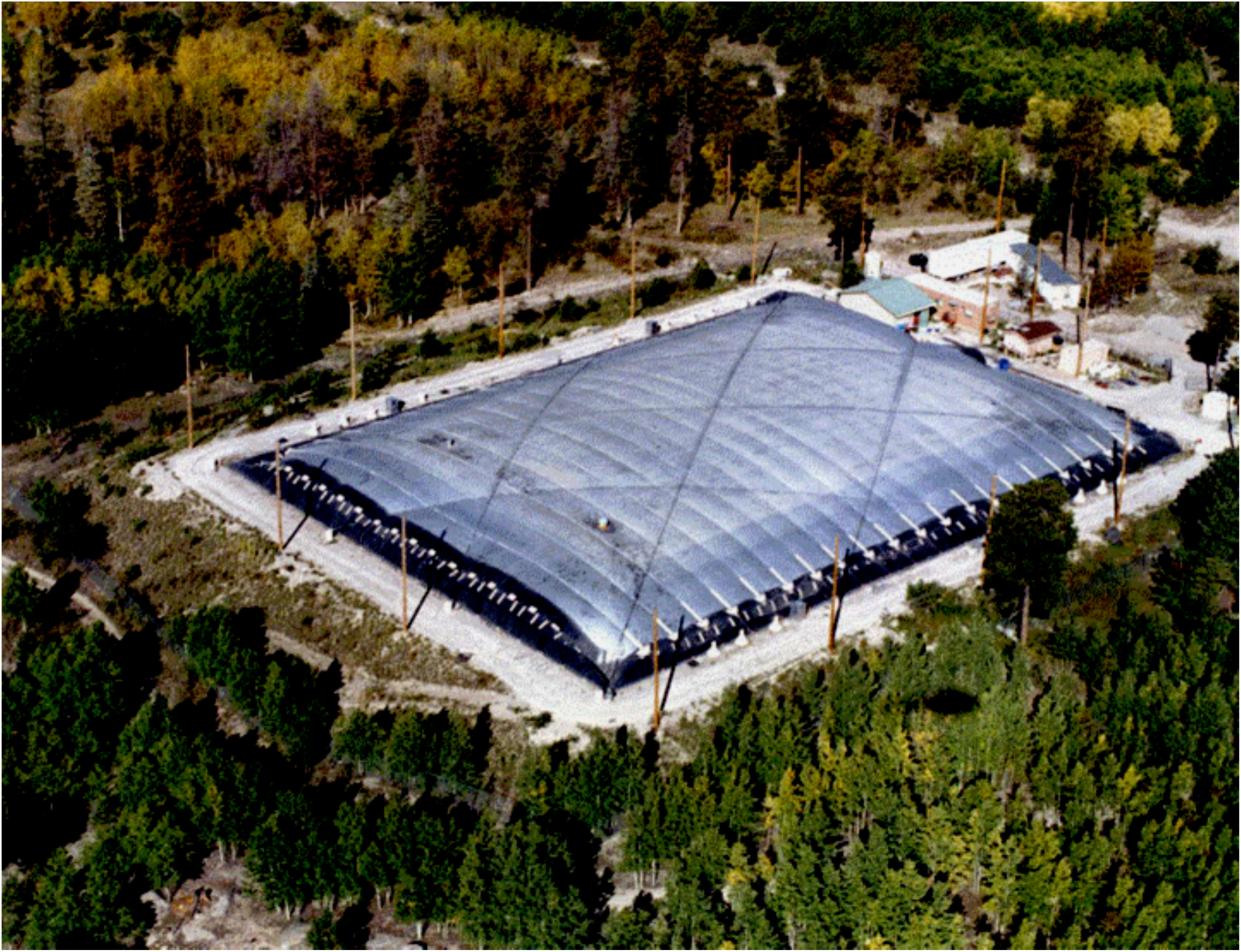}
\end{center}
\vspace{-0.3cm}
\caption{\label{hess_view} 
Examples of gamma-ray experiments: Cherenkov telescope H.E.S.S. (left) and  water pool Milagro (right)
}
\end{figure}

The minimal energy of a charged particle is
\begin{equation}
\gamma_{min}=\frac{E_{min}}{M} = \frac{n}{\sqrt{n^2-1}} ~.
\label{Cherenkov_energy}
\end{equation}

Particles with higher energy will produce a cone of Cherenkov light. This effect is used by Cherenkov telescopes
for air (H.E.S.S., MAGIC, Veritas, CTA) and by ground experiments in water (Milagro, HAWK). Detection of the shower in air and in water 
is illustrated in Fig.~\ref{cherenkov}.

We present examples of such experiments in Fig.~\ref{hess_view}.  On the left panel we show a view of the H.E.S.S.
experiment. This experiment made the most significant contribution to the development of TeV gamma-ray astrophysics in recent years.
On the right panel we show the Milagro experiment, a pioneering experiment in water Cherenkov techniques.

\subsection{Point sources of TeV gamma rays}

\begin{figure}[ht]
\begin{center}
\includegraphics[width=0.5\textwidth,angle=0]{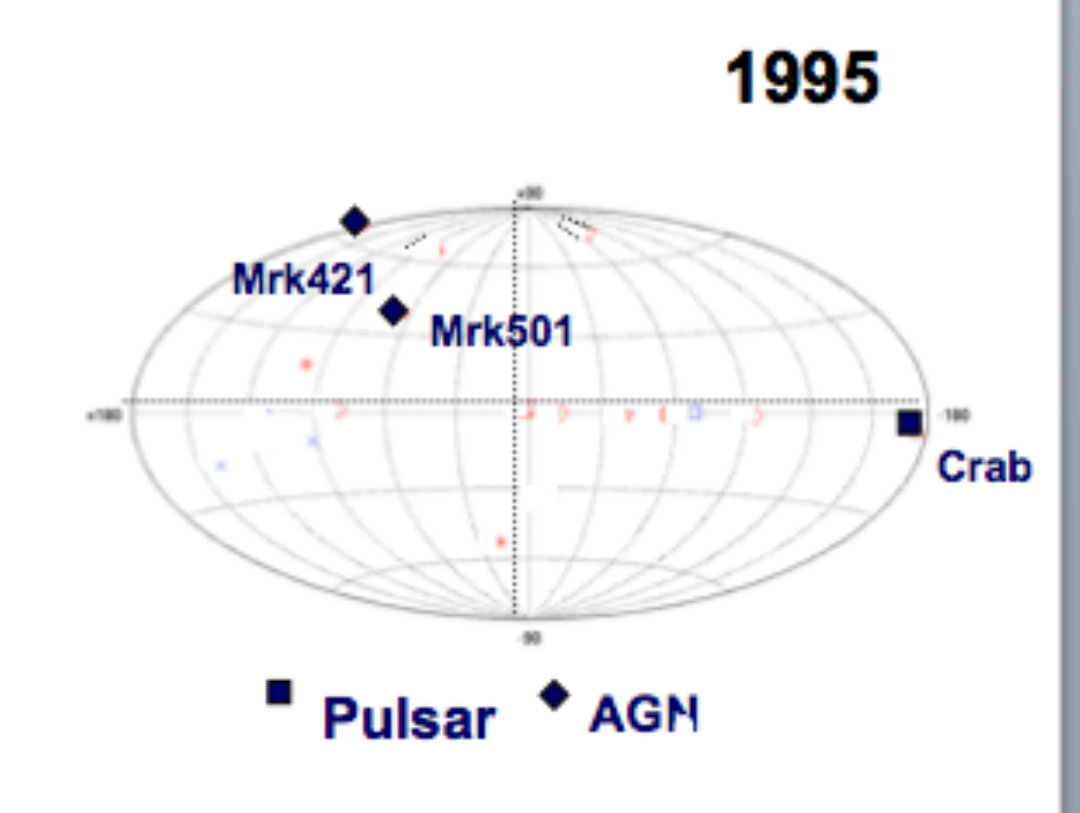}
\includegraphics[width=0.38 \textwidth,angle=0]{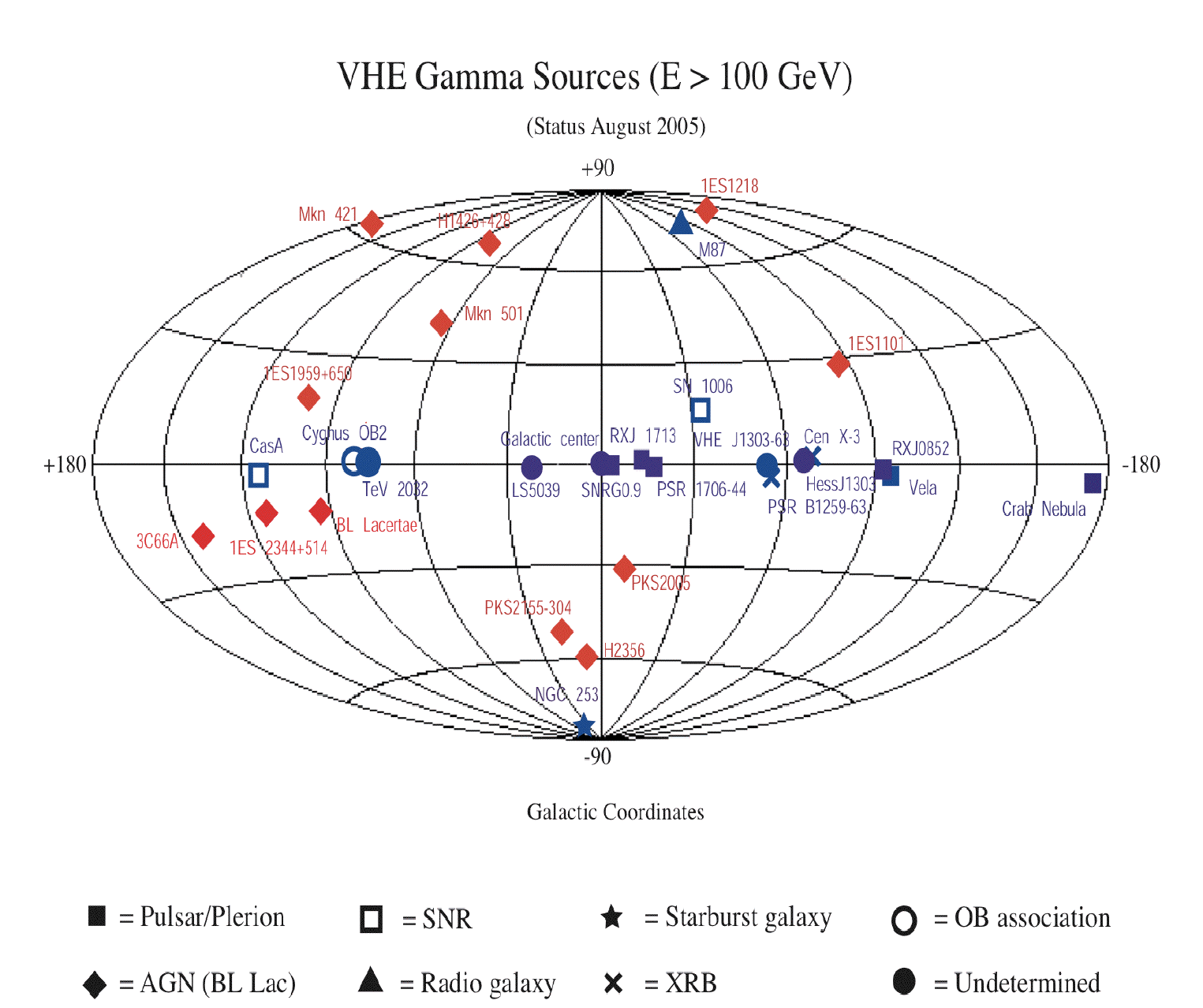}
\includegraphics[width=0.5\textwidth,angle=0]{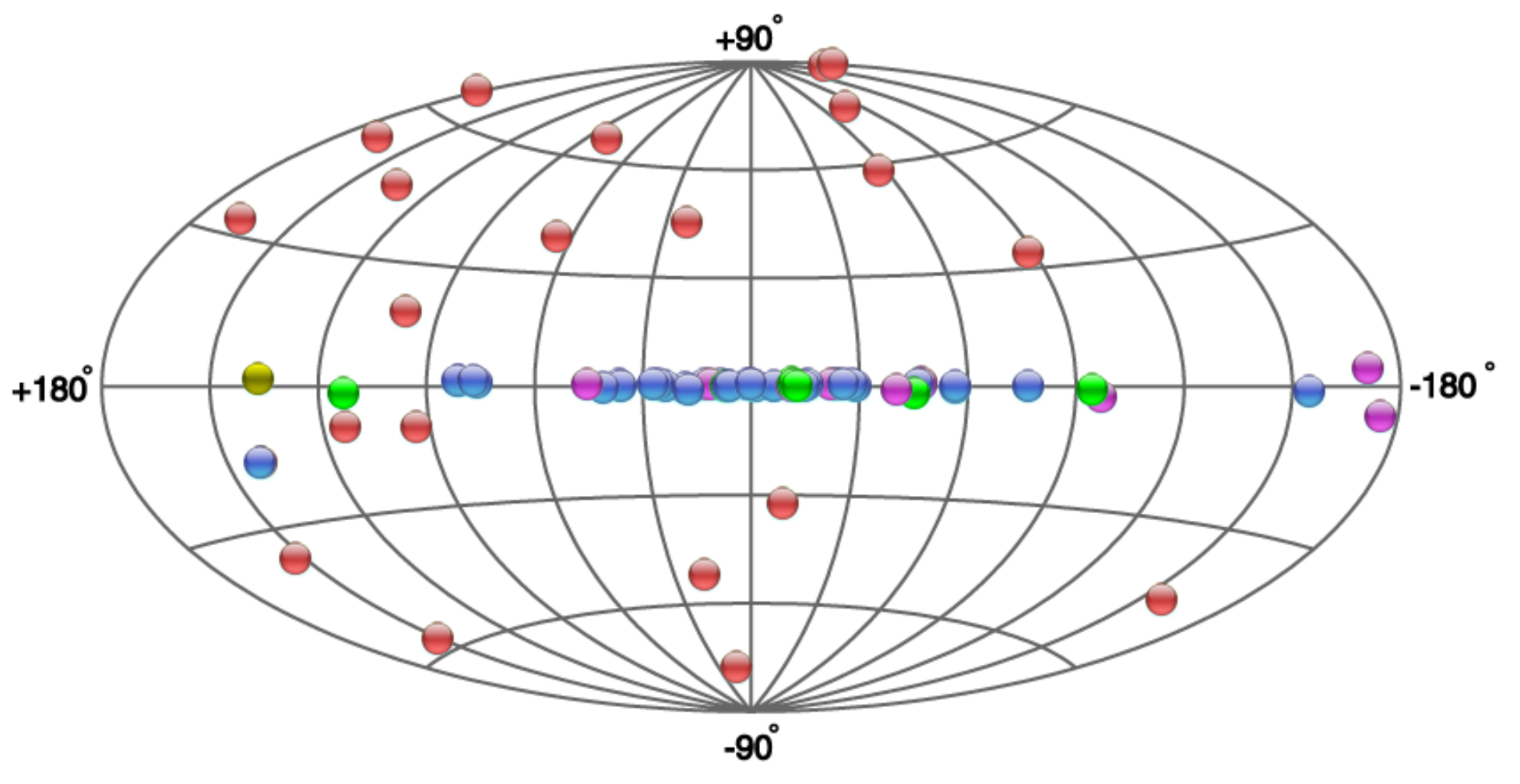}
\includegraphics[width=0.18 \textwidth,angle=0]{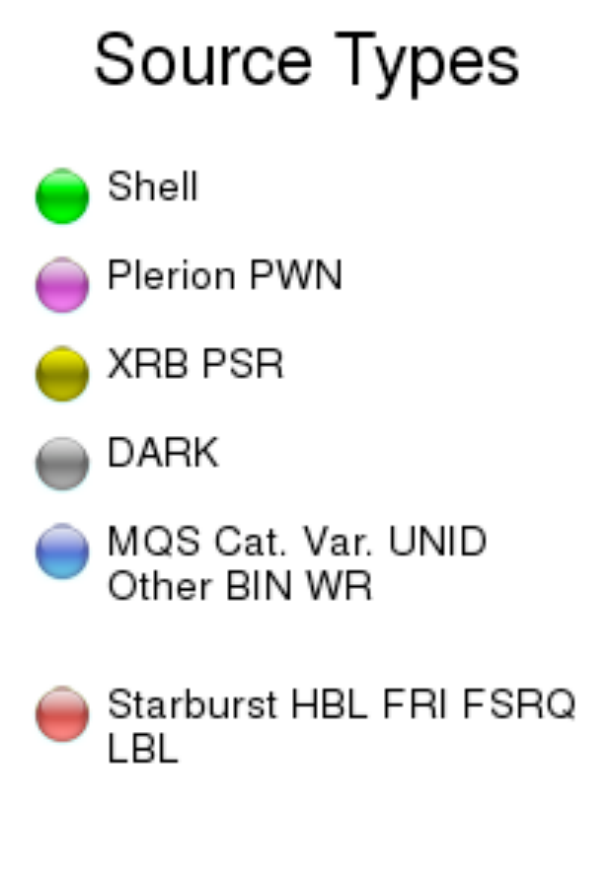}
\end{center}
\vspace{-0.3cm}
\caption{\label{sky_view} 
Sky in the TeV gamma rays with 3 sources in 1995 (top left), 32 sources in 2005 (top right), and 80 sources in 2009 (bottom)
}
\end{figure}

TeV gamma-ray astrophysics is developing  very quickly. One can see the number of detected sources in the sky as a function of time in  Fig.~\ref{sky_view}.
From 3 sources in 1995 one has 32 sources in 2005 and 80 sources in 2008.  In addition not only does the number of observed sources grow,
but also the number of different populations of sources. This is a very important fact for future experiments with better sensitivity like the Cherenkov Telescope  Array (CTA). 
They would have a very large potential for detecting many different classes of sources. 

In particular, in Fig.~\ref{sky_view} on the bottom panel,
red circles show extragalactic sources which contain BL Lac objects, radio galaxies, and starburst galaxies. Also in the galactic plane there
are many different classes of objects, which include supernova shells, pulsar wind nebulas, pulsars, binary systems and dark objects. 
Dark objects mean they were detected in gamma rays, but there is no corresponding source in other wavebands.

\begin{figure}[ht]
\begin{center}
\includegraphics[width=0.6\textwidth,angle=0]{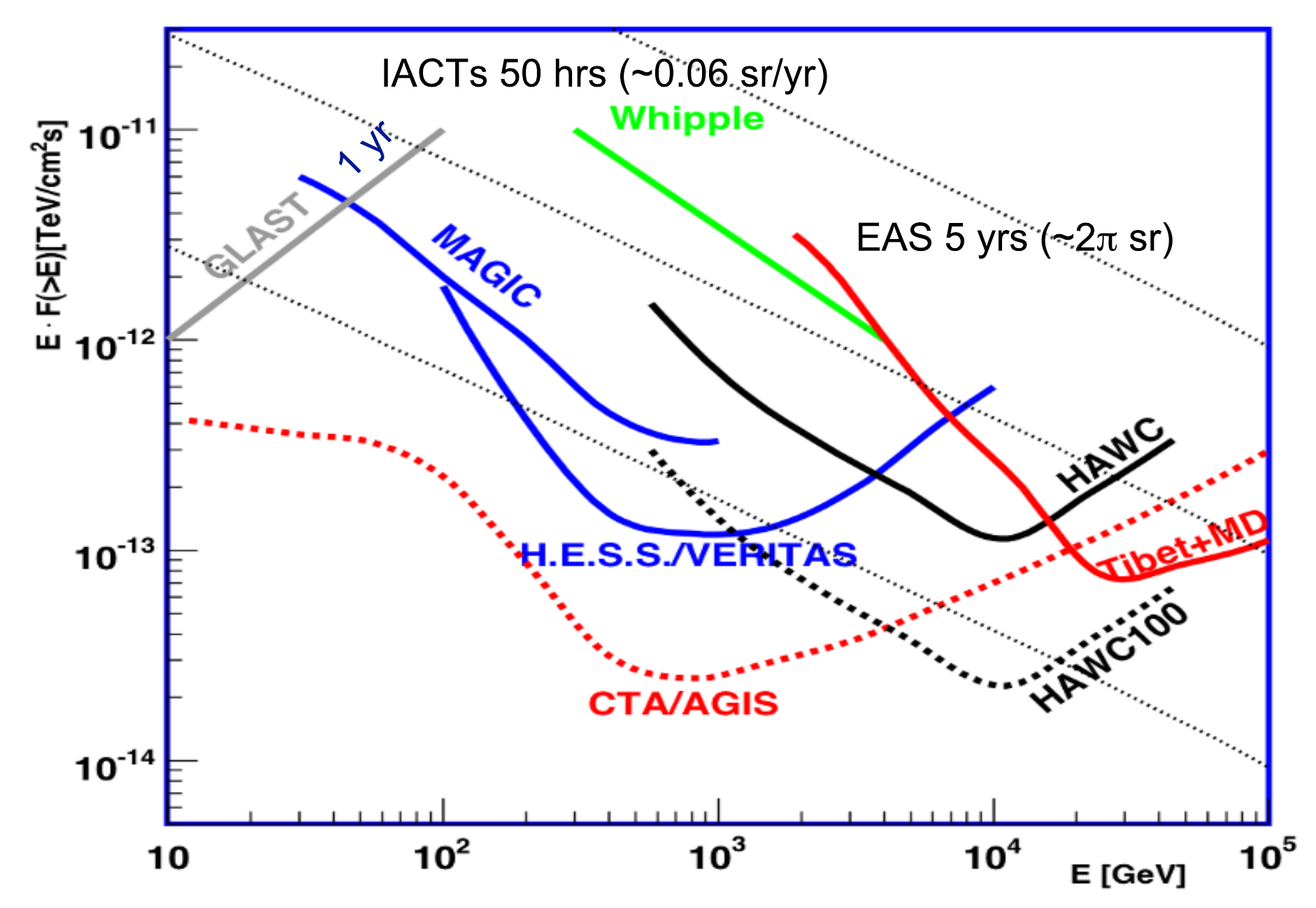}
\end{center}
\vspace{-0.3cm}
\caption{\label{point_source} 
Sensitivity of gamma-ray detectors to point sources, from Ref.~\cite{aharonian_review}}
\end{figure}

\begin{figure}[ht]
\begin{center}
\includegraphics[width=0.4\textwidth,angle=0]{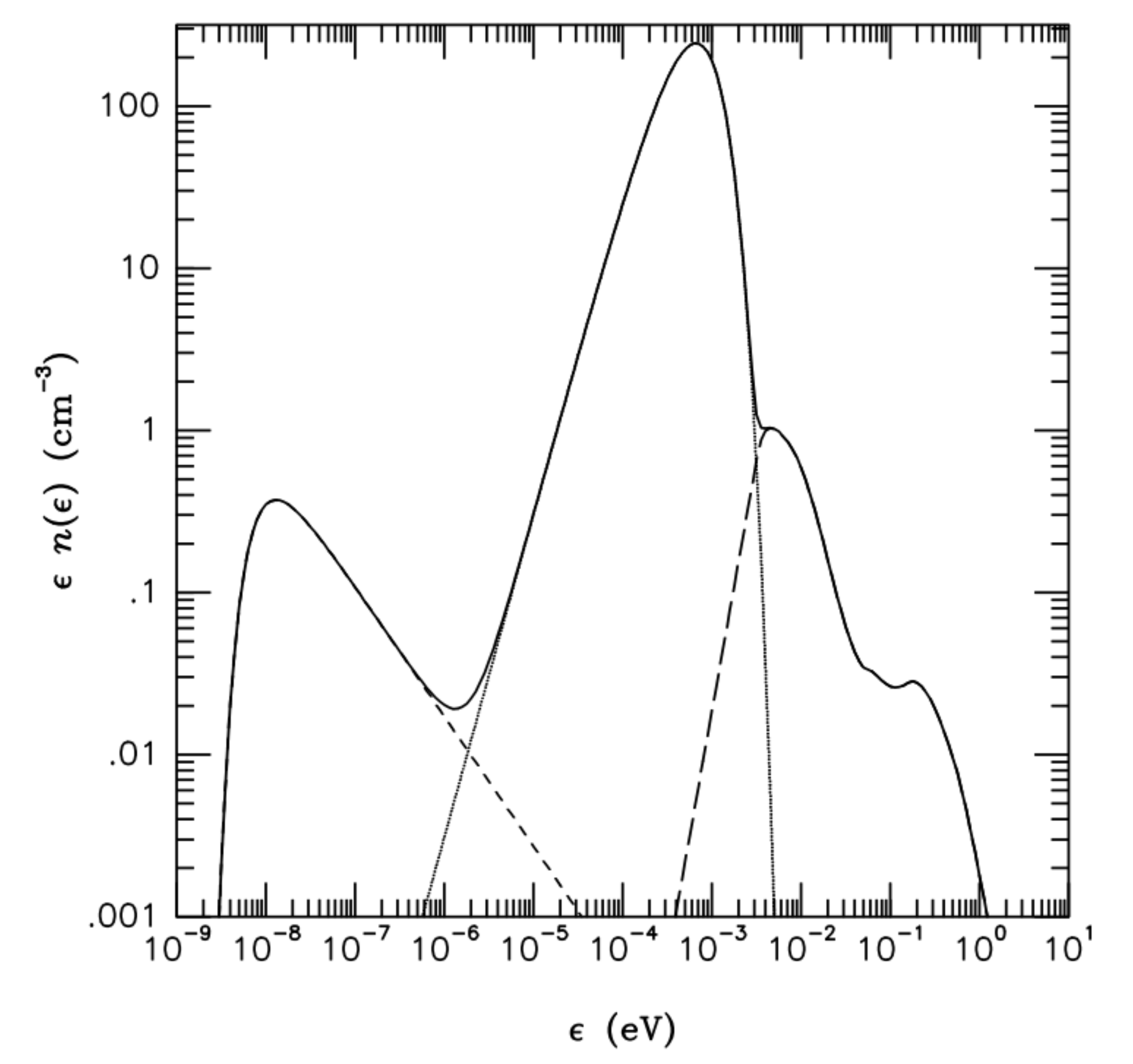}
\end{center}
\vspace{-0.3cm}
\caption{\label{background} 
 Redshift  for gamma rays as a function of energy. Lines show constant optical depth in two models of IR/O background.}
\end{figure}

\begin{figure}[ht]
\begin{center}
\includegraphics[width=0.8\textwidth,angle=0]{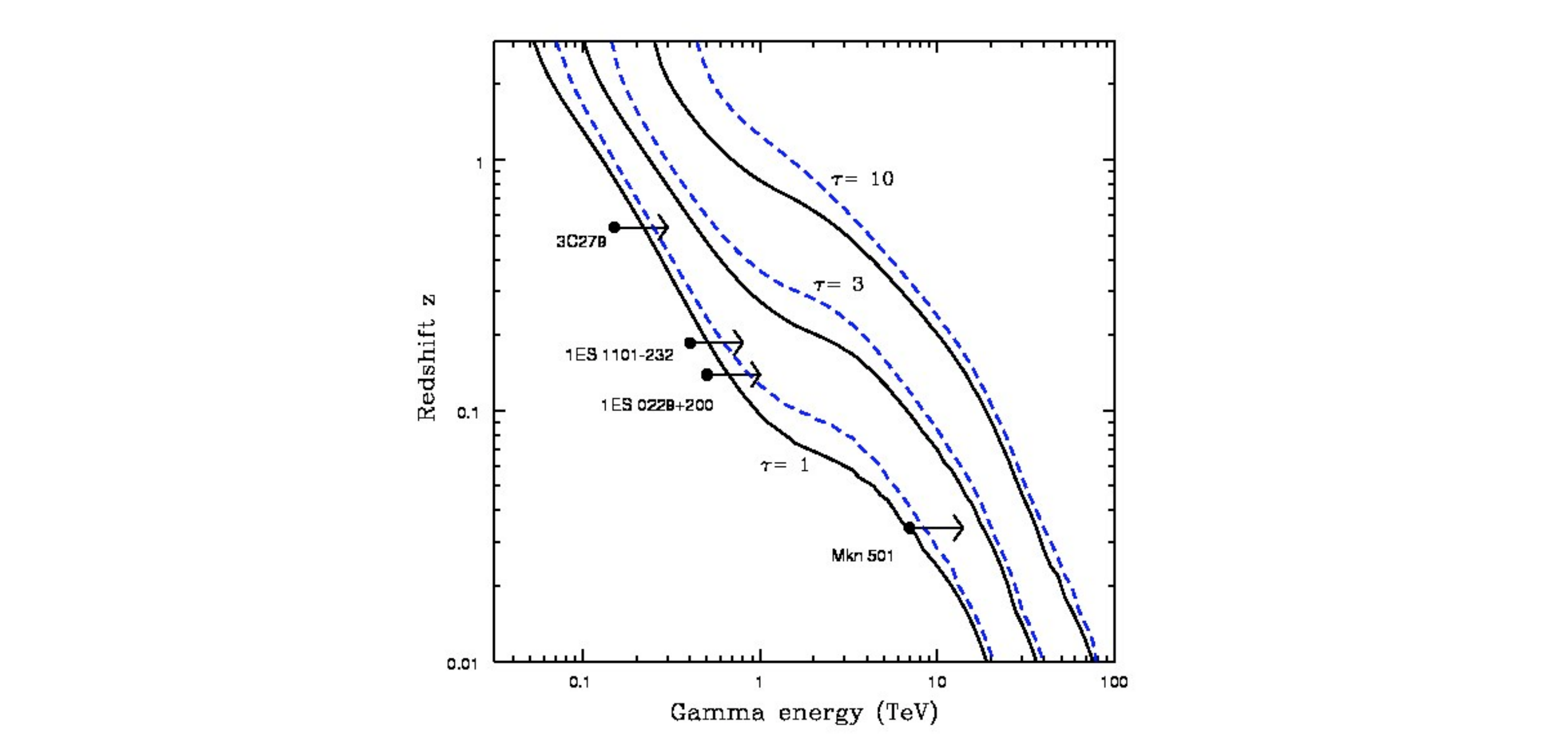}
\end{center}
\vspace{-0.3cm}
\caption{\label{tau} 
 Redshift  for gamma rays as a function of energy. Lines show constant optical depth in two models of IR/O background.}
\end{figure}

The sensitivity of gamma-ray detectors to point sources as a function of energy is shown in Fig.~\ref{point_source}.
The sensitivity of air telescopes is shown for 50 hours of observation for one source. The sensitivity for ground experiments  
is shown for 5 years, but they observe all the sky ($2\pi $sr).  At low energies  $E<10$~GeV the sensitivity of the GLAST (Fermi) satellite is the best.
One year of observations are shown. At large energies $E>10$ TeV the ground air-shower experiments (Tibet) have the best sensitivity. 
Future CTA projects will be orders of magnitude better than present-day experiments
from 10 GeV to 10 TeV energies.

Another important fact is that gamma rays cannot travel freely in the intergalactic space. They interact with optical/infrared 
background photons and disappear producing pairs of electrons and positrons. In Fig.~\ref{background} one can see the main backgrounds 
for gamma-ray propagation. They are shown in units of photon density per cm$^3$.
 The largest contribution comes from the CMB background with 400 photons per  cm$^3$. However, owing to the small energy of CMB photons,
 this background is important only for $E>1000$ TeV. For the experimentally interesting energy range $E<100$ TeV the main backgrounds 
 are infrared and optical. Since those backgrounds are created by galaxies and partly by dust they are strongly model dependent both as a function of energy and as a function of redshift.

Optical depth can be defined as 
\begin{equation}
\tau(E) = R \cdot \sigma_{\gamma\gamma}(E) \cdot n_{back}(z,\epsilon)~,
\label{tau_gamma}
\end{equation}
where $R$ is the distance travelled by photons, $\sigma_{\gamma\gamma}(E)$ is the pair-production cross section,
and $n_{back}(z,\epsilon)$ is the density of background photons. Distances on the cosmological scale are often expressed in terms of redshift.
One can express it through the Hubble law  $R = z \cdot c/H_0 $, where $H_0=70$ km/s/Mpc is the Hubble constant. 
In Fig.~\ref{tau}  contours of constant optical depth $\tau(E)$ are shown on the plane redshift versus energy for $\tau(E) = 1, 3 ,10$
in two different models of IR/O background.

\begin{figure}[htb]
\begin{center}
\includegraphics[width=\textwidth,angle=0]{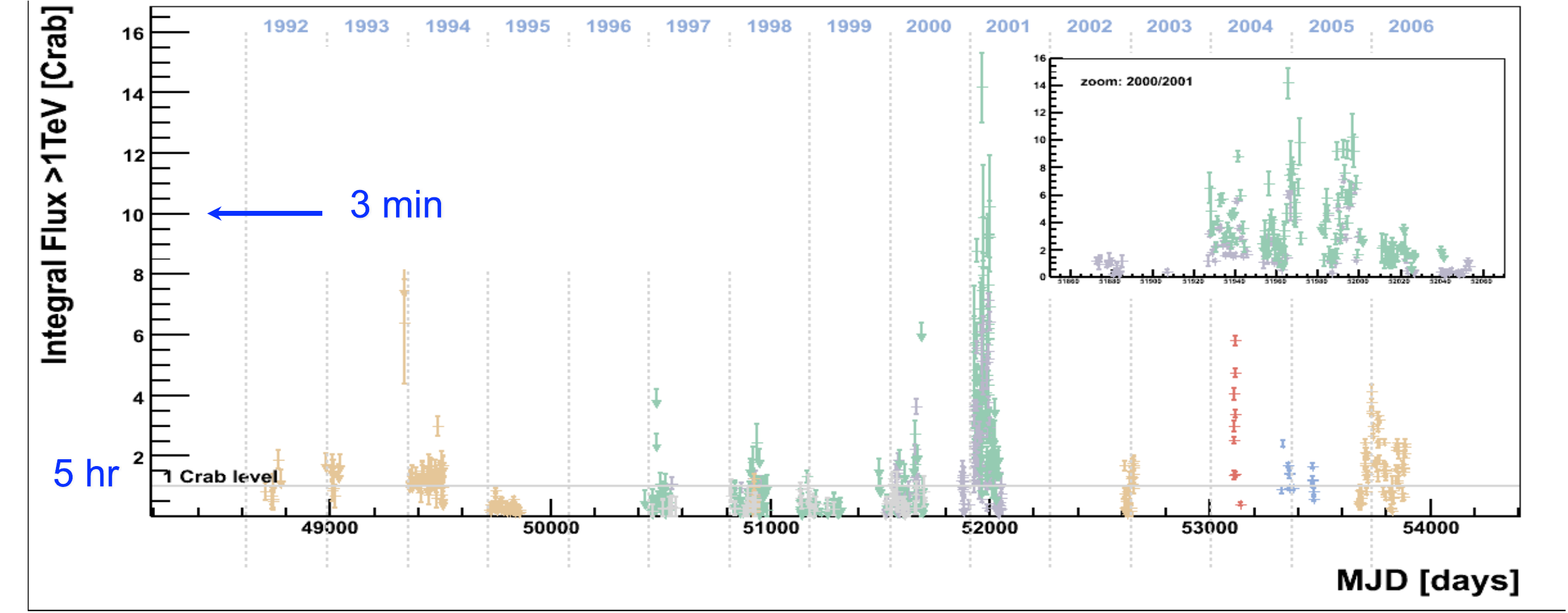}
\end{center}
\vspace{-0.3cm}
\caption{\label{Mkn421} 
Observation of Mkn 421 as a function of time}
\end{figure}

There is one important difference between air Cherenkov telescopes and water Cherenkov detectors. 
In Fig.~\ref{Mkn421} we plot world-wide monitoring of the nearby BL Lac  object Mkn 421 as a function of time.
One can see that air Cherenkov telescopes  can see a signal only on moonless nights, which restricts 
their operation to the corresponding intervals of time. On the contrary, water Cherenkov telescopes operate 
all the time they can see a source, which will allow source activity to be detected all the time. On the other hand, the problem of 
water Cherenkov experiments is poor sensitivity, which will prevent them from detection of 
relatively low fluxes and very fast variations in time. Thus both techniques are complementary to each other.

\begin{figure}[htb]
\begin{center}
\includegraphics[width=\textwidth,angle=0]{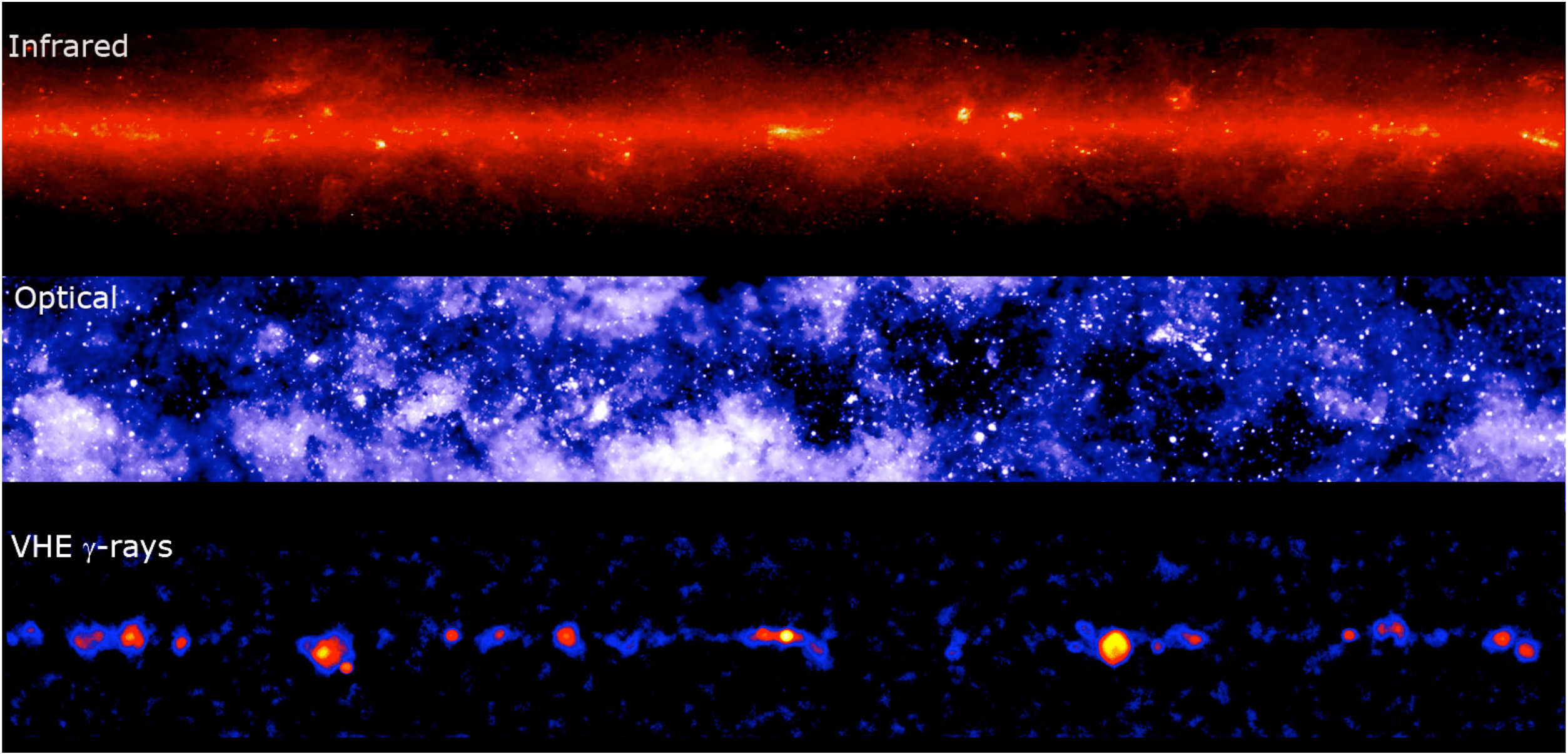}
\end{center}
\vspace{-0.3cm}
\caption{\label{galaxy_hess} 
Central part of the Milky Way galaxy in infrared, optical, and in TeV gamma rays. The TeV gamma-ray sky from H.E.S.S. observations
with a large number of sources. 
}
\end{figure}

\begin{figure}[htb]
\begin{center}
\includegraphics[width=\textwidth,angle=0]{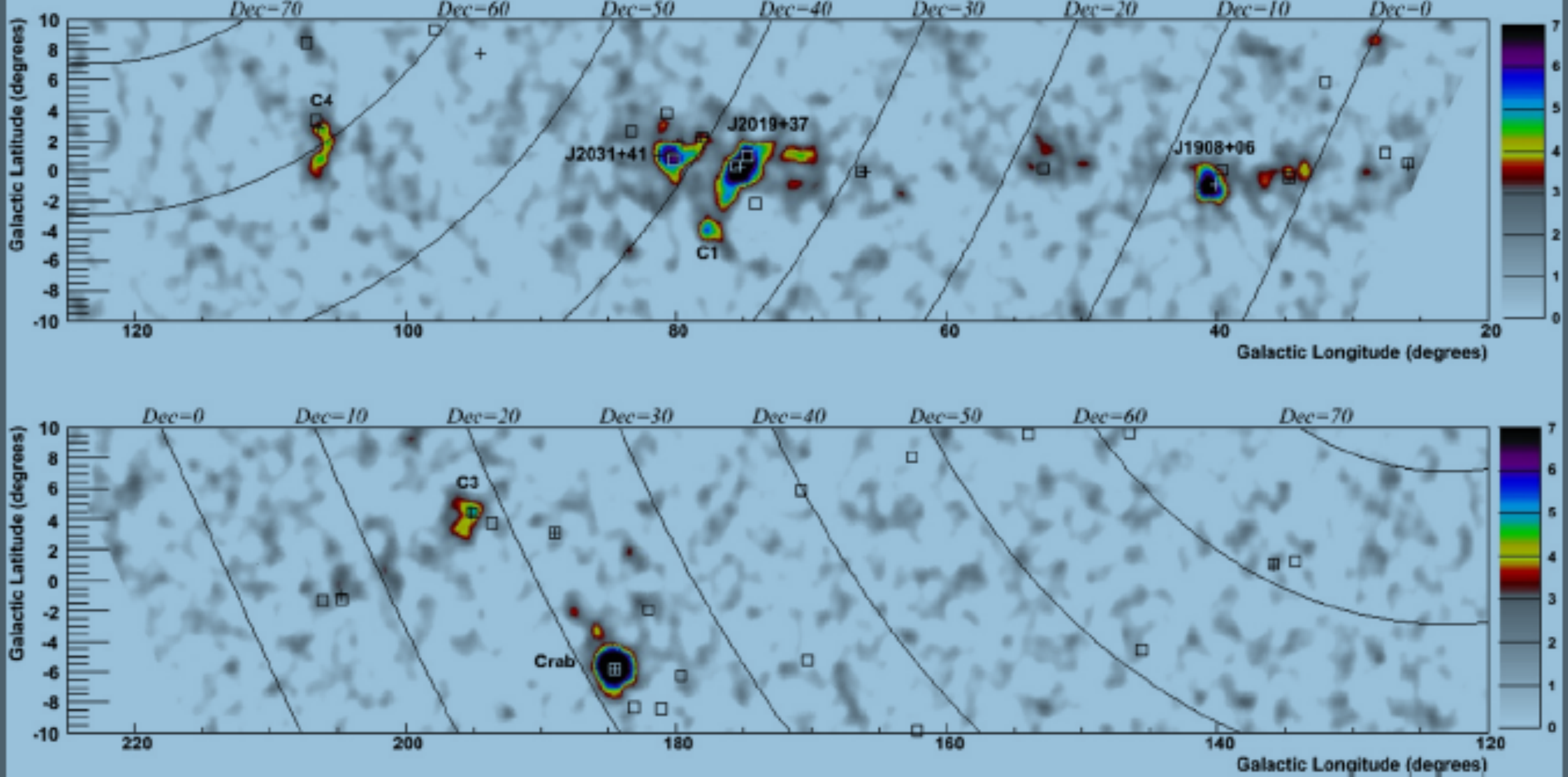}
\end{center}
\vspace{-0.3cm}
\caption{\label{milagro} 
The Milky Way galaxy in TeV gamma rays from galactic longitude 20$^\circ$ to 220$^\circ$ 
and galactic latitude from $-10^\circ$ to $10^\circ$. The image is the
culmination of a seven-year exposure by the Milagro instrument.}
\end{figure}

In Fig.~\ref{galaxy_hess} one can see a view of the central part of the Milky Way galaxy in three energy bands:
optical, infrared, and TeV gamma rays.
At least three astronomical source populations: supernova
remnants (SNRs), pulsar wind nebulae (PWNe), and binary
systems (BSs) are represented in this figure.
In addition, the H.E.S.S. observations of the central region of
our Galaxy revealed a diffuse TeV $\gamma$-ray emission component
which is apparently dominated by contributions from
giant molecular clouds (GMCs). These massive complexes
of gas and dust most likely serve as effective targets for
interactions of relativistic particles from nearby active or recent
accelerators. Thus one may claim that four galactic
source populations are already firmly established as effective TeV  $\gamma$-ray
emitters. Meanwhile, many sources discovered by H.E.S.S. in
the galactic plane remain unidentified. Although some of
these sources might have direct or indirect links to SNRs,
PWNe, and GMCs, one cannot exclude that a fraction of the
H.E.S.S. unidentified sources are related to other source classes.

The Milagro telescope has made the first measurement of
the diffuse TeV gamma-ray flux from the Galactic Disk.
Figure ~\ref{milagro}  shows the Galaxy (as visible from the Northern
Hemisphere) in TeV gamma rays. In addition to the individual
sources discussed above, the image (compiled from Milagro
data) shows the existence of a diffuse TeV gamma-ray flux
between galactic longitudes of 30$^\circ$ and 90$^\circ$.

\subsection{Extragalactic magnetic fields}

\begin{figure}[ht]
\begin{center}
\includegraphics[width=\textwidth,angle=0]{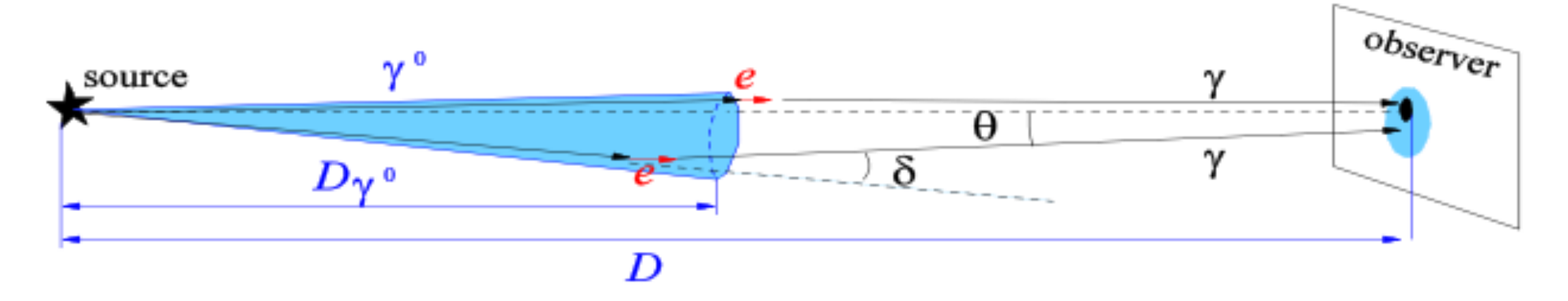}
\end{center}
\vspace{-0.3cm}
\caption{\label{methodEGMF} 
Detection of EGMF through observation of secondary emissions around a point source~\cite{neronov07}}
\end{figure}

Another very important field which will benefit  in the near future from TeV gamma rays 
is Extragalactic Magnetic Fields.  

Indeed, as discussed above, TeV gamma rays emitted by astrophysical sources can be measured  by detectors
on Earth. Practically all TeV gamma rays from galactic sources come directly to the detectors. However, this is not true for
extragalactic sources. As one can see from Fig.~\ref{tau}, even for nearby sources like Mkn 501, gamma rays 
with $E>10$ TeV cannot come freely to the detector. 
 The pair production on Extragalactic Background Light (EBL) reduces the flux of  $\gamma $-rays from the source by
\begin{equation}
\label{absorb}
F(E_{\gamma_0})=F_0(E'_{\gamma_0}(z_E))e^{-\tau(E_{\gamma_0},z_E)},
\end{equation}
where 
$F(E_{\gamma_0})$ is the detected spectrum, $F_0(E'_{\gamma_0})$ is the initial spectrum of the source, and $\tau(E_{\gamma_0},z_E)$ is the optical depth Eq.~(\ref{tau_gamma}).  
The typical distance which a primary gamma ray travels is
\begin{equation}
D_{\gamma_0} = D_\gamma(E_{\gamma_0}',z) = 40\frac{\kappa }{(1+z)^2} \left[\frac{E_{\gamma_0}'}{20\mbox{ TeV}}\right]^{-1}\mbox{ Mpc}~,
\label{Dgamma0}
\end{equation}
where a numerical factor $\kappa=\kappa(E_{\gamma_0},z)\sim 1$ accounts for the model uncertainties. 

The cascade electrons lose their energy via Inverse Compton (IC) scattering of the CMB photons within the distance
\begin{equation}
D_e=\frac{3m_e^2c^3}{4\sigma_TU_{\rm CMB}'E_e'}\simeq  10^{23}(1+z_{\gamma\gamma})^{-4}\left[\frac{E_e'}{10\mbox{ TeV}}\right]^{-1}\mbox{ cm}
\end{equation}
The deflection angle of the $e^+e^-$ pairs, accumulated over the cooling distance, depends on the correlation length of the magnetic field, $\lambda_B$.  Note also that electrons and positrons travel much shorter distances than primary  photons.

The $e^+e^-$ pairs 
 produced in interactions of multi-TeV $\gamma$-rays with EBL photons produce secondary $\gamma$-rays via IC scattering of the Cosmic Microwave Background (CMB) photons. Typical energies of the IC photons reaching the Earth are
\begin{equation}
E_{\gamma} = \frac{4}{3} (1+z_{\gamma\gamma})^{-1}\epsilon_{CMB}'\frac{E_e'^2}{m_e^2}\simeq 0.32
\left[\frac{E'_{\gamma_0}}{20 \mbox{ TeV}}\right]^2\mbox{ TeV}
\label{Esec}
\end{equation}
where $\epsilon_{CMB}'=6\times 10^{-4}(1+z_{\gamma\gamma})$~eV is the typical energy of CMB photons. In the above equation we have assumed that the energy of a primary $\gamma$-ray is $E'_{\gamma_0}\simeq 2E_e'$ with $E_{\gamma_0}'$ being the energy of the primary $\gamma$-rays at the redshift of the pair production. Upscattering of the infrared/optical background photons gives a sub-dominant contribution to the IC scattering spectrum because the energy density of CMB is much higher than the density of the infrared/optical background.

 Deflections of $e^+e^-$ pairs produced by the $\gamma$-rays, which were initially emitted slightly
away from the observer, could lead to `redirection' of the secondary cascade photons toward the observer. This effect leads to the appearance of two potentially observable effects: extended emission around an initially point source of $\gamma$-rays \cite{neronov07,elyiv09,kachelriess09} and delayed `echo' of $\gamma$-ray flares of extragalactic sources \cite{plaga,japanese}.

\begin{figure}[ht]
\begin{center}
\includegraphics[width=\textwidth,angle=0]{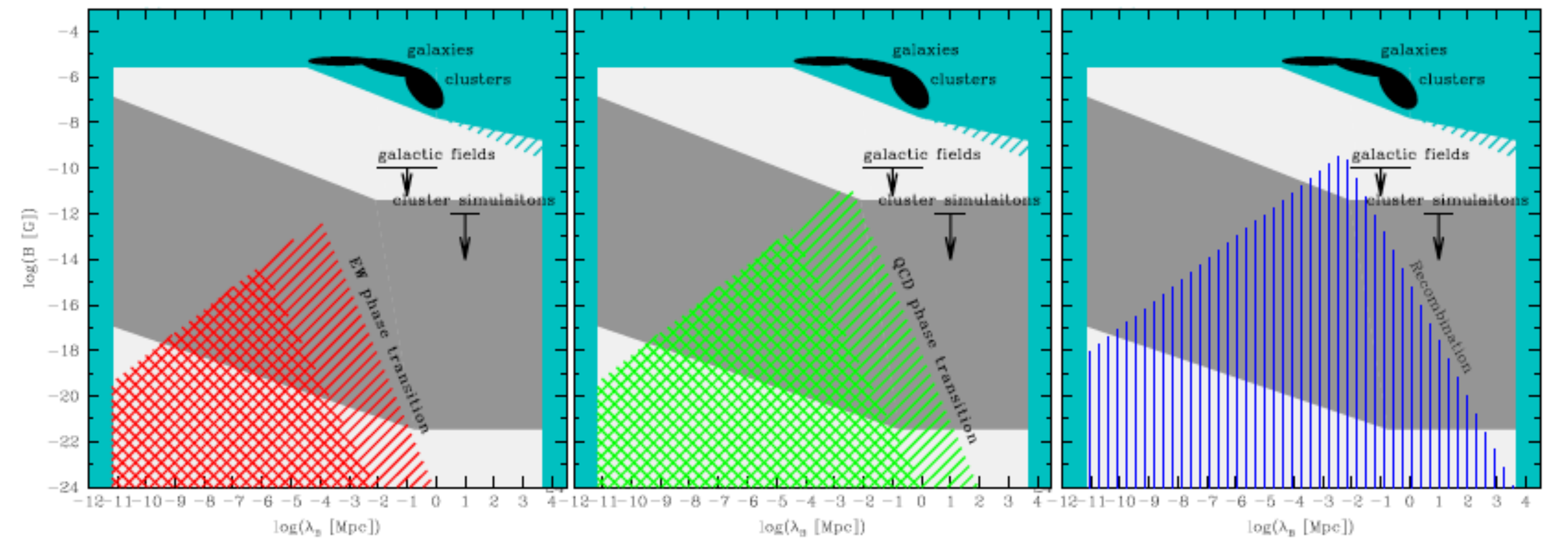}
\end{center}
\vspace{-0.3cm}
\caption{\label{egmf_sens} 
Model predictions and estimates for the EGMF strength. Cyan shaded region excluded by present day measurements.
Black ellipses show measurements of the field in the Galaxy and galaxy clusters.
{\bf Left panel:} left and right hatched regions show theoretically allowed range of values of ($\lambda_B$,B)
for non-helical and helical fields generated at the epoch of electroweak phase transition during radiation-dominated era. {\bf Middle
panel:} left and right hatched region show ranges of possible ($\lambda_B$,B) for non-helical and helical magnetic fields produced during the
QCD phase transition. {\bf Right panel:} hatched region is the range of possible ($\lambda_B$,B) for EGMF generated during recombination
epoch. Dark grey shaded region shows the range of ($\lambda_B$,B) parameter space accessible for the 
$\gamma$-ray measurements via 
$\gamma$-ray observations~\cite{Neronov:2009gh}.}
\end{figure}

The above processes are illustrated in Fig.~\ref{methodEGMF}.  Electron deflection $\delta$ depends on the magnetic field in the region of deflection.
Note, that, in principle, EGMF depends on the redshift, $B'=B'(z)$. In the simplest case, when the magnetic field strength changes only as a result of expansion of the Universe, $B'(z)\sim B_0(1+z)^2$, where $B_0$ is the present epoch EGMF strength.
This gives
\begin{eqnarray}
\delta&=&\frac{D_e}{R_L}\simeq 3\times 10^{-6}(1+z_{\gamma\gamma})^{-4}\left[\frac{B'}{10^{-18}\mbox{ G}}\right]\left[\frac{E_e'}{10\mbox{ TeV}}\right]^{-2}\nonumber\\
&\simeq& 3\times 10^{-6}(1+z_{\gamma\gamma})^{-2}\left[\frac{B_0}{10^{-18}\mbox{ G}}\right]\left[\frac{E_e'}{10\mbox{ TeV}}\right]^{-2}
\label{delta1}
\end{eqnarray}

Knowing the deflection angle of electrons, one can readily find the angular extension of the secondary IC emission from the $e^+e^-$ pairs 
\begin{equation}
\Theta_{\rm ext}\simeq
\left\{
\begin{array}{ll}
0.5^\circ(1+z)^{-2} \left[\frac{\tau_\theta}{10}\right]^{-1} 
&\\
\left[\frac{\displaystyle E_\gamma}{\displaystyle 0.1\mbox{ TeV}}\right]^{-1}\left[\displaystyle \frac{B_0}{\displaystyle 10^{-14}\mbox{ G}}\right],& \lambda_B'\gg D_e\\
&\\
0.07^\circ (1+z)^{-1/2} \left[\frac{\tau_\theta}{10}\right]^{-1}& \\
\left[\frac{\displaystyle E_\gamma}{\displaystyle 0.1\mbox{ TeV}}\right]^{-3/4}
\left[\frac{\displaystyle B_0}{\displaystyle 10^{-14}\mbox{ G}}\right]
\left[\frac{\displaystyle \lambda_{B0}}{\displaystyle 1\mbox{ kpc}}\right]^{1/2},&\lambda_B'\ll D_e
\end{array}
\right.
\label{Thetaext}
\end{equation}
This is a key point for detection of the field, since extended emission depends on energy in a well-defined way
and can be reconstructed using independent  measurements at different energies.

The possible ranges of the ($\lambda_B$,B) 
parameter space are shown in Fig. ~\ref{egmf_sens} for the cases when
magnetogenesis proceeds during electroweak or QCD
phase transitions or at the moment of recombination.

It is interesting to note that predictions for the
strength and correlation length of the primordial magnetic fields  fall in a region of ($\lambda_B$,B) parameter space which
is not accessible for the existing measurement techniques,
such as Faraday rotation or Zeeman splitting methods.
However, it turns out that this region of ($\lambda_B$,B) parameter space is accessible for the measurement techniques
which exploit the potential of the newly opened field of
very-high-energy (VHE) $\gamma$-ray astronomy ~\cite{Neronov:2009gh}.

\subsection{Summary}

Gamma-ray astronomy works, hundreds of sources have been detected in the GeV energy range and about one hundred in 
TeV energies.

There are several major questions to be answered in the near future:

\begin{itemize}
 \item One needs to understand the hadronic component  in a variety of astrophysical sources.

\item Extragalactic IR/O backgrounds have already been  constrained by observations 
of TeV sources to factor two uncertainty. The next step is precision determination of  those backgrounds
using measurements of many sources at different redshifts.

\item For the first time one has a possibility to study primordial magnetic fields
through TeV gamma-ray measurements.  We can test models of primordial magnetic fields in the near future.

\end{itemize}

There are several other important issues which were not discussed in this Lecture due 
to lack of time. The corresponding questions are:
\begin{itemize}

\item  Good measurements of blazar flairs can help to understand gravity near black holes. 

\item TeV gamma rays give one more constraint/signature on Dark Matter.     

\item Constraints on exotic physics (LIV, etc.) will be improved.

\end{itemize}

%
%
%
%
%

%
%
%

\section{High-energy neutrinos}
\label{chapter:neutrinos}

\subsection{Introduction}
\label{sec:nu_intro}

In this lecture we discuss theoretical predictions and experimental efforts to detect Ultra-High Energy 
neutrinos. In Section~\ref{sec:nu_exp} we discuss possible ways to detect UHE neutrinos and their
corresponding experiments.
 In Section~\ref{sec:nu_cosmo}  we show theoretical predictions for UHE neutrino fluxes and present the 
 status of experimental searches for such fluxes.
 In Section~\ref{sec:nu_point} we discuss another possibility to detect Galactic neutrino sources 
 at  multi-TeV energies. In Section~\ref{sec:nu_sum} we summarize all the results of this lecture.
 
\subsection{High-energy neutrino experiments}
\label{sec:nu_exp}

There are three types of ultra-high energy (UHE) neutrino experiments. 

First, neutrinos  can be detected 
by UHECR experiments. There are two possibilities for this. First, one can use the fact that the atmosphere horizontally 
has depth 36 times the vertical depth. Relatively   young electromagnetic horizontal showers can be caused by neutrinos only. 
Hadronic showers at such a depth consist of  muons only.  Second, one can look for events penetrating the Earth 
in the tau-neutrino channel, i.e., look for upward-going events.   This was used by the Auger experiment (see   Fig.~\ref{PAO}).
The resulting
limit on neutrino flux is shown in Fig.~\ref{cosmogenic}. Also less significant limits were presented by previous UHECR
experiments including Fly's Eye, AGASA, and HiRes (the HiRes limit is also shown in Fig.~\ref{cosmogenic}).

\begin{figure}[htb]
\begin{center}
\includegraphics[width=0.35\textwidth,angle=0]{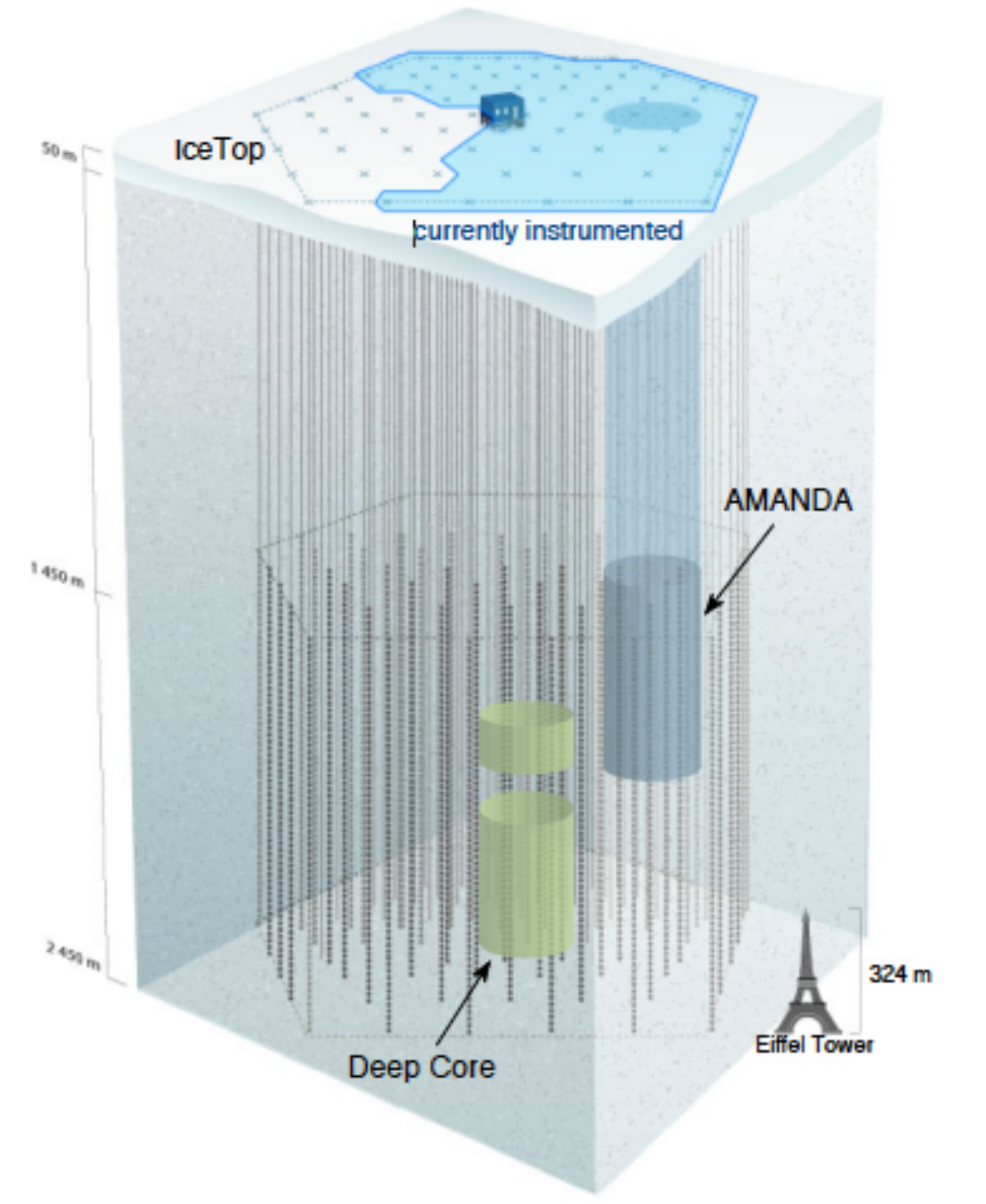}
\hspace{1cm}
\includegraphics[width=0.30\textwidth,angle=0]{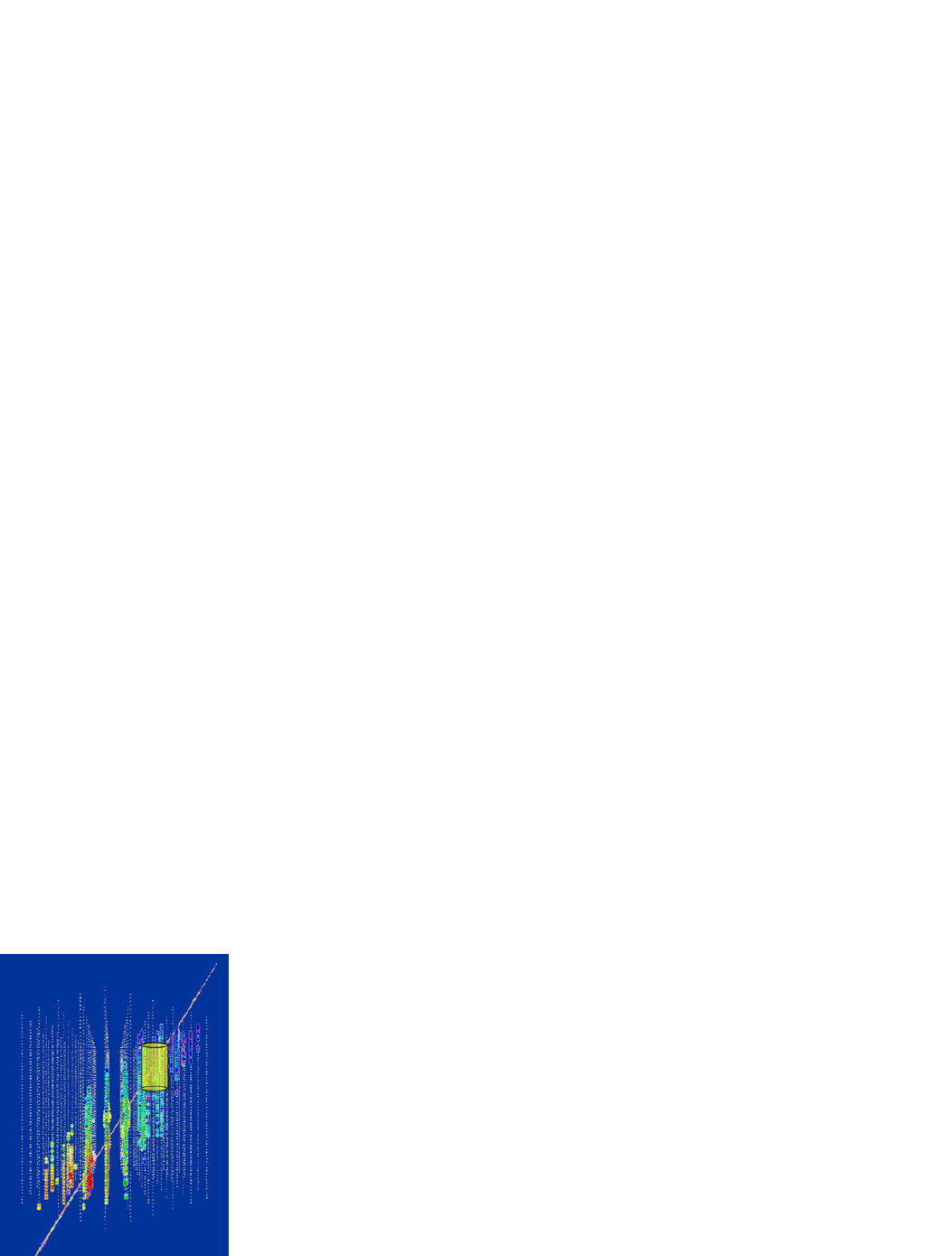}
\end{center}
\vspace{-0.3cm}
\caption{\label{icecube}
IceCube detector. {\bf Left:} Configuration of the IceCube detector. Eighty strings will be located at a depth of 1.5 km in the Antarctic ice
filling a volume of one cubic kilometre. The present construction stage is also shown~\cite{icecube_status}. {\bf Right:} Simulation of a high-energy neutrino event in the IceCube detector~\cite{Halzen:2008zj}.
}
\end{figure}

Second, one can detect neutrinos in the water or in the ice by detecting  
 Cherenkov light created by corresponding leptons after neutrino interaction in the medium. 
There are two important backgrounds for such measurement. First, secondary leptons,
mostly muons,  should not be confused with secondary muons from extensive air showers in the atmosphere.
In order to reduce the background of atmospheric muons one has to put the detector at a depth greater than one kilometre 
from the surface. Second, there are atmospheric neutrinos created by the same cosmic rays, which would produce 
isotropy in the space energy-dependent background.  In order to fight this background, one either has to go to high energies $E>10^{15-16}$ eV,
where it is small, or look for point sources on top of this background.

 Experiments that  worked with these techniques in the past were 
Baikal and ANTARES in water and AMANDA in ice. All those experiments had a volume $0.1$ km$^3$ or less.
The new-generation experiment IceCube  with a volume of 1~km$^3$ is in the construction stage at the moment. In Fig.~\ref{icecube} 
in the left panel one can see the configuration of this experiment, which consists of 80 strings, filling a cubic kilometre volume 
in the Antarctic ice at a depth of 1.5 km from the surface. Strings already implemented are shaded blue 
on top of the picture (see Ref.~\cite{icecube_status} for more details).  Also, as shown in the figure the top of the detector is covered by an array of ice tanks (ice top). 
In the right panel one can see a Monte Carlo simulation of a high-energy neutrino event, detected 
by the IceCube experiment. First results of this experiment will be discussed in 
Section~\ref{sec:nu_point}.

\begin{figure}[htb]
\begin{center}
\includegraphics[width=0.6\textwidth,angle=0]{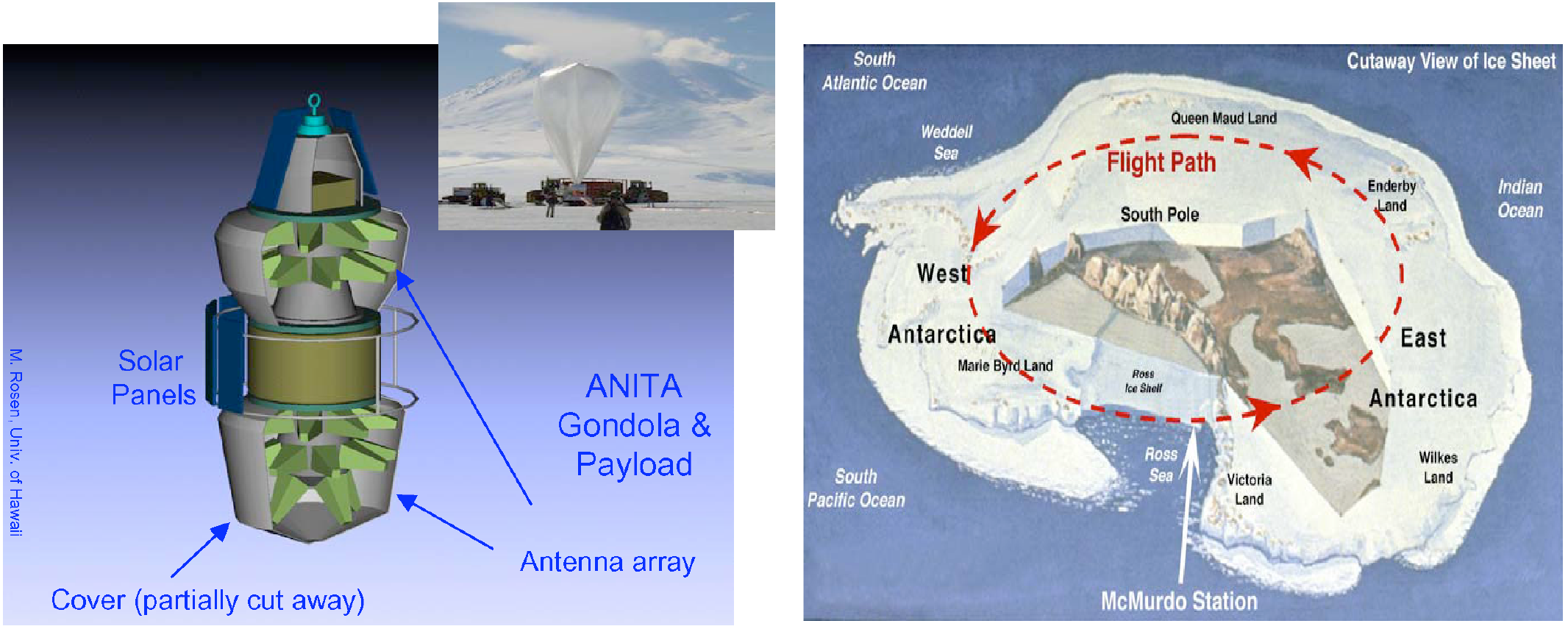}
\end{center}
\vspace{-0.3cm}
\caption{\label{anita_view}  ANtarctic Impulsive Transient Array
 (ANITA) radio balloon experiment. Array of radio antennas flying in the  ballon,
as shown on the left panel. It flies in circles over the Antarctic ice at a  height of 37 km (see right panel) and looks for radio signals which 
UHE neutrinos create in the ice.
}
\end{figure}

Finally, radio neutrino experiments exploit the Askaryan effect in which strong
coherent radio emission arises from electromagnetic showers
in any dielectric medium. High-energy neutrinos trigger a cascade of electromagnetic particles in the medium, 
which has net charge and can emit an analogue of Cherenkov light in the radio energy range. 
The main point of this effect is that the length of the radio wave is macroscopic (tens of centimetres) and is bigger than the size of the cascade itself. 
This in turn means that all 
electrons in the cascade emit coherently.     
The effect was first observed in 2000 at SLAC. 
Recently  the Askaryan effect has been clearly confirmed and characterized
for ice as the medium, as part of the pre-flight calibration
of the ANITA-1 payload.  The Askaryan effect can be seen only at high energies $E>10^{17-18}$ eV.
 Experiments using this effect benefit from the absence of atmospheric neutrino flux at 
 such high energies, but they also have to look over a huge effective volume in order 
 to see tiny neutrino fluxes at highest energies.

Experiments that used this effect to search for UHE neutrinos
are FORTE ~\cite{FORTE}, RICE~\cite{RICE2006},  and ANITA~\cite{ANITAlite,ANITA2008}.
FORTE is a satellite experiment, which, in
 particular, looked over the Greenland ice. Unfortunately, the threshold of this experiment was very high,
 $E_\nu>10^{22}$ eV, so it could test only exotic top-down models.
  RICE was an array  of radio antennas located in the ice at the South Pole
at the same place as the AMANDA experiment. This experiment presented its final results in 2006.  
 Finally, the most advanced for the moment of this kind of experiment is the  ANtarctic Impulsive Transient Array
 (ANITA) radio balloon experiment, see Fig.~\ref{anita_view}.  In the left panel one can see an array of radio antennas 
 in the balloon.  In the right panel one can see a schematic map of flight over the Antarctic at a height of 37 km.

\subsection{Search for cosmogenic  neutrinos}
\label{sec:nu_cosmo}

\begin{figure}[htb]
\begin{center}
\includegraphics[width=0.6\textwidth,angle=0]{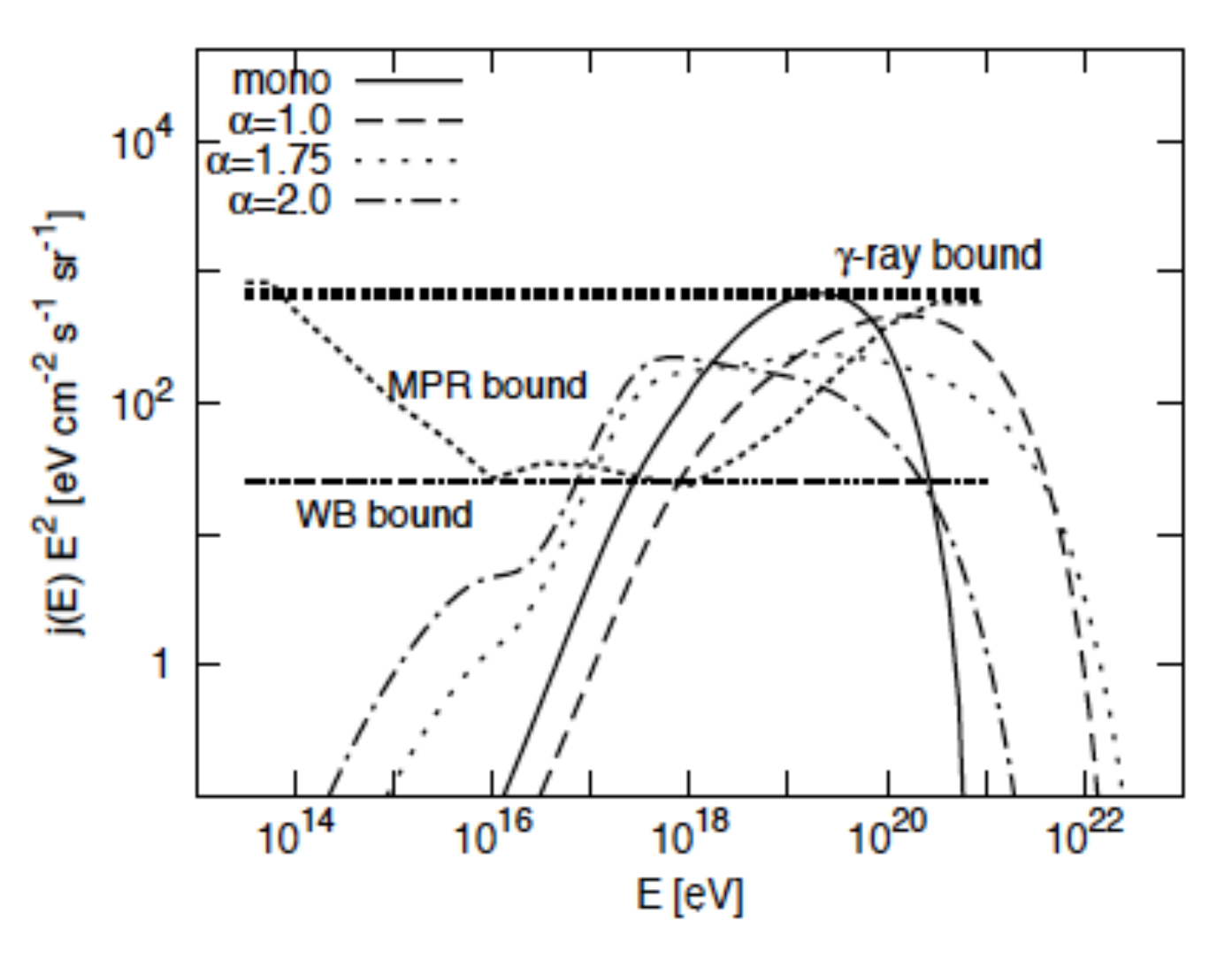}
\end{center}
\vspace{-0.3cm}
\caption{\label{cosmogenic}
Predictions of cosmogenic neutrino fluxes and theoretical bounds on them~\cite{Kalashev:2002kx,Semikoz:2003wv}
}
\end{figure}

As discussed in Section~\ref{sec:propag} UHECR protons lose their energy in interactions with CMB photons and produce
pions at energies above threshold $E>6 \cdot 10^{19}$ eV.  This GZK threshold was found in 1966~\cite{GZK}. 
As long ago as 1969 Berezinsky and Zatsepin suggested that one can try to observe secondary neutrinos from pion decays and called them
cosmogenic neutrinos~\cite{BZ}. 
Recently the ANITA Collaboration proposed to call such neutrinos Berezinsky--Zatsepin neutrinos, or BZ neutrinos~\cite{ANITA2008}.  Below we
follow this suggestion.

One can calculate the flux of BZ neutrinos theoretically, after fitting the corresponding proton spectrum to the experimental flux 
above some energy. The absolute limit for neutrino flux comes from the fact that gamma rays unavoidably produced from $\pi^0$ decays and from electrons $\pi^\pm$ decays cascade down to GeV energies and the maximum flux of such gamma rays cannot overshoot the EGRET measuremen
shown in Fig.~\ref{photons_EGRET}. This bound on the BZ neutrino flux  is called ``gamma-ray bound'' in Fig.~\ref{cosmogenic}.
Note that there are many additional ways to create photons in the EGRET energy range, including electron--positron pair production discussed in the previous section, so the real BZ neutrino  flux is always lower than this region.

\begin{figure}[htb]
\begin{center}
\includegraphics[width=0.4\textwidth,angle=0]{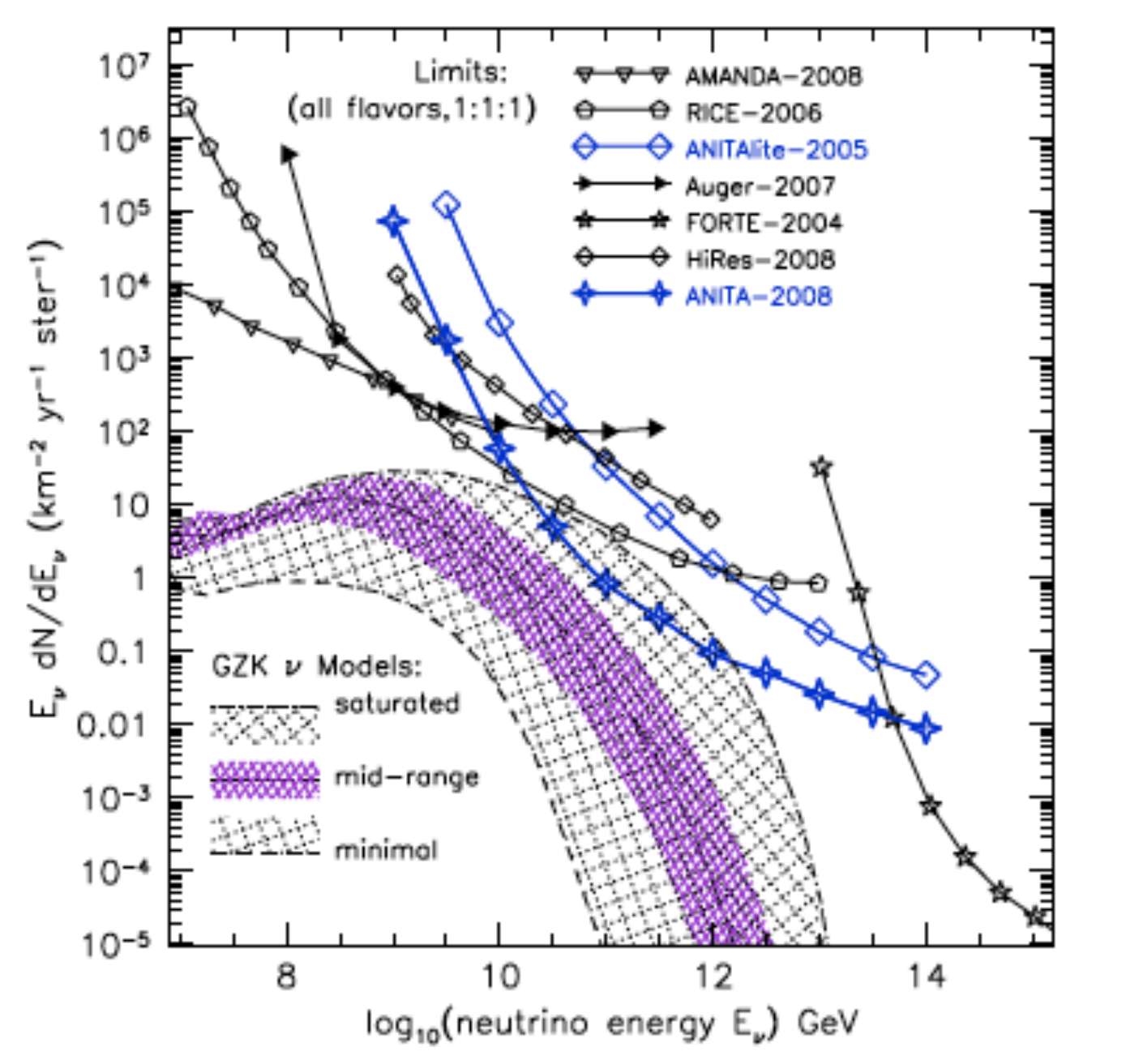}
\end{center}
\vspace{-0.3cm}
\caption{\label{cosmogenic_exp}
Experimental limits on cosmogenic neutrino flux.
Best up-to-date ANITA-1 limits based on no surviving candidates for 18 days of live time shown as ANITA-2008~\cite{ANITA2008}.
Also limits from Auger~\cite{nu_Auger2007}, HiRes~\cite{nu_HiRes}, FORTE~\cite{FORTE}, Anita prototype ANITAlite~\cite{ANITAlite}, RICE~\cite{RICE2006}, and AMANDA II~\cite{AMANDAII}
are shown.
}
\end{figure}

Also in  Fig.~\ref{cosmogenic} we plot two theoretical limits derived under a set of theoretical assumptions.
One is called the Waxman--Bahcall (WB) bound and the other the MPR bound. On the same figure we show several examples
of theoretical neutrino fluxes which violate both WB and MPR bounds, but all 
of them are consistent with the experimental gamma-ray bound.

In Fig. \ref{cosmogenic_exp} we show present-day experimental bounds confronting theoretical predictions for BZ neutrinos
from Ref.~\cite{ANITA2008}. One can see that the best up-to-date experimental bounds come from the ANITA experiment.
ANITA-1 was  able to  view
a volume of ice of $\sim 1.6$ Mkm$^3$ during 17.3 days,  however, volumetric acceptance to
a diffuse neutrino flux, accounting for the small solid angle of
acceptance for any given volume element, is several hundred
km$^3$ water-equivalent steradians at $E_\nu = 10^{19}$ eV. This allowed them for the first time a tough 
theoretically interesting region, excluding part of the parameter space with highest neutrino fluxes.   

On the same figure one can see existing limits on diffuse neutrino flux from the   Auger~\cite{nu_Auger2007}, HiRes~\cite{nu_HiRes}, FORTE~\cite{FORTE}, Anita prototype ANITAlite~\cite{ANITAlite}, RICE~\cite{RICE2006}, and AMANDA II~\cite{AMANDAII} experiments.

Let us note also that in Fig.~\ref{cosmogenic_exp} the composition is assumed to be proton-dominated. If recent Auger results presented in 
Fig.~\ref{composition2009} are confirmed, theoretical expectations for neutrino flux in Fig. \ref{cosmogenic_exp} will be
strongly reduced. This will make observations of the diffused flux of UHE neutrinos an even more complicated issue.
However, at lower energies one still can have a hope of seeing point sources with neutrinos, as will be discussed in the next section.

\subsection{Point sources of UHE neutrinos}
\label{sec:nu_point}

\begin{figure}[htb]
\begin{center}
\includegraphics[width=0.6\textwidth,angle=0]{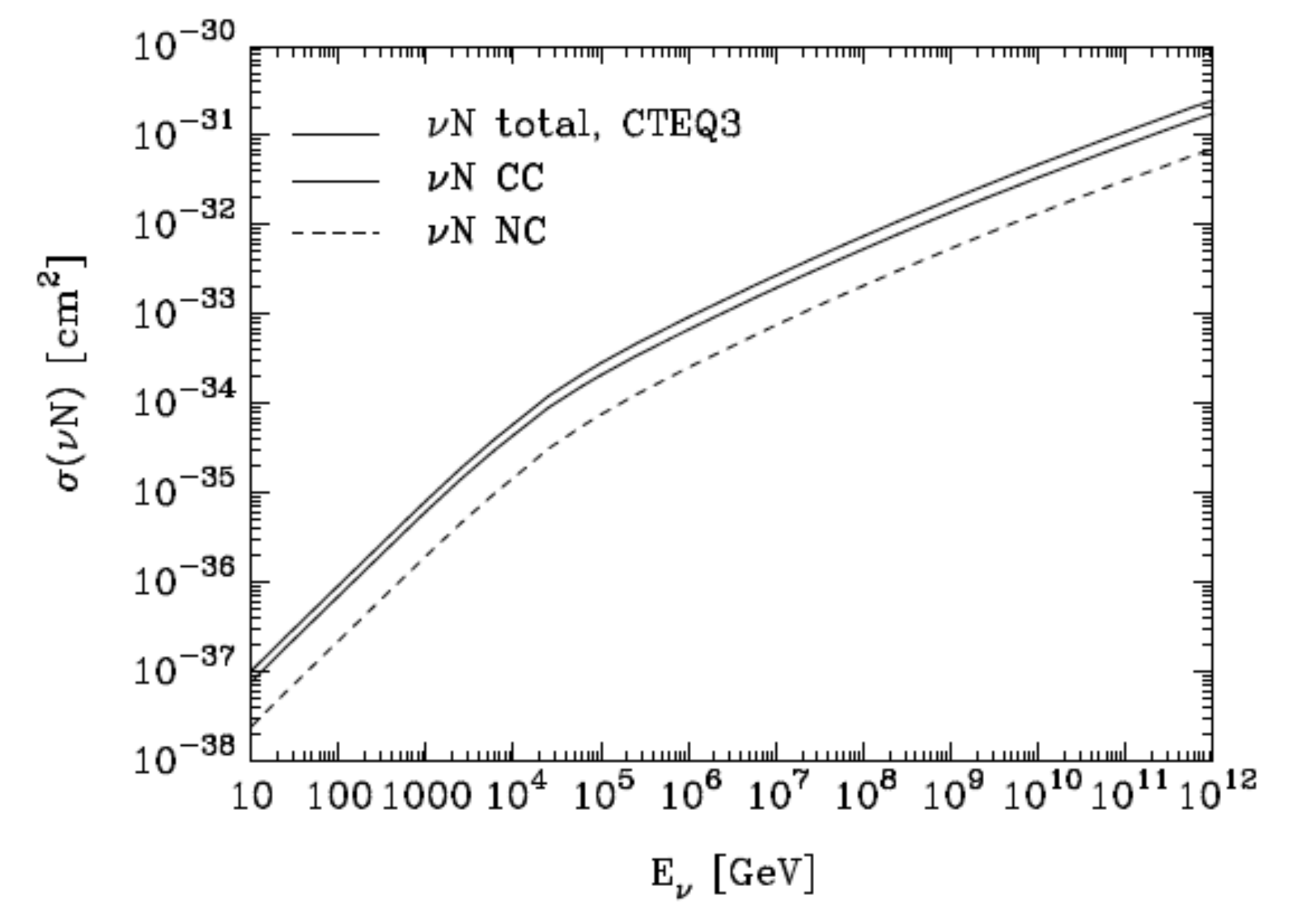}
\end{center}
\vspace{-0.3cm}
\caption{\label{nu_cross_section}
Neutrino--nucleon cross-section as a function of the neutrino energy. Charge-current and neutral-current contributions to the cross-section are shown with thin solid and dashed lines. 
 The total cross-section is presented by a thick solid line. See Ref.~\cite{nu_cross} for details. 
}
\end{figure}

At highest energies the neutrino flux is too low to detect one single source of neutrinos,
but at lower energies $E<1000$ TeV the flux from a single source can be high  enough to 
detect it. Indeed, in Fig.~\ref{nu_cross_section} the neutrino--nucleon cross-section is shown as a function of energy.
This cross-section is proportional to $E$ at low energies $E< 1$ TeV and to $E^{0.4}$ at high energies $E>10^6$ GeV.
Good candidates for neutrino sources in the Galaxy are objects emitting TeV gamma rays. They can produce neutrinos in the proton--proton collisions in objects in the case of binary systems and in the interaction with molecular clouds in the Galaxy.
In the 10 TeV energy range
\begin{equation}
\sigma_{p\nu}(10 \textrm{ TeV})= 10^{-34} ~\mbox{cm}^2~.
\label{sigma_pnu}
\end{equation}

In the IceCube detector only a small fraction of neutrinos will produce a signal:

\begin{equation}
\tau_{\nu}= \sigma_{p\nu} n_{ICE} R  \sim  ~ 10^{-5}~,
\label{tau_nu}
\end{equation}
where $n_{ICE} \sim 10^{24}/\mbox{cm}^3$ is the density of the ice and $R=1$ km is the height of the IceCube detector.

The expected flux of neutrinos produced in the proton--proton collisions in the Galactic sources 
is 
\begin{equation}
F_{\nu} \sim F_\gamma  = 10^{-12} \frac{1}{\mbox{cm}^2 \mbox{s}} \approx 3 \cdot 10^5 \frac{1}{\mbox{km}^2 \mbox{yr}}~.
\label{F_nu}
\end{equation}
Thus in the IceCube detector one can expect three events per year for a $10$ TeV neutrino flux.  

\begin{figure}[htb]
\begin{center}
\includegraphics[width=0.4\textwidth,angle=0]{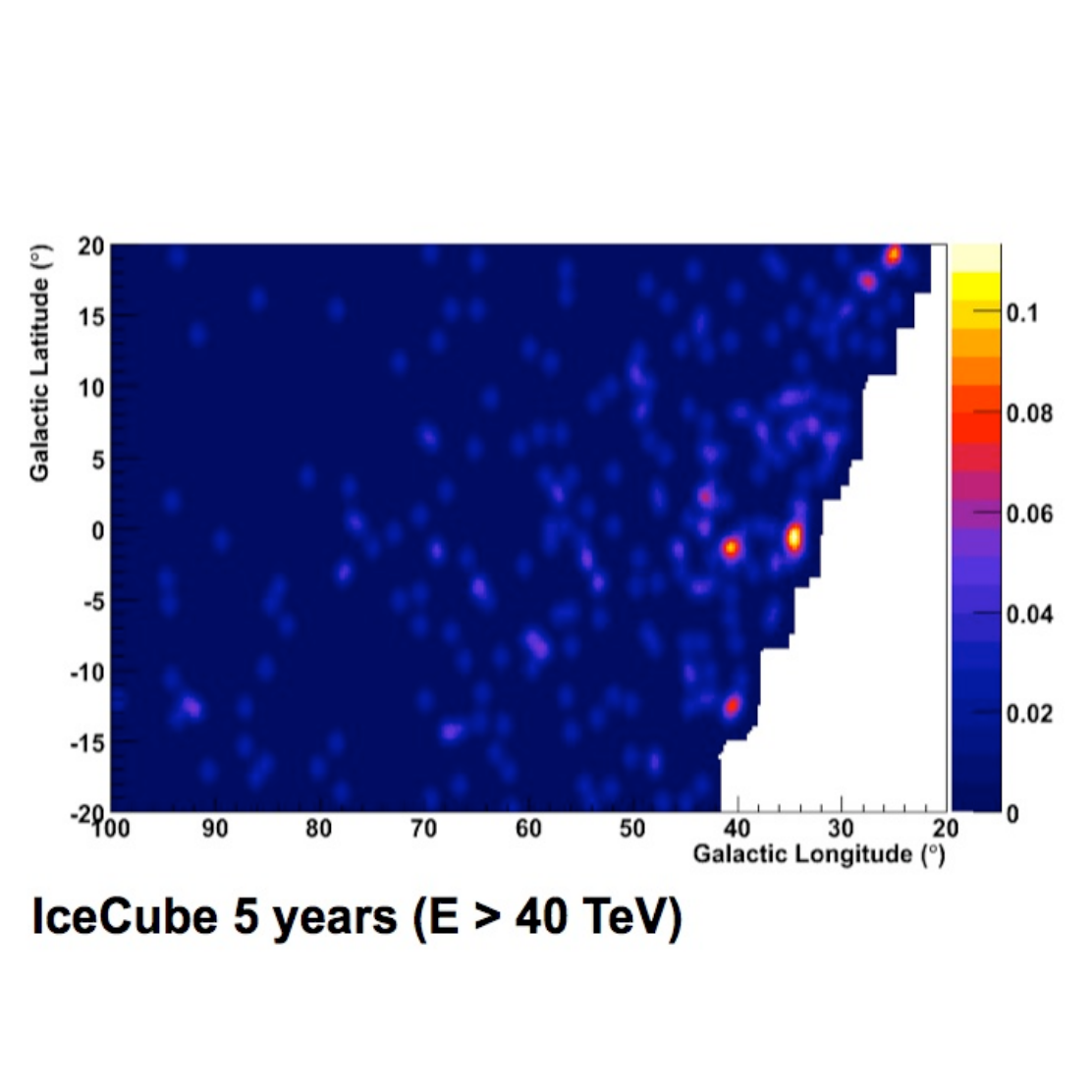}
\includegraphics[width=0.4\textwidth,angle=0]{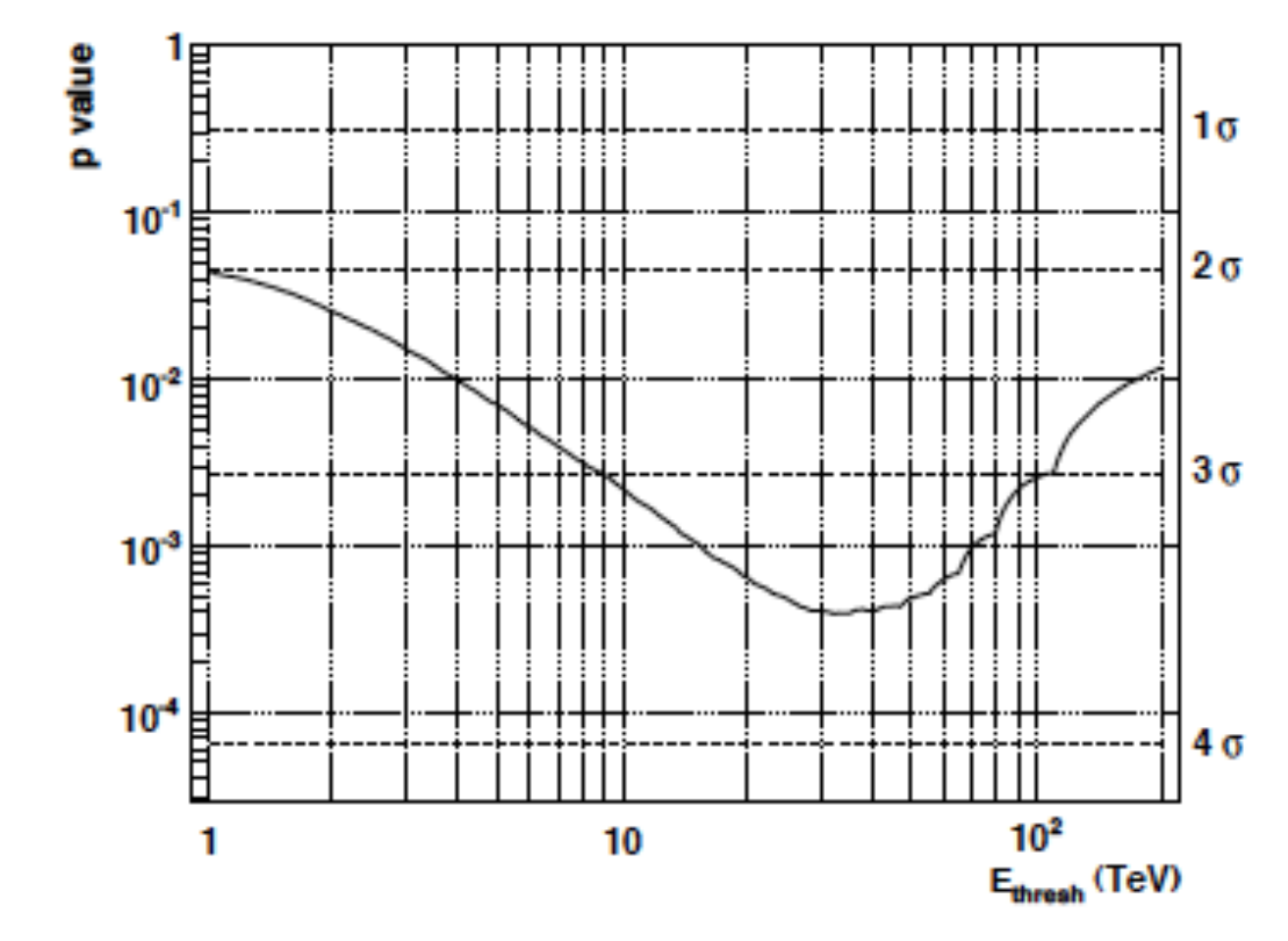}
\end{center}
\vspace{-0.3cm}
\caption{\label{milagro_icecube}
{\bf Left:} Simulated detection of Milagro TeV galactic sources by IceCube. {\bf Right:} 
Significance of Milagro hotspots after five years of observation of IceCube.
}
\end{figure}

In Fig.~\ref{milagro_icecube} one can see a simulation of Milagro sources from Fig.~\ref{milagro} after five years of
working of the IceCube detector.

\begin{figure}[htb]
\begin{center}
\includegraphics[width=\textwidth,angle=0]{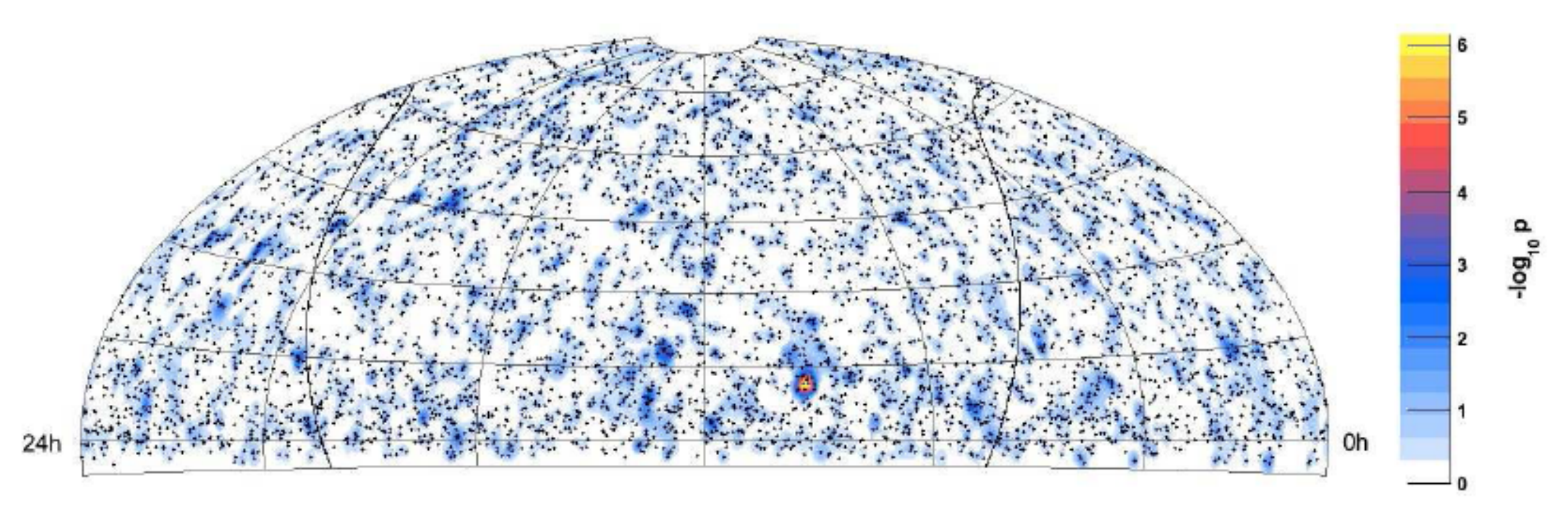}
\end{center}
\vspace{-0.3cm}
\caption{\label{icecube22}
Equatorial sky-map of events (points) and pre-trial significances ({\it p}-value) of the all-sky point source search in the 22-string IceCube detector~\cite{Abbasi:2009iv}. The solid curve is the galactic plane. The most significant spot arrives in a random sky with probability $P \sim 1 \%$.
}
\end{figure}

We now present recent results for point-source searches  using  data recorded during 2007--08 with
22 strings of IceCube (1/4 of the detector). An all-sky search within the declination
range $-5^\circ$ to $+85^\circ$ found the most significant
deviation from the background at $153.4^\circ$ r.a., $11.4^\circ$ dec.
Accounting for all trials in the point-source search, the
final {\it p}-value for this result is 1.34\%, consistent with the
null hypothesis of background-only events at the 2.2$\sigma$ level. No obvious source candidates
are near this location, and an analysis of the timing of the
events did not find any evidence of a burst in time. The
location can be added to the a priori source candidate
list for analysis using future IceCube data, in which case
a similar excess would be identified with much higher
significance~\cite{Abbasi:2009iv}.

\subsection{Summary}
\label{sec:nu_sum}

IceCube is half-complete.  If it observes first sources, a new field of 
astroparticle physics will be started: neutrino astrophysics. 
If not, much bigger detectors are needed with a size of at least 10 km$^3$.
Secondary neutrino flux from UHECR protons can be detected 
by future radio experiments, like ANITA. Neutrinos from some bright galactic sources 
can be detected by IceCube.  Extragalactic sources can be observed  during bright flair activity.
 In order to detect continuous flux from sources like Cen A   one needs detectors much larger than 1~km$^3$.                                                                                                                                                                                                                                                                                                        
Galactic SN can be detected with neutrinos at low and high energies.
Cubic-kilometre water detectors will be constructed if IceCube gives positive results.

\subsection*{Acknowledgements}
I would like to thank the Organizing Committee of the 5th CERN Latin American School for giving me the opportunity 
to present lectures there and for the excellent organization of the School.

\end{document}